\documentclass[11pt]{article}
\PassOptionsToPackage{table}{xcolor}

\usepackage{makecell}
\usepackage[utf8]{inputenc}
\usepackage[T1]{fontenc}
\usepackage{lmodern}
\usepackage{adjustbox}
\usepackage[margin=1in]{geometry}
\usepackage{tikz}
\usepackage{pgfplots}
\pgfplotsset{compat=1.18}
\usepackage{subcaption}
\usepackage{amsmath,amssymb}
\usepackage{bm}
\usepackage{graphicx}
\usepackage{booktabs}
\usepackage{array}
\usepackage[table]{xcolor}
\usepackage[most]{tcolorbox}
\usepackage{caption}
\usepackage{setspace}
\usepackage{longtable}
\usepackage{listings}
\usepackage{placeins}   %
\usepackage[numbers,sort&compress]{natbib}
\usepackage{authblk}
\usepackage[hidelinks]{hyperref}

\newcommand{\indep}{\mathpalette{\independenT}{\perp}}
\newcommand{\independenT}[2]{\mathrel{\rlap{$#1#2$}\mkern2mu{#1#2}}}

\setkeys{Gin}{width=0.9\linewidth,keepaspectratio}
\graphicspath{{./}{supplement/}}
\providecommand{\passthrough}[1]{#1}
\providecommand{\tightlist}{\setlength{\itemsep}{0pt}\setlength{\parskip}{0pt}}
\providecommand{\subsubsubsection}[1]{\par\medskip\noindent\textbf{#1}\par\nobreak\smallskip}
\usepackage{etoolbox}
\AtBeginEnvironment{longtable}{\scriptsize}

\title{\bfseries Does your cost-effectiveness model answer the question of interest? Marginal versus conditional inputs and transportability across populations}

\author[1,2,3,4]{Jeroen P. Jansen}
\author[4,5]{Harlan Campbell}
\author[6]{Shannon Cope}
\author[7]{David M Phillippo}
\author[8,9]{Antonio Remiro-Azócar} 
\affil[1]{Department of Clinical Pharmacy, School of Pharmacy, University of California, San Francisco}
\affil[2]{Associate Member in Cancer Control, UCSF Helen Diller Family Comprehensive Cancer Center, University of California, San Francisco}
\affil[3]{The Philip R. Lee Institute for Health Policy Studies, University of California, San Francisco}
\affil[4]{Precision Medicine Group, Health Economics \& Outcomes Research}
\affil[5]{Department of Statistics, University of British Columbia, Vancouver, Canada}
\affil[6]{Independent Researcher, Vancouver, Canada}
\affil[7]{University of Bristol, UK}
\affil[8]{External Collaboration and Experimentation, Novo Nordisk Pharma, Madrid, Spain}
\affil[9]{Department of Statistical Science, University College London, UK}
\date{\today}

\begin{document}
\maketitle

\begin{abstract}
\noindent\textbf{Objectives:} There has been increased appreciation of the differences between marginal and conditional estimates and different types of effect measures regarding their applicability to different target populations. This issue of transportability is of concern in model-based cost-effectiveness analysis (CEA) when treatment effects from (international) trials are applied to (country-specific) baseline risk estimates. The objective of this paper is to create awareness regarding the issues that arise when using different types of treatment effect and baseline risk estimates in a model-based CEA to inform health technology assessment (HTA).

\noindent\textbf{Methods:} We clarify collapsibility, marginal versus conditional estimation, and transportability; derive the ideal modeling approach implied by a marginal cost-effectiveness estimand; and characterize the issues of common modeling approaches, illustrated with a fictitious state-transition model.

\noindent\textbf{Results:} An individual-level simulation that predicts outcomes from conditional inputs and averages them over the target population targets the marginal cost-effectiveness estimand. Cohort-model approaches that marginalize inputs early, use conditional inputs at mean covariate values, or combine a conditional effect with a marginal baseline (or vice versa) can misstate cost-effectiveness results even with correct-population inputs; inputs from the wrong population add further error.

\noindent\textbf{Conclusions:} The most rigorous approach is an individual-level simulation that carries conditional inputs (baseline risk, treatment effect, prognostic effects and effect modifiers) and marginalizes late. Cohort approaches instead marginalize early, relying on aggregated inputs, and do not necessarily target the marginal cost-effectiveness estimand of interest for HTA.
Model developers should document, for each input, whether it is marginal or conditional and its population.
\end{abstract}

\section{Introduction}

Health Technology Assessment (HTA) is about making decisions for populations. As
such, we are interested in the expected effect of a new intervention in the
target population for decision-making. In jurisdictions with economic evaluations, cost-effectiveness analysis (CEA) is frequently used to inform
HTA. In CEA, we trade off health gains in the target population when
replacing standard of care (SoC) with the new intervention versus the losses in
health due to opportunity costs associated with the new intervention~\citep{drummond2015,neumann2017}.

In HTA, CEA is often performed with health economic decision or simulation
models where multiple sources of evidence related to treatment effects, baseline
risk or mean outcomes with SoC, utilities and costs are integrated and translated
into relevant outcomes for decision-making. For CEA based on country-specific models, 
it is common practice (and recommended) to use country-specific estimates for 
expected (short term) clinical outcomes with SoC (or baseline risk), and thus apply treatment effects of (international) randomized controlled trials (RCTs) to
predict expected outcomes with the new intervention in the country-specific
target population~\citep{welte2004,drummond2009,sculpher2006}.

There has been increased recognition regarding the
differences between marginal and conditional estimates for different types of
effect measures, particularly regarding their applicability to different target populations~\citep{phillippo2025effect, remiro2025marginal, remiro2024transportability}. This
issue of \textit{transportability} is inherent in model-based CEA when
treatment effects from international trials are applied to (country-specific) baseline risk
estimates in the cost-effectiveness model. However, this issue has not been
sufficiently recognized and evaluated~\citep{degtiar2023,dahabreh2020}.

The objective of this paper is to create awareness of the issues that arise when
using different types of treatment effect and baseline risk estimates as 
inputs for model-based CEA, focusing on marginal and conditional estimates of
collapsible and non-collapsible effect measures. To understand the issues, we first highlight key concepts that are fundamental to
the discussion. Next, we define the cost-effectiveness estimand of interest for HTA decision-making and infer implications for the ``ideal model'' and
required clinical input parameter estimates. We then outline
and illustrate the theoretical issues with commonly used cohort approaches for
model-based CEA and the nature of the input parameter estimates. We conclude with
some topline recommendations for developers of model-based CEA.

\section{Marginal and
conditional estimates, collapsible and non-collapsible effect measures, and transportability}\label{sec:marg_vs_cond}

Here we provide a summary of marginal and conditional treatment effect estimates, collapsible and non-collapsible effect measures, and implications for the transportability of the effect measure to different target populations; for an in-depth review see~\citep{colnet2023risk, phillippo2025effect, remiro2024transportability, remiro2025marginal}.

It is important to appreciate the difference between effect modifiers and prognostic factors~\citep{phillippo2016tsd18}. For a given treatment contrast, effect modifiers are characteristics that impact the comparative effect of treatment on the outcome for an individual, on a selected measurement scale. Within the context of said contrast, prognostic factors are characteristics that impact the outcome for an individual independent of treatment. Accordingly, differences in baseline risk across subgroups are driven by differences in the distribution of prognostic factors~\citep{webster2023choice}. Nevertheless, the distinction between effect modifiers and prognostic factors is not always apparent~\citep{webster2023choice}. While effect modifiers are not necessarily prognostic factors, a prognostic factor will inherently act as an effect modifier on at least one measurement scale~\citep{rothman2008modern}. Effect modification is therefore specific to the chosen measurement scale: when treatment has an effect, a prognostic factor that does not act as an effect modifier on the additive risk difference (RD) scale will inherently act as an effect modifier on the multiplicative relative risk (RR) scale~\citep{phillippo2016tsd18, rothman2008modern}.

It is equally important to recognize the difference between marginal and conditional treatment effects. Consider the analysis of an individual comparative study. Marginal estimates of the treatment effect are population-averaged estimates that reflect the expected effect across the entire study population. When treatment is unconfounded (e.g., in a randomized trial), marginal estimates can be directly obtained from a crude unadjusted contrast in outcome means or from an outcome regression without covariates. More generally, covariate-adjusted marginal estimates can be obtained; for instance, by marginalization: fitting a multivariable outcome regression, averaging its predictions under different treatment conditions across the covariate distribution of the study and contrasting the averages. Covariate adjustment is useful to gain precision when treatment is randomized and necessary to control for confounding when it is not. 

Conditional estimates of the treatment effect are estimates evaluated at specific covariate values. An estimate conditioning on a few binary or categorical covariates can be interpreted as a subgroup-specific effect. As a greater number of relevant effect modifiers are conditioned on, it comes closer to an individualized or subject-specific effect. Conditional estimates are obtained by stratification, or from an outcome regression that adjusts for effect modifiers and/or prognostic factors, evaluating the treatment effect at the covariate values of interest — the treatment main effect plus any treatment-by-covariate interaction terms. For covariates that are purely prognostic (no interaction), this reduces to the treatment coefficient itself; for effect modifiers, the conditional effect varies with the covariate values. 

Finally, we make a distinction between collapsible and non-collapsible effect measures~\citep{greenland1999,didelez2022}. Among collapsible effect measures, we further distinguish between those that are directly and indirectly collapsible. \textit{Directly collapsible} effect measures, such as the mean difference (MD) for continuous outcomes and the RD for binary outcomes, have the property that the marginal treatment effect in the overall population can always be expressed as a simple weighted average of subgroup-level conditional effects using population shares as weights \cite{colnet2023risk}. The RR is collapsible but not directly collapsible: the marginal RR is a weighted average of the conditional subgroup-level RRs, but the weights depend on baseline risks rather than population shares \cite{huitfeldt2019}. In contrast, non-collapsible effect measures, such as the odds ratio (OR) and hazard ratio (HR), lack this property~\citep{daniel2021,huitfeldt2019}: even in the absence of effect modification and confounding, when subgroups differ in purely prognostic factors, the constant conditional subgroup-level effects will not equal the overall marginal effect (Online Supplement~\ref{app:collapstransport}).

Accordingly, for non-collapsible effect measures, differences in purely prognostic factors or baseline risk -- in addition to differences in effect modifiers, or in the absence of effect modification -- may undermine transportability of a marginal treatment effect from the study population to a target population for decision-making. Of note, when there is treatment effect heterogeneity, differences in the joint distribution of purely prognostic factors and effect modifiers may also hamper transportability for effect measures that are collapsible but not directly collapsible, such as the RR \cite{colnet2023risk, remiro2024transportability, huitfeldt2019}. For directly collapsible measures, only differences in effect modifiers hamper the transportability of marginal treatment effects~\citep{remiro2024transportability, remiro2025marginal}.

Covariate adjustment can be used to transport marginal treatment effects from a study population to an external target population. Such \textit{population adjustment} approaches typically follow at least one of the following adjustment mechanisms: weighting (e.g., matching-adjusted indirect comparison)~\citep{signorovitch2010, phillippo2018,phillippo2016tsd18} or outcome regression-based standardization (g-computation)~\citep{robins1986,hernan2020book}, with recent extensions developed in the context of network meta-analyses and indirect treatment comparisons~\citep{phillippo2020mlnmr,phillippo2020sim, remiroazocar2022gcomp}. The latter approaches are akin to the marginalization approach described in an earlier paragraph but marginalizing over the covariate distribution of the target population, as opposed to that of the original study itself. 

Transportability analyses for conditional treatment effects are less common than for marginal effects. It is generally recognized that estimates conditioning on specific covariate values can be applied to other individuals or subgroups with the same covariate values. This is provided that the relevant effect modifiers have been conditioned on. It further requires that, within each subgroup, any unmeasured effect modifiers are distributed similarly in the original study population, where the conditional estimates were obtained, and in the target population. For the collapsible effect measures, including the RR, subgroup-level conditional effects are invariant to the distribution of purely prognostic factors, even if these are correlated with the effect modifiers. Conditioning on all effect modifiers is sufficient for the transportability of conditional estimates. Conversely, for the non-collapsible OR and HR, the requirement extends to prognostic factors, since an estimate conditional on the effect modifiers alone remains marginal with respect to said factors. The subgroup-level conditional effects still generally depend on the distribution of purely prognostic factors, which may lead to different baseline risk compositions within each subgroup. As such, conditioning on all effect modifiers between the study and target populations is insufficient: the conditional OR and HR may still fail to transport. Accordingly, non-collapsibility complicates the transportability of conditional effects.

Hence, whether an effect measure is marginal or conditional, and collapsible, directly collapsible or neither, has important implications for transportability~\citep{phillippo2025effect, remiro2024transportability, remiro2025marginal}. For decision modeling, however, these properties may not automatically favor directly collapsible measures. A decision model needs the event probability or expected outcome under each intervention in the target population. Whatever the effect measure, these outcomes depend on the same underlying covariate structure; the choice of measure only determines how that dependence is split between baseline risk and treatment effect. Yet decision models apply treatment effects at baseline risks other than those at which they were estimated, across patients, over time and between settings. For binary outcomes, an OR always yields a valid probability that is bounded between zero and one, whereas an RD may not. What matters is therefore whether the baseline risk and treatment effect inputs, marginal or conditional, match what the model requires, are compatible with each other, and reflect the target population.
\bigskip
\bigskip

\section{Ideal approach for relevant model-based cost-effectiveness analysis:
individual simulation modeling}\label{sec:ind_sim}

In HTA, the objective of decision-making based on cost-effectiveness is to select the
treatment with the greatest expected value for improving population health. This
decision is typically based on the net health benefit (NHB) (or net monetary
benefit) of each intervention under consideration~\citep{stinnett1998,weinstein1977}. The NHB of an intervention is defined as the health it generates minus the health that could have been
generated if the resources spent on that intervention were instead allocated to
other healthcare activities, with both components expressed in health units
such as quality-adjusted life years (QALYs). Formally, for intervention $k$, we define $NHB_{k} = E_{k} - C_{k}/\lambda$, where $E_{k}$ and $C_{k}$ are the total QALYs
and total costs accrued, respectively, and $\lambda$ is the
willingness-to-pay threshold, which converts the opportunity cost of resources
into forgone health~\citep{claxton2015,culyer2016}. Note that NHB is thus an absolute, intervention-specific quantity; the comparison between interventions is
made by ranking these values (equivalently, by their differences). Because total QALYs ($E_{k}$) and total costs accrued ($C_{k}$) may depend on individual patient characteristics, so
too does $NHB_{k}$. Healthcare decision-makers, however, must choose whether to
adopt an intervention for an entire population, not for individual patients with
specific characteristics. When a
health system decides to reimburse or recommend a treatment, that decision applies
broadly to all eligible patients, who have diverse baseline risks and
treatment responses. As such, the cost-effectiveness estimand of interest is
marginal, meaning we seek the average NHB by treatment across the entire target
patient population~\citep{welton2015,remiroazocar2021comment,remiroazocar2022estimands,remiroazocar2022gcomp}.

To obtain this marginal estimate, we must average individual-level predictions of
the NHB -- a function of patient-specific covariate values -- for each of
the compared treatments over the covariate distribution in the target population, yielding population-level marginal NHB estimates by treatment. As
such, the purpose of the health economic model is two-fold: (1) to predict the
individual-level NHB for each individual in the target population by treatment given their covariate values, individual
baseline risk and treatment effects, as well as utility and cost parameters; and
(2) to marginalize the individual NHB predictions over the target population. Of note, the term ``baseline risk'' in this section and Section \ref{issues-cohort} does not necessarily refer to event probabilities in the natural outcome scale, but can refer to baseline mean outcomes across the linear predictor scale determined by a link function.

We can formalize this with the following expression:
\begin{equation}\label{eq:1}
{\overline{NHB}}_{k(P)} = \int_{\mathfrak{X}} \varphi\!\left( g^{-1}\!\left(
\mu + \mathbf{x}\,\left( \beta_{1} + \beta_{2,k} \right) + \gamma_{k}
\right),\ \bm{\theta}_{k} \right) f_{(P)}\!\left( \mathbf{x} \right)\, d\mathbf{x}
\end{equation}
where ${\overline{NHB}}_{k(P)}$ is the average NHB with intervention \emph{k} for
population \emph{P}. $\varphi$ is a non-linear accumulation operator (the
``model structure'') carrying (period-specific) individual event probabilities or (clinical) outcomes through the simulation over the model horizon to reflect the course of disease (e.g. generating the trace of health state occupancy over time as a function of multiple transition rates). $\mu$ represents the baseline value on the scale of the
link function $g()$ for such a probability or outcome, when all covariates
equal zero and under a reference comparator (``no treatment'' or SoC), $\gamma_{k}$ is the treatment effect for intervention \emph{k} versus the reference comparator when all
covariates equal zero, $\mathbf{x}$ represents the vector of covariate values for
each individual, $\beta_{1}$ captures the prognostic effects showing how baseline
risk changes with covariates, $\beta_{2,k}$ represents the treatment-by-covariate, effect-modifying,
interaction terms that describe how treatment effects vary across covariate
values, and $g^{-1}()$ is the inverse link function that transforms the linear predictor  
back to the natural (e.g.\ probability) scale. 
$\mu$, $\gamma_{k}$, $\beta_{1}$ and $\beta_{2,k}$ are conditional parameters,  and $\mu +
\mathbf{x}\,\left( \beta_{1} + \beta_{2,k} \right) + \gamma_{k}$ represents the linear
predictor (akin to what is used in a multivariable generalized linear regression). Throughout this manuscript, we will generally assume that such a linear predictor has a correctly specified functional form that is linear in the covariates, with conditional treatment effects on the linear predictor scale varying linearly with any effect modifiers. Note that eq.~\ref{eq:1} is schematic: it is written for a single event probability or (clinical) outcome, with $\varphi$ taking the corresponding vector of multiple (period-specific) event probabilities and outcomes (that define the model structure) in the general case. None of the arguments that follow depend on the number of event probabilities or outcomes, so this single-outcome notation is retained throughout. 
$\bm{\theta}_{k}$ reflects a broad set of other (conditional) cost-effectiveness model input parameters, such as utility
values, resource use, and unit costs that are used to obtain the QALYs $E_{k}(\mathbf{x})$ and costs $C_{k}(\mathbf{x})$ accrued by an individual with
covariates $\mathbf{x}$ given its course of disease, and returns $NHB_{k}(\mathbf{x}) = E_{k}(\mathbf{x}) -
C_{k}(\mathbf{x})/\lambda$. 
$f_{(P)}\!\left( \mathbf{x} \right)$
represents the covariate distribution in the target population, over which the individual-level $NHB_{k}(\mathbf{x})$ is integrated to obtain the marginal quantity ${\overline{NHB}}_{k(P)}$.

This formalization reveals several essential requirements for implementing an
ideal model-based CEA that produces relevant estimates for population-level
decision-making. First, we require an individual-level simulation model structure
capable of generating patient-specific predictions~\citep{brennan2006,krijkamp2018}. Such a model simulates hypothetical patients one at a time, carrying each patient's covariate values through the model so that outcomes are obtained as a function of individual characteristics. We emphasize that this is a requirement on model \textit{structure}, not on data access: the belief that individual-level simulation presupposes individual participant data (IPD) is mistaken, conflating the level at which the model operates with the level at which its inputs are estimated. An individual-level simulation model requires a (synthetic) population $P$ and appropriate model input parameter estimates. Second, we need model input parameter estimates that appropriately characterize conditional baseline risk
corresponding to covariate values equal to zero under the reference comparator, treatment effects versus the comparator conditional on
covariates at zero, prognostic effects that describe how the baseline value varies with
covariates, and treatment-by-covariate interactions that capture how treatment
effects vary across patient characteristics. Importantly, recognizing that these
parameter estimates are conditional rather than marginal is essential because they
represent the building blocks of our individual outcome predictions. Third, we
must explicitly define the target population $P$ and its covariate distribution $f_{(P)}\!\left( \mathbf{x} \right)$ to ensure our averaging process reflects the actual patient mix that would receive the interventions. Fourth, we average the individual-level model outputs over the
target population to obtain the marginal cost-effectiveness estimate that is
relevant for population-level decision-making. This averaging (carrying conditional
inputs through the model and marginalizing only at the end) is standardization, the
g-computation formula~\citep{robins1986,hernan2020book,remiroazocar2022gcomp} applied
to the NHB rather than to a single treatment effect contrast. This approach ensures that CEA produces marginal estimates that reflect the expected value
of the interventions when applied to the target population.

The use of conditional baseline mean outcomes and treatment effect estimates as health
economic model inputs, combined with explicit modeling of prognostic effects and
treatment effect modifiers, offers substantial advantages for adapting CEA across
different country settings. The key to this lies in the assumption that
conditional parameter estimates hold across different populations. Specifically,
we assume that the conditional baseline outcome when covariates equal zero under no treatment or SoC, the prognostic effects describing how baseline risk varies with patient characteristics, the
treatment effects when covariates equal zero, and the treatment-by-covariate
interactions all remain constant across the study population used to estimate
these and the country-specific target populations. Under this assumption, when we
need to adapt a CEA to a new country-specific target population, we can use, i.e.\
transport, these conditional parameter estimates directly without re-estimation.
The country-specific marginal NHB estimates are then obtained simply by updating
the covariate distribution in the marginalization step to reflect the specific mix
of patient characteristics in the new country-specific target population.

However, this transportability assumption requires careful consideration. The
assumption holds only if there are no unmeasured differences between the study
population used to estimate the model input parameters and the target population
that would affect the baseline or treatment effects beyond those captured by the
covariates and explicitly included in the model. If these populations differ
systematically with respect to other factors not adjusted for in the regression
analysis then it may not be valid to transport these conditional parameter estimates to
the target population. The conditional inputs are conditional only on the covariates
included in the outcome regression, and so remain marginal over any prognostic factors or effect modifiers left unmeasured. The estimated baseline $\mu$ and the other coefficients carry the distribution of unmeasured factors in the study population, and transportability relies on the exchangeability of said distribution with that of the target population. For non-collapsible measures such as the HR or OR, omitting a prognostic factor not only shifts the baseline but also attenuates the treatment effect toward the null, so unmeasured between-population differences can compromise transportability even for a nominally conditional input. Accordingly, country-specific adaptations of the CEA that simply update the covariate distribution in the health economic simulation model may not be sufficient and can remain biased if the model input parameters do not account for all relevant between-population differences, or if the specification of the regression analysis lacks congeniality with the simulation model. When applying conditional estimates from (international) studies to inform country-specific CEA,
analysts must carefully evaluate whether the measured covariates adequately
capture the key sources of heterogeneity between populations, or whether
unmeasured differences are likely to threaten the validity of parameter
transportability.

\section{The issues with cohort-based cost-effectiveness models}\label{issues-cohort}

The most commonly used modeling approach in HTA is the
cohort model, in which expected costs and outcomes are calculated for a hypothetical group represented by a set of inputs, typically by tracking that group through health states over time~\citep{sonnenberg1993,siebert2012,omahony2015}. The defining feature of the cohort model is that, unlike individual-level simulation models (where we average individual-level model outputs to obtain the marginal NHB), it works directly with aggregated population-level -- \textit{population-average} -- inputs, bypassing the need for individual-level simulation. This approach has become widespread because it is computationally efficient and relatively straightforward to implement. However, when cohort models are used in CEA, important conceptual issues arise related to
how baseline risk and treatment effect estimates are incorporated. 

\subsection{The aggregation effect}\label{sec:aggregation}

Before considering whether the baseline risk and treatment effect inputs are marginal or conditional estimates, it is important to recognize a
limitation that applies to most cohort models. Writing
$\pi_{k}\!\left( \mathbf{x} \right) = g^{-1}\!\left( \mu + \mathbf{x}\left(
\beta_{1} + \beta_{2,k} \right) + \gamma_{k} \right)$ for the event probability or expected outcome of an individual
with covariates $\mathbf{x}$ under intervention \emph{k}, the marginal NHB that is
relevant for population-level decision-making averages the NHB over the target population (eq.~\ref{eq:1}) as represented with the left-hand side of the following expression, whereas a cohort model applies the
cost-effectiveness model once, to a single representative set of population-average inputs, as represented with the right-hand side:
\begin{equation}
\int_{\mathfrak{X}} \varphi\!\left( \pi_{k}\!\left( \mathbf{x} \right),\
\bm{\theta}_{k} \right) f_{(P)}\!\left( \mathbf{x} \right) d\mathbf{x}
\;\neq\;
\varphi\!\left( \int_{\mathfrak{X}} \pi_{k}\!\left( \mathbf{x} \right)
f_{(P)}\!\left( \mathbf{x} \right) d\mathbf{x},\ \bm{\theta}_{k} \right).
\label{eqn:aggregation}
\end{equation}

The two sides differ in whether the averaging over the target population takes
place before or after the input parameters are transformed according to the structure of the cost-effectiveness model, as represented by $\varphi()$, and they are not equal because $\varphi()$ tends to be non-linear in its inputs. For example, in a state-transition model this non-linearity lies in the recursion from per-cycle transition probabilities to state occupancy (See Online Supplement~\ref{app:cohorttransitionmodel}). We refer to the resulting difference as an \emph{aggregation effect}.

Three aspects are worth emphasizing. First, the aggregation effect is a property of the cohort structure itself rather than of the type of baseline risk and treatment effect input estimates: it arises whatever these inputs are, including marginal estimates that correctly reflect the target population, and it is unrelated to the choice of effect measure, arising even on a directly collapsible scale such as the RD. Second, it is driven by heterogeneity in the target population and disappears when that population is homogeneous. To illustrate this, note that if all individuals in the population share the same covariate values $\mathbf{x} = \mathbf{x}_0$, the distribution $f_{(P)}(\mathbf{x})$ collapses to a point mass at $\mathbf{x}_0$, and both sides of the inequality in eq. \ref{eqn:aggregation} become equal: $\int_{\mathfrak{X}} \varphi\!\left( \pi_{k}\!\left( \mathbf{x} \right),  \bm{\theta}_{k} \right) f_{(P)}\!\left( \mathbf{x} \right) d\mathbf{x} = \varphi\!\left( \pi_{k}\!\left( \mathbf{x}_0 \right), \bm{\theta}_{k} \right) = \varphi\!\left( \int_{\mathfrak{X}} \pi_{k}\!\left( \mathbf{x} \right) f_{(P)}\!\left( \mathbf{x} \right) d\mathbf{x}, \bm{\theta}_{k} \right)$. Intuitively, when there is no heterogeneity to average over, the order in which the averaging and the transformation are applied is irrelevant. When there is heterogeneity in the target population, the magnitude of the aggregation effect increases with the degree of heterogeneity. Third, the direction of the aggregation effect is determined by the convexity (or concavity) of $\varphi()$ and can be understood through Jensen's inequality. If $\varphi()$ is convex in its inputs, the cohort model underestimates the true marginal NHB: $\varphi\!\left( \int_{\mathfrak{X}} \pi_{k}\!\left( \mathbf{x} \right) f_{(P)}\!\left( \mathbf{x} \right) d\mathbf{x}, \bm{\theta}_{k} \right) \leq \int_{\mathfrak{X}} \varphi\!\left( \pi_{k}\!\left( \mathbf{x} \right), \bm{\theta}_{k} \right) f_{(P)}\!\left( \mathbf{x} \right) d\mathbf{x}$. Conversely, if $\varphi()$ is concave in its inputs, the cohort model overestimates the true marginal NHB: $\varphi\!\left( \int_{\mathfrak{X}} \pi_{k}\!\left( \mathbf{x} \right) f_{(P)}\!\left( \mathbf{x} \right) d\mathbf{x}, \bm{\theta}_{k} \right) \geq \int_{\mathfrak{X}} \varphi\!\left( \pi_{k}\!\left( \mathbf{x} \right), \bm{\theta}_{k} \right) f_{(P)}\!\left( \mathbf{x} \right) d\mathbf{x}$. In practice, the convexity of $\varphi()$ in a cost-effectiveness model and the relationship between $\varphi()$ and its inputs are likely to be complex, so that the overall impact on NHB depends on the specific values of transition probabilities, utilities, costs and time horizon, and may vary across different regions of the input space. 

One exception of the aggregation effect is instructive. NHB is a linear functional of state occupancy, so a cohort model constructed directly from the true marginal survival or state-occupancy curves, as in a partitioned survival cost-effectiveness model, avoids the aggregation effect. Those curves are themselves obtained by averaging individual trajectories (Online Supplement~\ref{app:aggreffectpsm}). 

The issues described in the following sections, which arise from whether the baseline risk and treatment effect inputs are marginal or conditional estimates, are incurred in addition to the aggregation effect rather than instead of it.

\subsection{Cohort models with marginal treatment effects and baseline risk}\label{sec:marg_cohort}

A cohort model with marginal estimates for the baseline risk and treatment effect
input parameters can be expressed as follows:
\begin{equation}\label{eq:2}
{\overline{NHB}}_{k(P)} = \varphi\!\left( g^{-1}\!\left( g\!\left(
{\overline{\pi}}_{0(P)} \right) + \Delta_{k(P)} \right),\ \bm{\theta}_{k} \right)
\end{equation}
where ${\overline{\pi}}_{0(P)}$ is the marginal estimate of the baseline risk under the reference comparator (no treatment or SoC) on
the natural scale in population \emph{P}, and $\Delta_{k(P)}$ is the marginal
estimate of the treatment effect for intervention \emph{k} versus the reference comparator on the linear predictor
scale in population \emph{P}. Please note, in eq.~\ref{eq:2}, the model input parameters $\bm{\theta}_{k}$ for utility values, resource use and costs are also marginal. 

For this cohort model to provide relevant  estimates for the target population \emph{P} (setting aside the aggregation effect described in Section~\ref{sec:aggregation}), both the baseline risk and treatment effect estimates used as model inputs need to reflect population \emph{P}. 

For the baseline risk, this means that ${\overline{\pi}}_{0(P)}$ can only be obtained from a study  -- or pool of studies -- that do not differ from the target population  \emph{P} in any factors that are associated with the outcome (under the reference comparator). Alternatively, with estimates for $\mu$ and the vector of
prognostic effects $\beta_{1}$, the marginal estimate for the baseline risk in the
target population \emph{P} can be obtained according to:
\begin{equation}\label{eq:3}
{\overline{\pi}}_{0(P)} = \int_{\mathfrak{X}} g^{-1}\!\left( \mu +
\mathbf{x}\beta_{1} \right) f_{(P)}\!\left( \mathbf{x} \right)\, d\mathbf{x}.
\end{equation}

Estimates $\mu$ and $\beta_{1}$ need to be obtained from studies that do not differ from the target population \emph{P} regarding any factors that are associated with the outcome \textit{not} captured by
$\mu$ and $\beta_{1}$. Any differences may result in biased estimation of the marginal
baseline risk parameter ${\overline{\pi}}_{0(P)}$ in target population \emph{P}. 

Appropriate treatment effect estimates for  $\Delta_{k(P)}$ can only be obtained from studies that do not differ from the target population \emph{P} regarding the joint distribution of any (purely) prognostic factors and effect modifiers. This is due to the dependence of marginal measures on such joint distribution, which can only be relaxed to a dependence on effect modifiers for directly collapsible absolute difference measures (the MD and RD). Alternatively, with estimates for $\mu$, $\gamma_{k}$, $\beta_{1}$, and $\beta_{2,k}$, the marginal treatment effect estimate can be obtained for target population \emph{P} according to:
\begin{equation}\label{eq:4}
\begin{aligned}
\Delta_{k(P)} &= g\!\left( {\overline{\pi}}_{k(P)} \right) - g\!\left(
{\overline{\pi}}_{0(P)} \right) \\
&= g\!\left( \int_{\mathfrak{X}} g^{-1}\!\left( \mu + \mathbf{x}\,\left(
\beta_{1} + \beta_{2,k} \right) + \gamma_{k} \right) f_{(P)}\!\left( \mathbf{x}
\right) d\mathbf{x} \right) \\
&\quad - g\!\left( \int_{\mathfrak{X}} g^{-1}\!\left( \mu + \mathbf{x}\beta_{1}
\right) f_{(P)}\!\left( \mathbf{x} \right) d\mathbf{x} \right).
\end{aligned}
\end{equation}
Estimates $\mu$, $\gamma_{k}$, $\beta_{1}$, and $\beta_{2,k}$ need to come from studies that do not differ from the target population regarding the joint distribution of prognostic factors and effect modifiers \textit{not} captured by these parameters. For directly collapsible measures, only differences in effect modifiers -- not captured within $\beta_{2,k}$ -- would be problematic. 

Because marginal estimates are inherently population-specific, combining a marginal treatment effect from an international trial with a marginal baseline taken from a local registry, as is common practice for country-specific CEA, may result in a model that provides an estimate of limited validity, even when each input is a sound marginal estimate within its own source population. 

A specific difficulty to note with this approach for time-to-event outcomes (e.g.\
progression free survival and overall survival in a cancer model) is that the
marginal HR is time-varying whenever covariate effects are present~\citep{phillippo2025effect} (see Online Supplement \ref{app:tvhr}). That is, even if the conditional HR is constant such that the proportional hazards assumption holds at the individual or subgroup level. Consequently, a single marginal HR resulting from an unadjusted Cox proportional hazards model, as frequently reported in the literature, cannot necessarily be applied as a constant treatment effect input parameter in a cost-effectiveness model~\citep{hernan2010,aalen2015,martinussen2013,jansen2011,jansen2012,ouwens2010}.

\subsection{Cohort models with population-average conditional treatment effects
and baseline risk}\label{sec:42}

A cohort model with conditional estimates for the baseline risk and treatment
effect input parameters can be expressed as:\begin{equation}\label{eq:5}
{\overline{NHB}}_{k(P)} = \varphi\!\left( g^{-1}\!\left( m_{0(P)} + d_{k(P)}
\right),\ \bm{\theta}_{k} \right)
\end{equation}
where $m_{0(P)}$ is the population-average conditional baseline risk and $d_{k(P)}$
the population-average conditional treatment effect, both on the linear predictor scale imposed by link function $g()$. Here, by \textit{population-average}, we are referring to the average conditional measure across all the subjects or subgroups in population $P$, with the averaging taking place on the linear predictor scale  (as opposed to the natural outcome scale used to calculate population-average marginal quantities)~\citep{remiro2025marginal}. Because we have assumed that the linear predictor is a linear function of the covariates, these population-average conditional quantities are equal to the corresponding conditional measures evaluated at the mean covariate values~\citep{remiro2025marginal}. Namely, the model input parameters can be obtained according to: $m_{0(P)} = \mu +
{\overline{\mathbf{x}}}_{(P)}\beta_{1}$ and $d_{k(P)} =
{\overline{\mathbf{x}}}_{(P)}\beta_{2,k} + \gamma_{k}$, respectively, using the
estimates for $\mu$, $\beta_{1}$, $\gamma_{k}$, and $\beta_{2,k}$ (see Section \ref{sec:marg_cohort})
in combination with the covariate mean values ${\overline{\mathbf{x}}}_{(P)}$ in the target population
\emph{P}. Note that where the linear predictor is itself a non-linear function of the covariates (e.g., through spline or polynomial terms) -- more specifically, where conditional treatment effects on the linear predictor scale vary non-linearly with any effect modifiers -- the population-average conditional quantities and the conditional measures evaluated at the mean covariate values generally differ as well, even under an identity link. 

The issue with the cohort modeling approach according to Equation \ref{eq:5}, using population-average conditional estimates for the baseline and treatment effect input parameters (and setting aside the aggregation effect), is that there is a mismatch between these conditional quantities and the marginal inputs required by a cohort-based cost-effectiveness model. 

Firstly, the population-average conditional baseline value $m_{0(P)}$ does not generally correspond to the marginal baseline value on the linear predictor scale, i.e., $m_{0(P)} \neq g\!\left({\overline{\pi}}_{0(P)}\right)$; or equivalently on the natural outcome scale, $g^{-1}\!\left(m_{0(P)}\right) \neq {\overline{\pi}}_{0(P)}$. If the inverse link function $g^{-1}()$ is non-linear, the risk obtained by transforming the average linear predictor differs from the average of the risks obtained by transforming each individual's linear predictor:
\begin{equation}
g^{-1}\!\left( \int_{\mathfrak{X}} \left(\mu + \mathbf{x}\beta_{1}\right) f_{(P)}\!\left( \mathbf{x} \right) d\mathbf{x} \right) \neq \int_{\mathfrak{X}} g^{-1}\!\left( \mu + \mathbf{x}\beta_{1} \right) f_{(P)}\!\left( \mathbf{x} \right) d\mathbf{x}.
\label{eqn:convex-effect}
\end{equation}
As for the aggregation effect in Section~\ref{sec:aggregation}, this divergence arises from Jensen's inequality, but here it concerns the outcome regression model rather than the cost-effectiveness model, and can therefore be incurred in addition to the aggregation effect. Similarly, the direction of impact of the divergence in eq.~\ref{eqn:convex-effect} is determined by the convexity of $g^{-1}()$. If this is convex in its inputs, the population-average conditional baseline risk underestimates the marginal baseline risk: $g^{-1}\!\left( m_{0(P)} \right) \leq {\overline{\pi}}_{0(P)}$. Conversely, if it is concave in its inputs, the population-average conditional baseline risk overestimates the marginal baseline risk: $g^{-1}\!\left( m_{0(P)} \right) \geq {\overline{\pi}}_{0(P)}$. For example, for a logistic outcome regression model, $g^{-1}()$ is the inverse logit function converting log odds on the linear predictor scale to probabilities on the natural outcome scale. This function is convex for negative values of the linear predictor (probability below 0.5) and concave for positive values of the linear predictor (probability above 0.5). The mismatch between population-average conditional and marginal baseline risks for the log and logit link functions is demonstrated in Online Supplement~\ref{sec:baseline}. 

Secondly, the population-average conditional treatment effect $d_{k(P)}$ does not generally equal the marginal treatment effect $\Delta_{k(P)}$ in eq.~\ref{eq:4}. As discussed above for the baseline risk, the two quantities differ in their order of operations, or the scale on which the averaging takes place~\citep{phillippo2025effect}. The marginal treatment effect takes the unconditional expectation of outcomes on their natural scale, then contrasts the transformed averages on the linear predictor scale imposed by the link function. Conversely, the population-average conditional treatment effect contrasts conditional outcome expectations on the linear predictor scale, then takes the average of the contrasts. The two measures only commute when $g()$ is the identity link; otherwise the changing order of operations is relevant, and $d_{k(P)} \neq \Delta_{k(P)}$ in general. Here we have Jensen's inequality operating across outcomes for two different treatment arms, such that the overall impact depends on the differential convexity (or concavity) between the two arms. 

The mismatch between population-average conditional and marginal treatment effects for the log and logit link functions is further explained in Online Supplement~\ref{sec:trt_effect}. In brief, under the log link function and the (log) RR scale: without effect modification, the population-average conditional and marginal treatment effects are identical, owing to collapsibility, but this is not necessarily the case where there is effect modification. Under the logit link function and the non-collapsible (log) OR scale: without effect modification, population-average conditional treatment effects overestimate the benefit of interventions relative to marginal treatment effects, but the discrepancy can go in either direction when there is effect modification. Similar findings have been demonstrated in the time-to-event setting for proportional hazards models and the non-collapsible (log) HR scale (and illustrated in the context of an example case study in Section \ref{sec:example}),  with recent research highlighting that effect modification can result in conflicting treatment rankings between population-average conditional and marginal estimates for non-collapsible effect measures~\citep{phillippo2025effect}. Of note, conflicting treatment recommendations may also arise for the (log) RR, which is collapsible, under effect modification (Online Supplement \ref{sec:trt_effect_log}). 

Both aspects described in this section -- for the baseline risk and the treatment effect -- manifest a common underlying problem: incompatibilities between the population-average conditional output of the outcome regression model and the marginal input required by the cohort-based cost-effectiveness model. We shall use the term \textit{cross-model scale mismatch} to refer to this systematic discrepancy. 

\subsection{Cohort models with population-average conditional treatment effects
and marginal baseline risk, or vice versa}\label{sec:43}

When partitioned survival models are used for model-based CEA, e.g.\ in oncology,
we often see that a conditional HR is applied to a marginal survival curve for the
reference treatment. Such a cohort-based CEA that combines a marginal baseline
estimate with a conditional treatment effect estimate can be
expressed as follows:
\begin{equation}\label{eq:6}
{\overline{NHB}}_{k(P)} = \varphi\!\left( g^{-1}\!\left( g\!\left(
{\overline{\pi}}_{0(P)} \right) + d_{k(P)} \right),\ \bm{\theta}_{k} \right).
\end{equation}
This combination arises naturally in practice because marginal survival curves (e.g., Kaplan-Meier curves) and conditional HRs from covariate-adjusted Cox proportional hazards models are routinely reported in clinical trial publications. Conversely, a cohort
approach in which a marginal treatment effect is applied to a population-average
conditional baseline results in the following formulation:
\begin{equation}\label{eq:7}
{\overline{NHB}}_{k(P)} = \varphi\!\left( g^{-1}\!\left( m_{0(P)} + \Delta_{k(P)}
\right),\ \bm{\theta}_{k} \right).
\end{equation}
Neither eq.~\ref{eq:6} nor eq.~\ref{eq:7} will produce valid effectiveness inputs for the cohort-based CEA because they combine incompatible marginal and conditional quantities, producing an ambiguous measure that does not correspond to any meaningful or interpretable estimate. 

To see this, note that generally for non-identity link functions, the ``effective risk'' under intervention $k$ implied by eq.~\ref{eq:6} is neither the risk obtained by applying the marginal treatment effect to the marginal baseline risk nor the risk obtained by applying the population-average conditional treatment effect to the conditional baseline risk: $g^{-1}\!\left( g\!\left({\overline{\pi}}_{0(P)}\right) + d_{k(P)} \right)
\neq
g^{-1}\!\left( g\!\left({\overline{\pi}}_{0(P)}\right) + \Delta_{k(P)} \right)$ because $d_{k(P)} \neq \Delta_{k(P)}$ and $g^{-1}\!\left( g\!\left({\overline{\pi}}_{0(P)}\right) + d_{k(P)} \right)
\neq g^{-1}\!\left( m_{0(P)} + d_{k(P)} \right)$ because $g({\overline{\pi}}_{0(P)}) \neq m_{0(P)} $, as per Section~\ref{sec:42}. The same applies for the ``effective risk'' implied by eq.~\ref{eq:7}. In both cases, the resulting input is an ambiguous quantity that is neither fully marginal nor fully conditional. 

Both formulations carry the aggregation effect of Section~\ref{sec:aggregation}, and propagate estimand incompatibility issues at an even earlier stage than the formulations in Section \ref{sec:42}. The cross-model scale mismatch still applies, either between whichever component of the inputs is a population-average conditional quantity (the treatment effect for eq.~\ref{eq:6} and the baseline for eq.~\ref{eq:7}), or between the overall ``effective risk''. Of note, while partitioned survival models avoid the aggregation effect, they typically combine parametric fits to Kaplan-Meier-based marginal survival curves with Cox model-based conditional HRs, and are particularly prone to the issues described in this section.

\section{Illustrative example}
\label{sec:example}

\subsection{Model structure, target populations, inputs, and modeling scenarios}

We illustrate the issues described so far using a cost-effectiveness model for advanced cancer,
comparing a new treatment versus SoC; full model details and R
code are provided in the Online Supplement (~\ref{app:handcoded}--\ref{app:hesim-indiv}).
The model is a state-transition model with three health states (Stable, Progressed,
Death) over a 30-year time horizon, by which time essentially all patients have died.
The progression rate is modeled using a Weibull proportional hazards model with a linear predictor that is linear in the covariates, age and
ECOG performance status, and a treatment indicator (Table~\ref{tab:example-coef}). The Weibull shape parameter is $\nu = \exp(0.150) \approx 1.16$, indicating a slightly increasing hazard over time.

Age is a prognostic factor; ECOG status is both a prognostic factor and an effect
modifier on the (log) HR scale, so that the treatment benefit is larger in ECOG~0 patients (conditional HR $= 0.33$) than in ECOG~1 patients (conditional HR
$= 0.52$). The treatment also reduces mortality while in the Stable state (conditional HR $= 0.45$), reflecting an overall survival benefit beyond progression-free survival; this treatment effect on mortality is assumed to be constant across
patients. Because this effect is not modified by any covariate, its population-average
conditional counterpart also equals 0.45; the marginal counterpart
drifts modestly, to at most $0.469$, because age is prognostic and the two arms
deplete at different rates. Mortality after progression is three times the
Stable-state background rate and carries no treatment effect.

We consider two populations: a trial population (Population~A: mean age 60, $30\%$
ECOG~1) and an older population (Population~B: mean age 71, $70\%$ ECOG~1). Each
can serve as the \emph{target population} for a decision. Inputs derived from the
target population itself are reported as ``matched''; inputs derived from trial
Population~A and applied to a decision in Population~B are reported as ``mismatched''.

Cost-effectiveness is computed for both populations using an individual-level simulation approach (eq.~\ref{eq:1}) and four alternative cohort-model approaches (eqs.~\ref{eq:2}, \ref{eq:5}, \ref{eq:6} and \ref{eq:7}), all with the same state-transition structure. These modeling scenarios also differ in whether the baseline risk -- here ``risk'' refers to an event probability -- and treatment effect inputs are conditional or marginal.

The conditional estimates in Table~\ref{tab:example-coef} are the building blocks describing how baseline risk and the treatment effect depend on covariates in any individual, with all other factors held constant, and are the inputs used directly by the individual-level model (Scenario~1) for target populations A and B. The subgroup-level conditional progression HRs are assumed identical and constant over time in both populations ($0.33$ for ECOG~0 and $0.52$ for ECOG~1). The population-average conditional and marginal quantities used in the cohort scenarios are population-specific (Table~\ref{tab:example-inputs}).

For a cohort-based state-transition model, marginal inputs for the baseline risk can be constructed in two ways, and they are not equivalent.
One can average each cycle's transition probability over the covariate distribution,
holding the weights fixed at their baseline values; or one can average the survival
function over that distribution and difference it. The second is the more realistic
construction, since it corresponds to fitting a curve to a marginal Kaplan--Meier curve and
differencing it, and its weights are implicitly those of the patients still at
risk in each cycle. The first holds the baseline covariate mix fixed for the whole
horizon and so continues to charge the cohort with the hazard of the starting mix long after the
high-risk patients have progressed out. The two approaches agree only in the first cycle, and we
use the second approach throughout.

For the marginal measures, used by Scenario~2, the survival function is therefore
averaged over the covariate distribution of the target population and then differenced:
\begin{equation*}
\overline{S}_{0(P)}(t) = \int_{\mathfrak{X}} \exp\!\left(-\exp\!\left(\mu +
\mathbf{x}\beta_{1}\right)\left(\tfrac{t}{12}\right)^{\nu}\right)
f_{(P)}\!\left(\mathbf{x}\right) d\mathbf{x}, \qquad
\overline{\pi}_{0(P)}(t) = 1 - \frac{\overline{S}_{0(P)}(t)}{\overline{S}_{0(P)}(t-1)}.
\end{equation*}
This is eq.~\ref{eq:3} applied to the cumulative risk, evaluated as a weighted sum over
a discrete grid of age and ECOG profiles rather than by simulation. The resulting series
rises and then falls --- from $0.045$ to $0.060$ by cycle~6 and down to $0.044$ by
cycle~360 in Population~A, and from $0.159$ to $0.186$ by cycle~3 and down to $0.049$ in
Population~B --- because the patients at highest risk progress out of the Stable state
first, so the survivors are progressively enriched with low-risk patients. The marginal
progression HR varies over time for the same reason: it rises from $0.440$ to $0.454$ by
cycle~12 in Population~A and from $0.503$ to $0.579$ in Population~B, then falls back to
$0.406$ and $0.458$ respectively, as the two arms' at-risk sets deplete at different
rates. It is this per-cycle effect, rather than a single summary value, that the cohort
model uses (Online Supplement~\ref{app:tvhr}).

For the population-average conditional measures, used by Scenario~3, the conditional
model is instead evaluated at the mean covariates of the population:
\begin{equation*}
g^{-1}\!\left(m_{0(P)}(t)\right) = 1 - \exp\!\left(-\exp\!\left(\mu +
\overline{\mathbf{x}}_{(P)}\beta_{1}\right)\left(\left(\tfrac{t}{12}\right)^{\nu} -
\left(\tfrac{t-1}{12}\right)^{\nu}\right)\right).
\end{equation*}
Because the conditional log hazard ratio is assumed to vary linearly with the only effect modifier (ECOG), the population-average conditional hazard ratio and the conditional hazard ratio at the mean covariate values are equivalent. For the Weibull model here, $g()$ is the complementary log-log of the risk,
$g(\pi)=\log(-\log(1-\pi))$, so that $g^{-1}()$ returns the risk directly, as in
eq.~\ref{eq:3}, and $m_{0(P)}(t)$ is the complementary log-log of the conditional
progression probability in cycle $t$ at the mean covariate values. Unlike the marginal
inputs, there is no ambiguity in the construction of conditional inputs for the baseline risk; for a single initial covariate vector the
per-cycle probabilities chain exactly to the Weibull survival function. Because that vector is carried
unchanged throughout, however, there is no population to deplete, and the probability
rises monotonically with the Weibull hazard: from $0.038$ at cycle~1 to $0.109$ at
cycle~360 in Population~A, and from $0.138$ to $0.361$ in Population~B. It therefore
starts below the marginal series --- averaging over a heterogeneous population raises the
early event probability --- crosses above it within the first year, and diverges from it
steadily thereafter (Table~\ref{tab:example-inputs}). The population-average conditional
progression HR is constant over time and calculated using the linear predictor
$\exp(0.7 \times -1.10 + 0.3 \times (-1.10+0.45))=0.38$ versus
$\exp(0.3 \times -1.10 + 0.7 \times (-1.10+0.45))=0.46$, respectively.

As illustrated in Table~\ref{tab:example-inputs}, there are important differences between the population-average conditional and
marginal progression probabilities and between the population-average conditional and
marginal progression HRs, even within the same trial population.
Table~\ref{tab:example-scenarios} summarizes the scenarios, which estimands are targeted
by each scenario for the baseline risk and the treatment effect, and which issues each
scenario is subject to.

In Scenarios~4 and~5, the new treatment arm is obtained by applying the treatment effect to the
baseline on the hazard scale, cycle by cycle. In Scenarios~2 and~3 both arms are instead
derived directly from the underlying regression model, which is algebraically
equivalent. At cycles~1, 12 and~60 the marginal progression probability under the new
treatment in Scenario~2 is $0.020$, $0.027$ and $0.020$ in Population~A, and $0.083$,
$0.081$ and $0.039$ in Population~B. The corresponding population-average conditional
probabilities in Scenario~3 are $0.015$, $0.025$ and $0.032$, and $0.066$, $0.111$ and
$0.142$. The arms for the new treatment show the same divergence as the SoC ones: the marginal
series turns downward while the population-average conditional series continue to rise.
Both arms are then run through the same state-transition structure.

\begin{table}[!htb]
\centering
\caption{Conditional estimates related to disease progression and mortality for the individual-level simulation model inputs.} 
\label{tab:example-coef}
{\footnotesize
\begin{tabular}{@{}llrp{6.5cm}@{}}
\toprule
\textbf{Transition} & \textbf{Parameter} & \textbf{Coef.} & \textbf{Interpretation} \\
\midrule
Progression (Stable$\to$Prog.) & Intercept              & $-5.500$ & Baseline on the complementary log-log (log cumulative hazard) scale \\
Progression                    & Age                    & $0.080$  & Prognostic factor (HR = 1.08 per year) \\
Progression                    & ECOG 1                 & $1.100$  & Prognostic factor (HR = 3.00 vs.\ ECOG 0) \\
Progression                    & Treatment              & $-1.100$ & Treatment effect at ECOG 0 (HR = 0.33) \\
Progression                    & Treatment $\times$ ECOG 1 & $0.450$ & Effect modification: treatment effect at ECOG 1 (HR = 0.52) \\
Progression                    & $\ln(\nu)$             & $0.150$  & Weibull shape $\nu = \exp(0.150) \approx 1.16$ (slightly increasing hazard) \\
\addlinespace
Stable$\to$Death               & Treatment              & $-0.799$ & HR = 0.45 while on treatment (Stable) \\
Stable$\to$Death               & Background             & n/a      & Age-banded life table (risk rises with age) \\
Progressed$\to$Death           & n/a                    & n/a      & $3\times$ the Stable-state rate; no treatment effect \\
\bottomrule
\end{tabular}
}
\end{table}

\begin{table}[!htb]
\centering
\caption{Marginal and population-average conditional inputs used for the cohort models,
derived from the (shared) conditional estimates of Table~\ref{tab:example-coef}. The
baseline risk is the per-cycle progression probability under SoC; the treatment effect
is the progression hazard ratio (HR) for the new treatment versus SoC. All quantities
are specific to the population of the column group. Cycles are monthly, so cycle~12 is
one year and cycle~360 is thirty years. The models consume the full 360-cycle vectors;
the cycles shown span the whole time horizon, although few patients remain in the
Stable state at later cycles.}
\label{tab:example-inputs}
{\footnotesize
\begin{tabular}{@{}rrrrrrrrr@{}}
\toprule
& \multicolumn{4}{c}{\textbf{Population A (trial, target)}}
& \multicolumn{4}{c}{\textbf{Population B (target)}} \\
\cmidrule(lr){2-5}\cmidrule(l){6-9}
& \multicolumn{2}{c}{Baseline risk} & \multicolumn{2}{c}{Treatment effect\textsuperscript{a}}
& \multicolumn{2}{c}{Baseline risk} & \multicolumn{2}{c}{Treatment effect\textsuperscript{a}} \\
\cmidrule(lr){2-3}\cmidrule(lr){4-5}\cmidrule(lr){6-7}\cmidrule(l){8-9}
\textbf{Cycle}
& \makecell{Marg.\\$\overline{\pi}_{0}(t)$}
& \makecell{Cond.\\$g^{-1}(m_{0}(t))$}
& \makecell{Marg.\\$\exp(\Delta_{k}(t))$}
& \makecell{Cond.\\$\exp(d_{k})$}
& \makecell{Marg.\\$\overline{\pi}_{0}(t)$}
& \makecell{Cond.\\$g^{-1}(m_{0}(t))$}
& \makecell{Marg.\\$\exp(\Delta_{k}(t))$}
& \makecell{Cond.\\$\exp(d_{k})$} \\
\midrule
  1 & 0.045 & 0.038 & 0.440 & 0.381 & 0.159 & 0.138 & 0.503 & 0.456 \\
  3 & 0.057 & 0.051 & 0.445 & 0.381 & 0.186 & 0.182 & 0.524 & 0.456 \\
  6 & 0.060 & 0.057 & 0.450 & 0.381 & 0.172 & 0.204 & 0.559 & 0.456 \\
 12 & 0.058 & 0.064 & 0.454 & 0.381 & 0.135 & 0.226 & 0.579 & 0.456 \\
 24 & 0.054 & 0.072 & 0.435 & 0.381 & 0.104 & 0.250 & 0.512 & 0.456 \\
 36 & 0.052 & 0.077 & 0.414 & 0.381 & 0.090 & 0.265 & 0.496 & 0.456 \\
 60 & 0.050 & 0.083 & 0.401 & 0.381 & 0.075 & 0.285 & 0.517 & 0.456 \\
120 & 0.047 & 0.092 & 0.415 & 0.381 & 0.060 & 0.313 & 0.523 & 0.456 \\
240 & 0.045 & 0.103 & 0.417 & 0.381 & 0.052 & 0.343 & 0.487 & 0.456 \\
360 & 0.044 & 0.109 & 0.406 & 0.381 & 0.049 & 0.361 & 0.458 & 0.456 \\
\bottomrule
\end{tabular}
\par\smallskip
\begin{minipage}{\textwidth}
\footnotesize\raggedright
Marg.\ $=$ marginal; Cond.\ $=$ population-average conditional.\\
\textsuperscript{a}\,The subgroup-level conditional progression HRs are $0.33$ (ECOG~0)
and $0.52$ (ECOG~1) in both populations and at every cycle. Mortality from the Stable
state is taken from an age-banded life table, to which a constant conditional HR of
$0.45$ is applied; its marginal counterpart rises to at most $0.469$ over the horizon, since
age is prognostic and the two arms deplete at different rates. Mortality after
progression is three times the life-table hazard, with no treatment effect.
\end{minipage}
}
\end{table}

\begin{table}[htbp]
\centering
\caption{Different model scenarios, estimands used by each scenario for the baseline
risk and the treatment effect across the two target populations, and issues to which
each scenario is subject.}
\label{tab:example-scenarios}
{\small
\begin{adjustbox}{max width=\textwidth}
\begin{tabular}{@{}l
  >{\raggedright\arraybackslash}p{3.0cm}
  >{\raggedright\arraybackslash}p{2.1cm}
  >{\raggedright\arraybackslash}p{2.1cm}
  >{\raggedright\arraybackslash}p{4.6cm}
  >{\raggedright\arraybackslash}p{3.2cm}@{}}
\toprule
\textbf{\#} & \textbf{Scenario} & \textbf{Baseline} & \textbf{Treatment effect} &
\textbf{Inputs used}\textsuperscript{a} & \textbf{Issues}\textsuperscript{b} \\
\midrule
1 & Individual-level (eq.~\ref{eq:1}) & Conditional, per patient & Conditional, per patient &
Conditional coefficients, per patient (Table~\ref{tab:example-coef}) & - \\
\addlinespace
2 & Cohort, marginal (eq.~\ref{eq:2}) & Marginal ($\overline{\pi}_0$) & Marginal ($\Delta$)\textsuperscript{c} &
Marginal progression probability $+$ marginal progression HR (Table~\ref{tab:example-inputs}) & - \\
\addlinespace
3 & Cohort, avg.\ conditional (eq.~\ref{eq:5}) & Pop.-avg.\ cond.\ ($m_0$) & Pop.-avg.\ cond.\ ($d$) &
Conditional model at mean covariates: probability $+$ HR (Table~\ref{tab:example-inputs}) & Cross-model scale mismatch \\
\addlinespace
4 & Cohort, mixed (marg.\ baseline) (eq.~\ref{eq:6}) & Marginal ($\overline{\pi}_0$) & Pop.-avg.\ cond.\ ($d$) &
Marginal progression probability $+$ at-mean conditional HR (Table~\ref{tab:example-inputs}) &
Cross-model scale mismatch; input incompatibility \\
\addlinespace
5 & Cohort, mixed (cond.\ baseline) (eq.~\ref{eq:7}) & Pop.-avg.\ cond.\ ($m_0$) & Marginal ($\Delta$)\textsuperscript{c} &
At-mean conditional probability $+$ marginal progression HR (Table~\ref{tab:example-inputs}) & Cross-model scale mismatch; input incompatibility \\
\bottomrule
\end{tabular}
\end{adjustbox}
\par\smallskip
\begin{minipage}{\textwidth}
\footnotesize\raggedright
\textsuperscript{a}\,Only progression inputs are listed, as these are what separate the
scenarios. In each scenario, the mortality inputs follow the same assignment as the
progression inputs -- marginalized over the covariate distribution where marginal,
evaluated at the mean age where population-average conditional.\\
\textsuperscript{b}\,The aggregation effect that all cohort approaches are subject to has been addressed by averaging the inputs on
the scale at which state occupancy is linear, prior to propagating the averages through the cohort model (Online Supplement~\ref{app:aggreffectpsm}).\\ 
\textsuperscript{c}\,Applied per cycle rather than as a single summary value
(Table~\ref{tab:example-inputs}).
\end{minipage}
}
\end{table}

\FloatBarrier

\subsection{Results}

\begin{figure}[htbp]
\centering
\includegraphics[width=0.95\linewidth]{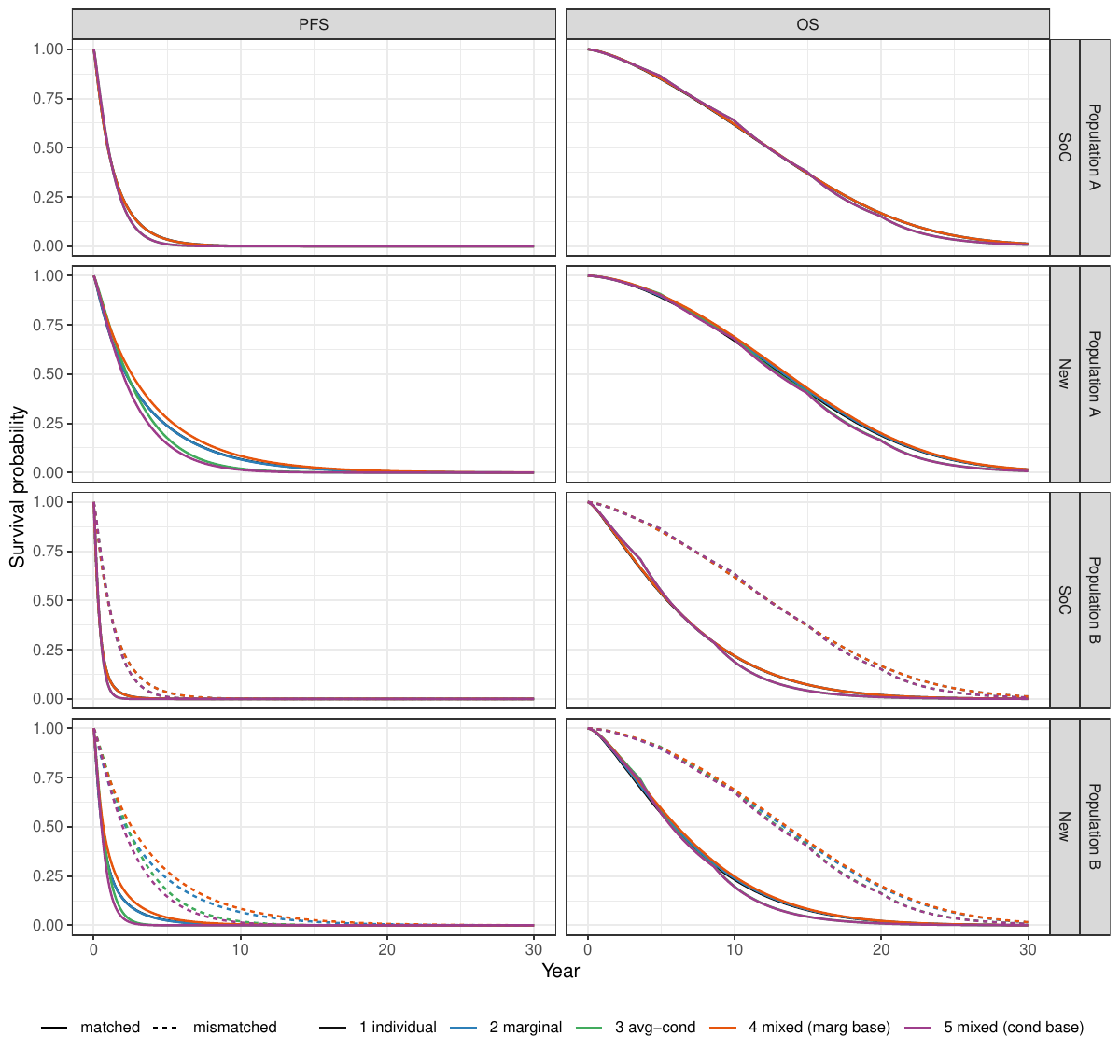}
\caption{Progression-free survival (PFS) and overall survival (OS) for standard of care (SoC)
and the new treatment, in target populations A and B. With the individual-level modeling approach, we obtain appropriate survival estimates for both target populations (solid black curves). Among the cohort approaches using correct-population (matched) inputs, the marginal-input scenario tracks the reference closely, while the scenarios built on population-average conditional inputs overestimate overall survival and substantially underestimate progression-free survival (solid colored curves). Using incorrect-population inputs with the cohort approaches (i.e., population A estimates used for a population B target) produces much larger departures (dashed colored curves).}
\label{fig:example-survival}
\end{figure}

\begin{figure}[htbp]
\centering
\includegraphics[width=0.95\linewidth]{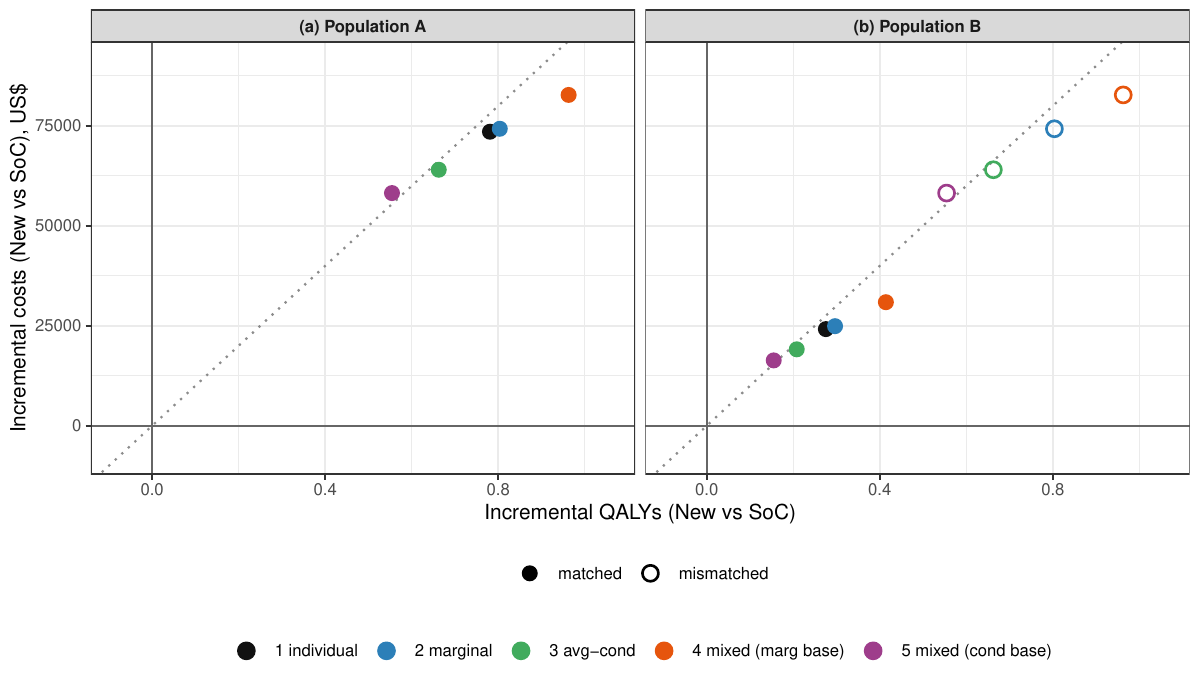}
\caption{Cost-effectiveness plane (incremental costs vs.\ incremental QALYs, for new treatment ``New'' vs.\ SoC),
as two panels sharing the same axes: (a)~Population~A and (b)~Population~B. Each point is
one modeling scenario (color), and the dotted line is a US\$100{,}000/QALY
willingness-to-pay threshold. With the individual-level modeling approach we get an appropriate estimate for each target population (black points). With the cohort approaches,
even using correct-population (matched) inputs, the estimates depart from the reference in
both directions (filled colored circles). For Population~B, using incorrect-population inputs (Population~A estimates for
a Population~B target) produces much larger departures (open colored circles), with resulting values sitting at
Population~A's values because the cohort estimates only vary with the population-average conditional and/or marginal clinical inputs in Table~\ref{tab:example-inputs} (already averaged over their
respective covariate distribution). Cohort scenario 5 falls on the opposite side of the willingness-to-pay threshold from
the individual-level reference, which can reverse the cost-effectiveness conclusions.}
\label{fig:example-ceplane}
\end{figure}

For each scenario, we summarize the modeling approach and describe the results observed for Population~A and the older Population~B. Figure~\ref{fig:example-survival}, Table~\ref{tab:example-results}, and Figure~\ref{fig:example-ceplane} show the survival curves and cost-effectiveness results for every scenario for both populations; the corresponding health-state occupancy distributions over time are shown in Online Supplement~\ref{illustrative-example-output} (Figure~\ref{fig:example-stateocc}).

\textbf{Scenario 1 (individual-level simulation, the reference, eq.~\ref{eq:1}).} For each
individual in the target population, the conditional progression and mortality rates are
computed from their age and ECOG status, an individual trace is run, and the outcomes
are averaged over the target population (``marginalize late''). This procedure targets the
marginal cost-effectiveness estimand of interest and is considered the reference, generating incremental QALYs of $0.781$ (Population~A) and
$0.275$ (Population~B). One can produce results for a new target population simply by updating the
covariate distribution.

\textbf{Scenario 2 (cohort, marginal inputs, eq.~\ref{eq:2}).} The conditional estimates
(Table~\ref{tab:example-coef}) are transformed into marginal progression and mortality
inputs (Table~\ref{tab:example-inputs}) by averaging the survival function over the
target covariate distribution and differencing it, and a
single cohort trace is run. While cohort models are generally subject to the aggregation effect of
Section~\ref{sec:aggregation}, we have addressed this by averaging the inputs on the
scale at which state occupancy is linear (Online Supplement~\ref{app:aggreffectpsm}), prior to propagating the averages through the cohort model. Hence the residual discrepancy with respect to the reference is small. Incremental QALYs are $0.804$
(Population~A, $+3\%$ with respect to the reference) and $0.297$ (Population~B, $+8\%$).

\textbf{Scenario 3 (cohort, population-average conditional inputs, eq.~\ref{eq:5}).} The population-average conditional treatment HR ($0.38$ for Population~A, $0.46$
for Population~B) is applied to the population-average conditional SoC baseline. This approach incurs the cross-model scale mismatch of Section~\ref{sec:42}: it feeds population-average conditional inputs to the cohort-based model that do not coincide with the marginal inputs that are required, and thus cannot target the marginal 
cost-effectiveness estimand of interest. The incremental QALYs are $0.663$ (Population~A, $-15\%$) and $0.208$ (Population~B, $-24\%$).

\textbf{Scenario 4 (cohort, marginal baseline + conditional effect, eq.~\ref{eq:6}).} The
population-average conditional treatment HR is now applied to the marginal SoC baseline. This approach incurs the cross-model scale mismatch and the additional input incompatibilities described in Section~\ref{sec:43}. The incremental QALYs are $0.963$ (Population~A, $+23\%$) and $0.414$
(Population~B, $+50\%$).

\textbf{Scenario 5 (cohort, conditional baseline + marginal effect, eq.~\ref{eq:7}).} The reverse
combination than Scenario 4: the marginal HR is applied to the population-average conditional SoC baseline. This approach also incurs the cross-model scale mismatch and the additional input incompatibilities described in Section~\ref{sec:43}. The incremental QALYs are $0.554$ (Population~A, $-29\%$) and
$0.155$ (Population~B, $-44\%$).

Even with inputs representing the correct target population, some cohort scenarios depart
from the individual-level reference in both directions, by up to $29\%$ in Population~A
and $50\%$ in Population~B. Scenario~2, the only one supplying the input type
eq.~\ref{eq:2} requires, comes closest ($+3\%$ and $+8\%$); every scenario that
substitutes a population-average conditional quantity for a marginal one departs further.
When inputs from Population~A are applied to Population~B (mismatched scenarios 2b--5b),
incremental QALYs are overestimated by between $+0.279$ and $+0.687$ QALYs relative to
the individual-level reference for Population~B --- larger than any of the matched
discrepancies. In some scenarios this is enough to change the cost-effectiveness
conclusions relative to the willingness-to-pay threshold, as illustrated in
Figure~\ref{fig:example-ceplane} with a US\$100{,}000/QALY willingness-to-pay threshold.

\begin{table}[htbp]
\centering
\caption{Incremental QALYs, costs, and incremental cost-effectiveness ratios (ICERs) for the new treatment versus SoC by target population, model scenario and whether input estimates are matched to reflect the target population. Costs and ICERs are in US dollars; the difference relative to the individual-level reference is in incremental QALYs. The estimates for the cohort-based approaches only vary with the population-average conditional and/or marginal clinical inputs in Table \ref{tab:example-inputs}, already averaged over their respective covariate distribution. As such, the ``matched'' inputs for target population A and the ``mismatched'' inputs for target population B produce identical results.}
\label{tab:example-results}
{\footnotesize
\setlength{\tabcolsep}{5pt}
\begin{tabular}{@{}cllrrrr@{}}
\toprule
\textbf{\#} & \textbf{Scenario} & \textbf{Input source} & \textbf{$\Delta$QALYs} &
\textbf{Difference} &
\textbf{$\Delta$Costs (\$)} & \textbf{ICER (\$/QALY)} \\
\midrule
\multicolumn{7}{@{}l}{\emph{Target: Population A (matched inputs need no transport as the source trial population is the target)}}\\
1 & Individual-level (ref.)      & any            & 0.781 & n/a & 73{,}538 & 94{,}164 \\
2 & Cohort, marginal            & matched    & 0.804 & $+0.023$  & 74{,}277 & 92{,}426 \\
3 & Cohort, avg conditional     & matched    & 0.663 & $-0.118$ & 64{,}026 & 96{,}639  \\
4 & Cohort, mixed (marg.\ base) & matched    & 0.963 & $+0.182$ & 82{,}744 & 85{,}944  \\
5 & Cohort, mixed (cond.\ base) & matched    & 0.554 & $-0.227$ & 58{,}196 & 104{,}975  \\
\addlinespace
\multicolumn{7}{@{}l}{\emph{Target: Population B (older population than the source trial)}}\\
1 & Individual-level (ref.)      & any            & 0.275 & n/a & 24{,}167 & 87{,}777  \\
2 & Cohort, marginal            & matched    & 0.297 & $+0.021$ & 24{,}903 & 83{,}991  \\
2b & Cohort, marginal            & mismatched (A) & 0.804 & $+0.528$ & 74{,}277 & 92{,}426  \\
3 & Cohort, avg conditional     & matched    & 0.208  & $-0.067$ & 19{,}107 & 91{,}880 \\
3b & Cohort, avg conditional     & mismatched (A) & 0.663  & $+0.387$ & 64{,}026 & 96{,}639  \\
4 & Cohort, mixed (marg.\ base) & matched    & 0.414 & $+0.139$ & 30{,}902 & 74{,}632  \\
4b & Cohort, mixed (marg.\ base) & mismatched (A) & 0.963 & $+0.687$ & 82{,}744 & 85{,}944 \\
5 & Cohort, mixed (cond.\ base) & matched    & 0.155 & $-0.121$ & 16{,}339 & 105{,}580  \\
5b & Cohort, mixed (cond.\ base) & mismatched (A) & 0.554 & $+0.279$ & 58{,}196 & 104{,}975  \\
\bottomrule
\end{tabular}
}
\end{table}

\FloatBarrier

\clearpage

\clearpage

\subsection{Why the estimates differ}

The resulting estimates differ for several reasons: the type of baseline risk and treatment effect estimates used as inputs, even when drawn from the correct population, and the population those inputs reflect.

Consider first the scenarios where the baseline risk and treatment effect inputs are aligned with the target population. Due to addressing the aggregation effect, Scenario~2 
departs from the reference only by $+0.023$ QALYs in Population~A and by $+0.021$ in
Population~B. For the cohort models, the discrepancy between any conditional inputs and the required marginal inputs has a greater impact on the incremental QALYs, and this impact acts in opposite directions.

Replacing the marginal treatment effect with the
population-average conditional treatment effect raises incremental QALYs in Population~A by $0.159$ when the baseline is marginal (0.804 in Scenario 2 to 0.963 in Scenario 4) and by $0.108$ (0.554 in Scenario 5 to 0.663 in Scenario 3) when the baseline is conditional. This is because the
population-average conditional HR lies further from the null (HR=1) than all the marginal
HRs in Table~\ref{tab:example-inputs}, which is consistent with the observation that population-average conditional treatment effects tend to overstate the benefit of interventions relative to marginal treatment effects for non-collapsible measures (see Section~\ref{sec:42} and Online Supplement~\ref{sec:trt_effect} for the OR). Identical trends are observed in Population B. 

Conversely, replacing the marginal baseline with the population-average conditional baseline lowers incremental QALYs: in Population A, by $0.249$ with a marginal treatment effect (from Scenario 2 to Scenario 5) and by $0.300$ with a population-average conditional treatment effect (from Scenario 4 to Scenario 3). The same behavior is observed in Population B. The reason is that the conditional progression probability climbs with the Weibull hazard while the marginal one turns downward as the highest-risk patients progress out of the Stable state. Scenario~4 only substitutes the treatment effect and overshoots (0.963 incremental QALYs in Population A); Scenario~5 only substitutes the baseline risk and undershoots by more (0.554); Scenario~3 makes both substitutions, which partly cancel (0.663). In both populations, the four cohort scenarios rank by absolute error exactly as Table~\ref{tab:example-scenarios} would rank them by the issues they carry. 

Consider next the scenarios where the baseline risk and treatment effect inputs are misaligned with the target population. Because Population~A is younger and with a greater proportion of ECOG~0 patients, for whom the treatment is more
effective, its inputs imply a larger benefit: using them for a Population~B decision considerably inflates incremental QALYs from $0.297$ to $0.804$ (Scenario~2), an error an order of
magnitude larger than the $+0.021$ incurred with matched inputs,
against a Population~B reference of $0.275$ incremental QALYs. Using Population~A inputs for a Population~B decision similarly inflates incremental QALYs for the other cohort scenarios: from 0.208 to 0.663 (Scenario 3), from 0.414 to 0.963 (Scenario 4) and from 0.155 to 0.554 (Scenario 5). 

Similar trends are observed using an alternative measure for the treatment effect: the absolute difference in 5-year overall survival. The individual-level simulation model gives a gain of $4.2$ percentage points for the new treatment versus SoC in Population~A and a corresponding gain of $3.3$ percentage points in Population~B. While this measure is directly collapsible, it is not necessarily immune to transportability issues because the populations differ in their covariate mix, through age and ECOG status, both of which could be effect modifiers on the absolute difference scale. 

\FloatBarrier

\section{Some recommendations}

Based on the issues raised with different types of baseline risk and treatment effect estimates, we provide the following initial recommendations for population-level decision-making in model-based CEA, organized by the modeling approach adopted and the data available.

\subsection{Individual-level simulation models}

The most rigorous approach is to carry conditional inputs through an individual-level simulation model~\citep{welton2015}, as set out in
Section~\ref{sec:ind_sim}. Namely, the baseline and treatment effect at reference
covariate values together with prognostic effects and treatment-by-covariate
interactions, are combined in a conditional model that is marginalized over an explicitly defined covariate distribution for
the target population (``marginalize late'')~\citep{krijkamp2018,karnon2012,alaridescudero2023}.

Feasibility hinges upon the ability to estimate (or the availability of) a relevant conditional
baseline along with prognostic effects, as well as the conditional treatment
effect along with parameters that capture effect modification. In practice, these two
sets of inputs come from different sources: the baseline and prognostic effects
typically come from a registry, cohort, or other observational data sources, and the treatment
effect and effect modification parameters come from one or more clinical trials. The applicability of these conditional quantities to the target population depends on the transportability aspects set out in Section~\ref{sec:ind_sim}.

\subsubsection{When IPD are available}

Where two sets of inputs come from different sources, these must be derived using the same outcome model specification: that is, the conditional baseline component with prognostic effects and
the conditional treatment effect component with interactions need to be defined on the same scale, with the same
link function and conditional on the same covariates entered in the same functional
form. Conditioning on different covariate adjustment specifications in the regression
analysis leads to different conditional estimands, which may or may not be compatible
with the conditional estimands implied by the inputs of the individual-level
simulation model. With IPD from the relevant source studies (and assuming full covariate availability), this compatibility can be
imposed directly, by fitting identical outcome model specifications to each data source. Transportability between each source and the target population can then be checked by comparing their covariate distributions, assuming all influential covariates are measured. Of note, federated data systems that allow for fitting outcome regression models across multiple decentralized data sources -- without requiring transfer of the raw IPD to a single central repository -- show promise to further support this approach.

\subsubsection{When IPD are not available}\label{sec:noipd}

Model developers often do not have IPD for all relevant source studies. While it may be
feasible to estimate the baseline directly in the target population from registry or
other real-world data, access to IPD for all the clinical trials that are the source of
treatment effect estimates is unlikely. What is then required from the publications reporting these trials is the complete fitted outcome model rather than the treatment effect alone: the intercept and prognostic coefficients, the conditional treatment effect at the reference covariate values, any treatment-by-covariate interaction terms, and the outcome model specification under which these were estimated, together with the associated measures of uncertainty. When using parametric time-to-event models such as the Weibull, this extends to the parametric form and parameters of the baseline hazard, since a reported hazard ratio alone does not allow the underlying conditional model to be reconstructed. Alternatively, where IPD are held for at least one study in the evidence network of clinical trials informing treatment effects, multilevel network meta-regression~\citep{phillippo2020mlnmr, phillippo2023validating, phillippo2026multilevel, jansen2026multilevel} can be used. It requires only aggregate outcomes and covariate summaries from the remaining studies and recovers the conditional relationships from this combination of individual and aggregate-level data, subject to a shared effect modifier assumption across comparators. If IPD for the baseline are limited, they may be supplemented with aggregate-level
information on the same principles, extending to real-world baseline risk and prognostic
effects estimated from single-arm cohort studies.

When IPD are not available for any relevant source study and only reported summary estimates remain, the choice is between marginalizing any available conditional summaries, preferably as part of an individual-level simulation, or directly inputting any marginal summaries into the cohort model. Much depends on the availability of compatible conditional estimates that condition on the same covariate set, and on how much covariate structure the reported conditional estimates can
support. The individual-level simulation model can carry only those covariates for which estimates exist and leaves
the rest unmodeled, applying coarser estimands than intended. Bayesian model calibration may also help: matching the
marginal outcomes reported for each study to the marginal covariate distribution of its
population can help recover or refine the baseline risk, prognostic effects, treatment effect
and effect modification estimates required by the individual-level simulation model~\citep{jalal2021baycann}. Weighing against the cohort alternative is whether transition
rates depend on time in state or on accumulated history, which a cohort model cannot
represent without a proliferation of tunnel states. In the limiting case where no
covariate structure can be supported at all, the individual-level simulation would run on one average conditional input for carbon copies of a single representative
individual, retaining its handling of memory and history, but inheriting similar transportability limitations as
the cohort model, which are discussed in Section~\ref{Sec:62}. 

In the absence of IPD to define the target population, its covariate distribution can be defined based on published marginal covariate distributions combined with an assumed or externally sourced correlation structure.

\subsubsection{Implementation}

Individual-level simulation models for CEA can be implemented with the \texttt{hesim}
package for R (\url{https://hesim-dev.github.io/hesim/})~\citep{incerti2021hesim}. The \texttt{hesim}
workflow aligns well with the ``marginalize late'' principle and proceeds in three
steps: (1) Parameterization: statistical models for disease progression, utilities,
and costs are estimated using individual patient data or aggregate data from
multiple studies; (2) Simulation: the statistical models from Step 1 are combined
to construct an economic model, where disease progression, QALYs, and costs are
simulated over an \emph{explicitly defined target population} (specified by its
covariate distribution, i.e.\ the mix of patient characteristics) and the treatment
strategies of interest, with outcomes averaged over that population as the last step,
which is precisely how \texttt{hesim} operationalizes ``marginalize late''; and (3)
Decision analysis: simulated outcomes are used to perform CEA. Uncertainty in parameters is
propagated throughout using probabilistic sensitivity analysis. Since the statistical and simulation models can be integrated in a single R script, parameter uncertainty can be easily captured with \texttt{hesim}~\citep{incerti2019}.

\subsection{Cohort models}\label{Sec:62}

We recognize that cohort models remain common in current HTA practice.

\subsubsection{What the best cohort approach requires}\label{cohort-ingredients}

The most appropriate cohort-based approach uses
marginal estimates of baseline risk and treatment effects that are representative of the target population (Section~\ref{sec:marg_cohort}). With such inputs, eq.~\ref{eq:2} reproduces the marginal outcome probabilities of the target population by construction. Even then, cohort models have their own set of issues: carrying marginal inputs through a non-linear cohort model over a
heterogeneous population is subject to a residual aggregation effect
(Section~\ref{sec:aggregation}). We have addressed this in the illustrative example in Section~\ref{sec:example} by averaging the survival function -- the scale at which state occupancy is linear (Online Supplement~\ref{app:aggreffectpsm}) -- over the target covariate distribution and differencing it, rather than averaging each period's transition probability with fixed weights (Online Supplement~\ref{app:cohorttransitionmodel}). In preliminary analyses for the illustrative example (Online Supplement~\ref{hsoccupancy}), the latter approach demonstrated a considerable aggregation effect which markedly exacerbated the discrepancy with respect to the individual-level simulation model used as the reference. 

The ``best-case'' scenario is one where the source studies
do not differ from the target population in their covariate distribution, so
that their marginal estimates apply directly; or one where the
same information required for an individual-level model is in hand: the conditional
baseline, conditional treatment effect at reference covariate values, 
prognostic effects, treatment-by-covariate interactions, and the covariate
distribution of the target population. These are precisely the ingredients needed to
construct valid marginal inputs for the correct population for the decision, by marginalizing the conditional estimates over the corresponding covariate distribution. 

It follows that whenever the correct
marginal cohort inputs must be constructed rather than read off a matching source -- for instance, because the population of the source study is not directly relevant to the decision -- an
individual-level model can be run as well, and should arguably be preferred based on the ``marginalize late'' principle. Nevertheless, there are trade-offs to consider. While marginal treatment effects can be identified from randomized trials with minimal assumptions (assuming no missingness), the within-trial identification of conditional quantities requires additional statistical assumptions about model validity,  particularly with continuous covariates or where there are only a few individuals in some covariate subgroups~\cite{van2022estimands, van2024covariate}. The weaker statistical assumptions for marginal estimation do not translate from within-trial analysis to transportability across populations, due to marginal measures' general dependence on the distribution of purely prognostic factors, but the conditional quantities are not necessarily portable either, as they may depend on omitted or unobserved effect modifiers (and prognostic factors in some cases). 

\subsubsection{Deviations in practice}

When the ingredients in Section~\ref{cohort-ingredients} are missing, the required marginal inputs for the cohort model cannot be
constructed in a valid manner, and the model departs from the best-case by necessity. 
Deviations from this best-case are primarily driven by the availability of data and reported
estimates rather than by analyst choice. They can be
organized along two axes: whether the baseline risk and treatment
effect inputs are (i) marginal estimates, population-average conditional estimates, or a
mixture, and (ii) drawn from the correct target population or from a different (study)
population. In the illustrative example, marginal inputs (eq.~\ref{eq:2}) misestimate outcomes only modestly because we have accounted for the aggregation effect; population-average conditional inputs (eq.~\ref{eq:5}) incur the cross-model scale mismatch of Section~\ref{sec:42}; and a conditional treatment effect combined with a marginal baseline or vice versa (eq.~\ref{eq:6} and eq.~\ref{eq:7}, respectively) mixes incompatible quantities. The cross-model scale mismatch may lead to strongly overestimating or underestimating the benefit of an intervention, as illustrated in our applied example. Each of these errors can be compounded when the
inputs additionally come from the wrong population, as when marginal estimates
reported for the study population are applied to the target population without
adjustment for differences in influential covariates.

The critical question that emerges is: what matters most, the alignment of
estimand types (marginal vs.\ conditional) or alignment with the correct target
population? Our illustrative example is only suggestive, but it indicates that both
matter materially and that the second can dominate. For the correct population,
estimand misalignment alone produced divergences in incremental QALYs relative to the reference case of up to $29\%$ in the
trial population and up to $50\%$ in the older population. Discrepancies were in different directions across modeling scenarios. Using inputs from the
wrong population produced larger errors still, and did so even for the scenario in which estimands were correctly aligned. (Scenario~2 departs from the reference by $+0.021$ QALYs with matched inputs and by $+0.528$ with inputs transported from the trial population.) The example illustrates these orderings; it does not establish them in general. Errors of opposite sign can partly cancel, as in Scenario~3, and a simulation study across model structures, effect measures and degrees of heterogeneity would be needed to characterize their magnitude.

\subsubsection{Partitioned survival versus state-transition structure}

Partitioned survival models deserve comment as they are the
default in oncology. A partitioned survival model fits progression-free survival (PFS) and overall survival (OS) separately, with no structural link between them, so the implied state occupancies can be incoherent, with PFS exceeding OS or implausible post-progression survival~\citep{woods2017}. A state-transition model imposes the required transition structure and guarantees coherent occupancy, which is the
usual reason to prefer it. However, the corresponding cohort model is subject to aggregation bias by applying a single population-average transition matrix repeatedly across cycles, which is not equivalent to averaging the resulting state occupancies across individuals. Aggregation bias is avoided by the partitioned survival alternative because occupancy is a linear functional of the population-average curves. 

The best-case partitioned survival cohort model here is one which pairs a marginal baseline survival curve for the comparator with a marginal HR to recover the marginal survival curve for the intervention. Unfortunately, this is rarely the case in most applications, which are subject to the cross-model scale mismatches we have described. Practitioners will typically combine parametric fits to marginal survival curves with Cox model-based conditional HRs reported in publications, replicating Scenario~4 (eq.~\ref{eq:6}). Combining stratified (conditional) subgroup-level survival curves with conditional or marginal HRs, reported in publications or derived from IPD reconstructed using digitized survival curves, is also common and similar to Scenario~3 (eq.~\ref{eq:5}) and  Scenario~5 (eq.~\ref{eq:7}), respectively. Having this in mind, even the best-case partitioned survival model is not free from issues. Marginal HRs are time-varying when conditional HRs are constant (Figure~\ref{fig:example-mhr}) because proportional hazards cannot hold simultaneously on the marginal and conditional scales~\citep{daniel2021, phillippo2025effect}, so a single HR summary is often insufficient. Applying marginal HRs to PFS and OS separately also introduces incoherence, since endpoint-specific HRs can push PFS above OS. 

An individual-level state-transition model is the only option that resolves these tensions: it retains the structural coherence between state occupancies, targets the relevant marginal cost-effectiveness estimand by marginalizing late, and can be readily adapted to new target populations by carrying the conditional inputs and marginalizing over the relevant covariate distribution. 

\subsection{Recommendations for any modeling approach}

Analysts must thoroughly understand their baseline risk and treatment effect
estimates. This requires documentation of: (i) which population these estimates represent; (ii)
whether they are marginal or conditional estimates; (iii) for conditional measures derived from an outcome regression model, which covariates were
adjusted for in the original analysis; and (iv) for treatment effects, the effect measure
used and whether it is directly collapsible, collapsible but not directly so, or
non-collapsible. Analysts should also invest effort in understanding which factors are
important prognostic factors and effect modifiers in their specific decision
context. Without all these pieces of information, it is very challenging to
assess the appropriateness of combining different inputs in a CEA model, or to
anticipate the direction and magnitude of potential biases, due to estimand mismatches or transportability errors when combining data from different (study) populations.

A qualification is in order. The portability of conditional inputs, which our preferred individual-level simulation approach is based on (Section~\ref{sec:ind_sim}), is not always guaranteed. While marginal estimands strongly depend on the population, conditional estimands are defined relative
to a set of conditioning covariates, so may shift with that set as well as with the population. The
practical choice is therefore a trade-off between dependence on the population and
dependence on the conditioning covariate set. Developing the outcome regression and simulation model together as part of an integrated approach,
so that covariate sets are compatible across modules by construction, goes a long way in resolving the trade-off in favor of conditional inputs.  

\section{Conclusion}

HTA allocates resources across populations. As such, the
cost-effectiveness estimand of interest is marginal. To inform HTA decision-making, the most rigorous cost-effectiveness modeling approach is an individual-level simulation that carries conditional inputs (baseline, treatment effect, prognostic effects, and effect-modifiers) through the model and averages outcomes over an explicitly defined target population (``marginalize late''). The commonly used cohort-based approach instead represents the population by a single average cohort whose inputs are either marginal, population-average conditional, or a mixture. By construction, only marginal inputs corresponding to the target population for the decision can ultimately target the cost-effectiveness estimand of interest, and even then a residual aggregation bias may remain depending on implementation. Population-average conditional inputs and mixtures of inputs add an estimand mismatch, particularly when treatment effect measures are non-collapsible. Beyond these issues, the limited transportability of baseline risk and treatment effect inputs when these have been estimated in the wrong population seems to be an even greater threat. In our example, this produced discrepancies that were larger than those induced by estimand misalignment alone. Above all, model developers should document, for every model input related to baseline risk and treatment effects, which population it represents, whether it is marginal or conditional, and -- for conditional inputs which covariates these were adjusted for. 

\bibliographystyle{vancouver}
\bibliography{references}

\begin{thebibliography}{10}

\bibitem{drummond2015}
Drummond MF, Sculpher MJ, Claxton K, Stoddart GL, Torrance GW.
\newblock Methods for the Economic Evaluation of Health Care Programmes.
\newblock 4th ed. Oxford: Oxford University Press; 2015.

\bibitem{neumann2017}
Neumann PJ, Sanders GD, Russell LB, Siegel JE, Ganiats TG, editors.
\newblock Cost-Effectiveness in Health and Medicine.
\newblock 2nd ed. New York: Oxford University Press; 2017.

\bibitem{welte2004}
Welte R, Feenstra T, Jager H, Leidl R.
\newblock A decision chart for assessing and improving the transferability of
  economic evaluation results between countries.
\newblock PharmacoEconomics. 2004;22(13):857--876.

\bibitem{drummond2009}
Drummond M, Barbieri M, Cook J, Glick HA, Lis J, Malik F, et~al.
\newblock Transferability of economic evaluations across jurisdictions: ISPOR
  Good Research Practices Task Force report.
\newblock Value in Health. 2009;12(4):409--418.

\bibitem{sculpher2006}
Sculpher MJ, Claxton K, Drummond M, McCabe C.
\newblock Whither trial-based economic evaluation for health care decision
  making?
\newblock Health Economics. 2006;15(7):677--687.

\bibitem{phillippo2025effect}
Phillippo DM, Remiro-Az{\'o}car A, Heath A, Baio G, Dias S, Ades A, et~al.
\newblock Effect modification and non-collapsibility together may lead to
  conflicting treatment decisions: A review of marginal and conditional
  estimands and recommendations for decision-making.
\newblock Research synthesis methods. 2025;16(2):323--349.

\bibitem{remiro2025marginal}
Remiro-Az{\'o}car A, Phillippo DM, Welton NJ, Dias S, Ades AE, Heath A, et~al.
\newblock Marginal and conditional summary measures: transportability and
  compatibility across studies.
\newblock arXiv preprint arXiv:250721925. 2025;.

\bibitem{remiro2024transportability}
Remiro-Az{\'o}car A.
\newblock Transportability of model-based estimands in evidence synthesis.
\newblock Statistics in medicine. 2024;43(22):4217--4249.

\bibitem{degtiar2023}
Degtiar I, Rose S.
\newblock A review of generalizability and transportability.
\newblock Annual Review of Statistics and Its Application. 2023;10:501--524.

\bibitem{dahabreh2020}
Dahabreh IJ, Robertson SE, Steingrimsson JA, Stuart EA, Hern{\'a}n MA.
\newblock Extending inferences from a randomized trial to a new target
  population.
\newblock Statistics in Medicine. 2020;39(14):1999--2014.

\bibitem{colnet2023risk}
Colnet B, Josse J, Varoquaux G, Scornet E.
\newblock Risk ratio, odds ratio, risk difference... Which causal measure is
  easier to generalize?
\newblock arXiv preprint arXiv:230316008. 2023;.

\bibitem{phillippo2016tsd18}
Phillippo DM, Ades AE, Dias S, Palmer S, Abrams KR, Welton NJ.
\newblock NICE DSU Technical Support Document 18: Methods for
  Population-Adjusted Indirect Comparisons in Submissions to NICE.
\newblock NICE Decision Support Unit; 2016.
\newblock Available from:
  \url{https://www.sheffield.ac.uk/nice-dsu/tsds/population-adjusted}.

\bibitem{webster2023choice}
Webster-Clark M, Keil AP.
\newblock How choice of effect measure influences minimally sufficient
  adjustment sets for external validity.
\newblock American Journal of Epidemiology. 2023;192(7):1148--1154.

\bibitem{rothman2008modern}
Rothman KJ, Greenland S, Lash TL, et~al.
\newblock Modern epidemiology. vol.~3.
\newblock Wolters Kluwer Health/Lippincott Williams \& Wilkins Philadelphia;
  2008.

\bibitem{greenland1999}
Greenland S, Robins JM, Pearl J.
\newblock Confounding and collapsibility in causal inference.
\newblock Statistical Science. 1999;14(1):29--46.

\bibitem{didelez2022}
Didelez V, Stensrud MJ.
\newblock On the logic of collapsibility for causal effect measures.
\newblock Biometrical Journal. 2022;64(2):235--242.

\bibitem{huitfeldt2019}
Huitfeldt A, Stensrud MJ, Suzuki E.
\newblock On the collapsibility of measures of effect in the counterfactual
  causal framework.
\newblock Emerging Themes in Epidemiology. 2019;16:1.

\bibitem{daniel2021}
Daniel R, Zhang J, Farewell D.
\newblock Making apples from oranges: comparing noncollapsible effect
  estimators and their standard errors after adjustment for different covariate
  sets.
\newblock Biometrical Journal. 2021;63(3):528--557.

\bibitem{signorovitch2010}
Signorovitch JE, Wu EQ, Yu AP, Gerrits CM, Kantor E, Bao Y, et~al.
\newblock Comparative effectiveness without head-to-head trials: a method for
  matching-adjusted indirect comparisons applied to psoriasis treatment with
  adalimumab or etanercept.
\newblock PharmacoEconomics. 2010;28(10):935--945.

\bibitem{phillippo2018}
Phillippo DM, Ades AE, Dias S, Palmer S, Abrams KR, Welton NJ.
\newblock Methods for population-adjusted indirect comparisons in health
  technology appraisal.
\newblock Medical Decision Making. 2018;38(2):200--211.

\bibitem{robins1986}
Robins J.
\newblock A new approach to causal inference in mortality studies with a
  sustained exposure period---application to control of the healthy worker
  survivor effect.
\newblock Mathematical Modelling. 1986;7(9--12):1393--1512.

\bibitem{hernan2020book}
Hern{\'a}n MA, Robins JM.
\newblock Causal Inference: What If.
\newblock Boca Raton: Chapman \& Hall/CRC; 2020.

\bibitem{phillippo2020mlnmr}
Phillippo DM, Dias S, Ades AE, Belger M, Brnabic A, Schacht A, et~al.
\newblock Multilevel network meta-regression for population-adjusted treatment
  comparisons.
\newblock Journal of the Royal Statistical Society: Series A.
  2020;183(3):1189--1210.

\bibitem{phillippo2020sim}
Phillippo DM, Dias S, Ades AE, Welton NJ.
\newblock Assessing the performance of population adjustment methods for
  anchored indirect comparisons: a simulation study.
\newblock Statistics in Medicine. 2020;39(30):4885--4911.

\bibitem{remiroazocar2022gcomp}
Remiro-Az{\'o}car A, Heath A, Baio G.
\newblock Parametric G-computation for compatible indirect treatment
  comparisons with limited individual patient data.
\newblock Research Synthesis Methods. 2022;13(6):716--744.

\bibitem{stinnett1998}
Stinnett AA, Mullahy J.
\newblock Net health benefits: a new framework for the analysis of uncertainty
  in cost-effectiveness analysis.
\newblock Medical Decision Making. 1998;18(2 Suppl):S68--S80.

\bibitem{weinstein1977}
Weinstein MC, Stason WB.
\newblock Foundations of cost-effectiveness analysis for health and medical
  practices.
\newblock New England Journal of Medicine. 1977;296(13):716--721.

\bibitem{claxton2015}
Claxton K, Martin S, Soares M, Rice N, Spackman E, Hinde S, et~al.
\newblock Methods for the estimation of the National Institute for Health and
  Care Excellence cost-effectiveness threshold.
\newblock Health Technology Assessment. 2015;19(14):1--504.

\bibitem{culyer2016}
Culyer AJ.
\newblock Cost-effectiveness thresholds in health care: a bookshelf guide to
  their meaning and use.
\newblock Health Economics, Policy and Law. 2016;11(4):415--432.

\bibitem{welton2015}
Welton NJ, Soares MO, Palmer S, Ades AE, Harrison D, Shankar-Hari M, et~al.
\newblock Accounting for Heterogeneity in Relative Treatment Effects for Use in
  Cost-Effectiveness Models and Value-of-Information Analyses.
\newblock Medical Decision Making. 2015;35(5):608--621.

\bibitem{remiroazocar2021comment}
Remiro-Az{\'o}car A, Heath A, Baio G.
\newblock Conflating marginal and conditional treatment effects: comments on
  `Assessing the performance of population adjustment methods for anchored
  indirect comparisons: a simulation study'.
\newblock Statistics in Medicine. 2021;40(11):2753--2758.

\bibitem{remiroazocar2022estimands}
Remiro-Az{\'o}car A.
\newblock Target estimands for population-adjusted indirect comparisons.
\newblock Statistics in Medicine. 2022;41(28):5558--5569.

\bibitem{brennan2006}
Brennan A, Chick SE, Davies R.
\newblock A taxonomy of model structures for economic evaluation of health
  technologies.
\newblock Health Economics. 2006;15(12):1295--1310.

\bibitem{krijkamp2018}
Krijkamp EM, Alarid-Escudero F, Enns EA, Jalal HJ, Hunink MGM, Pechlivanoglou
  P.
\newblock Microsimulation modeling for health decision sciences using R: a
  tutorial.
\newblock Medical Decision Making. 2018;38(3):400--422.

\bibitem{sonnenberg1993}
Sonnenberg FA, Beck JR.
\newblock Markov models in medical decision making: a practical guide.
\newblock Medical Decision Making. 1993;13(4):322--338.

\bibitem{siebert2012}
Siebert U, Alagoz O, Bayoumi AM, Jahn B, Owens DK, Cohen DJ, et~al.
\newblock State-transition modeling: a report of the ISPOR-SMDM Modeling Good
  Research Practices Task Force--3.
\newblock Value in Health. 2012;15(6):812--820.

\bibitem{omahony2015}
O'Mahony JF, Newall AT, van Rosmalen J.
\newblock Dealing with time in health economic evaluation: methodological
  issues and recommendations for practice.
\newblock PharmacoEconomics. 2015;33(12):1255--1268.

\bibitem{hernan2010}
Hern{\'a}n MA.
\newblock The hazards of hazard ratios.
\newblock Epidemiology. 2010;21(1):13--15.

\bibitem{aalen2015}
Aalen OO, Cook RJ, R{\o}ysland K.
\newblock Does Cox analysis of a randomized survival study yield a causal
  treatment effect?
\newblock Lifetime Data Analysis. 2015;21(4):579--593.

\bibitem{martinussen2013}
Martinussen T, Vansteelandt S.
\newblock On collapsibility and confounding bias in Cox and Aalen regression
  models.
\newblock Lifetime Data Analysis. 2013;19(3):279--296.

\bibitem{jansen2011}
Jansen JP.
\newblock Network meta-analysis of survival data with fractional polynomials.
\newblock BMC Medical Research Methodology. 2011;11:61.

\bibitem{jansen2012}
Jansen JP, Cope S.
\newblock Meta-regression models to address heterogeneity and inconsistency in
  network meta-analysis of survival outcomes.
\newblock BMC Medical Research Methodology. 2012;12:152.

\bibitem{ouwens2010}
Ouwens MJNM, Philips Z, Jansen JP.
\newblock Network meta-analysis of parametric survival curves.
\newblock Research Synthesis Methods. 2010;1(3--4):258--271.

\bibitem{karnon2012}
Karnon J, Stahl J, Brennan A, Caro JJ, Mar J, M{\"o}ller J.
\newblock Modeling using discrete event simulation: a report of the ISPOR-SMDM
  Modeling Good Research Practices Task Force--4.
\newblock Value in Health. 2012;15(6):821--827.

\bibitem{alaridescudero2023}
Alarid-Escudero F, Krijkamp EM, Enns EA, Yang A, Hunink MGM, Pechlivanoglou P,
  et~al.
\newblock A tutorial on time-dependent cohort state-transition models in R
  using a cost-effectiveness analysis example.
\newblock Medical Decision Making. 2023;43(1):21--41.

\bibitem{phillippo2023validating}
Phillippo DM, Dias S, Ades A, Belger M, Brnabic A, Saure D, et~al.
\newblock Validating the assumptions of population adjustment: application of
  multilevel network meta-regression to a network of treatments for plaque
  psoriasis.
\newblock Medical Decision Making. 2023;43(1):53--67.

\bibitem{phillippo2026multilevel}
Phillippo DM, Dias S, Ades A, Welton NJ.
\newblock Multilevel network meta-regression for general likelihoods: synthesis
  of individual and aggregate data with applications to survival analysis.
\newblock Journal of the Royal Statistical Society Series A: Statistics in
  Society. 2026;189(3):1856--1875.

\bibitem{jansen2026multilevel}
Jansen JP.
\newblock Multilevel network meta-regression for multistate models:
  Population-adjusted joint synthesis of progression and survival data from
  individual and aggregate evidence.
\newblock arXiv preprint arXiv:260725120. 2026;.

\bibitem{jalal2021baycann}
Jalal H, Trikalinos TA, Alarid-Escudero F.
\newblock {BayCANN}: Streamlining {Bayesian} Calibration with Artificial Neural
  Network Metamodeling.
\newblock Frontiers in Physiology. 2021;12:662314.

\bibitem{incerti2021hesim}
Incerti D, Jansen JP. hesim: Health Economic Simulation Modeling and Decision
  Analysis; 2021.
\newblock arXiv:2102.09437.

\bibitem{incerti2019}
Incerti D, Thom H, Baio G, Jansen JP.
\newblock R you still using Excel? The advantages of modern software tools for
  health technology assessment.
\newblock Value in Health. 2019;22(5):575--579.

\bibitem{van2022estimands}
Van~Lancker K, Vo TT, Akacha M.
\newblock Estimands in heath technology assessment: a causal inference
  perspective.
\newblock Statistics in medicine. 2022;41(28):5577--5585.

\bibitem{van2024covariate}
Van~Lancker K, Bretz F, Dukes O.
\newblock Covariate adjustment in randomized controlled trials: General
  concepts and practical considerations.
\newblock Clinical Trials. 2024;21(4):399--411.

\bibitem{woods2017}
Woods B, Sideris E, Palmer S, Latimer N, Soares M.
\newblock NICE DSU Technical Support Document 19: Partitioned Survival Analysis
  for Decision Modelling in Health Care: A Critical Review.
\newblock NICE Decision Support Unit; 2017.

\bibitem{sjolander2016note}
Sj{\"o}lander A, Dahlqwist E, Zetterqvist J.
\newblock A note on the noncollapsibility of rate differences and rate ratios.
\newblock Epidemiology. 2016;27(3):356--359.

\end{thebibliography}

\appendix
\renewcommand{\thesection}{\Alph{section}}
\setcounter{figure}{0}
\renewcommand{\thefigure}{S\arabic{figure}}
\setcounter{table}{0}
\renewcommand{\thetable}{S\arabic{table}}
\captionsetup{list=false}
\clearpage
\section*{Supplementary Material}
\addcontentsline{toc}{section}{Supplementary Material}

\section{Collapsibility and transportability with a prognostic factor and an effect
modifier}\label{app:collapstransport}

This supplement separates two aspects that are easily conflated: the dependence of a
marginal effect measure on the covariate distribution in the absence of effect
modification, which is due to non-collapsibility; and its dependence on the covariate distribution when effect modification is present, which affects collapsible and non-collapsible measures alike. We take a marginally randomized trial and a binary
outcome throughout.

\medskip
\textbf{Collapsibility without effect modification.} The cleanest demonstration
removes effect modification entirely. Let a single binary prognostic factor $X$ split the trial population, with control risks $0.30$ at $X=0$ and $0.50$ at $X=1$ and treated risks
$0.50$ and $0.70$, respectively (Table~\ref{tab:collapsibility-nem}). The conditional treatment effect is constant
on both the OR and the RD scale: the conditional OR is $2.33$ and the conditional RD is
$0.20$ in each stratum. Here we assume a marginally randomized trial where $P(X \mid T)=P(X)$ because $X \indep T$, where $T$ denotes treatment. 

The marginal RD equals $0.20$ whatever the value of $P(X{=}1)$ but the marginal OR falls to $2.25$ at $P(X{=}1)=0.5$ and returns to $2.33$ only at the
extremes, where $P(X=1)\in \{0,1\}$ and the population is homogeneous. With no effect modification on either of
these scales, the marginal OR still depends on the covariate distribution while the
marginal RD does not; this is due to non-collapsibility. The RR, despite being collapsible, is not constant across strata ($1.67$ and $1.40$) due to the scale dependence of effect modification: a prognostic factor that does not act as an effect modifier on the additive risk difference scale will inherently act as an effect modifier on the multiplicative risk ratio scale~\citep{rothman2008modern}.

\medskip
{\footnotesize
\captionof{table}{Two strata defined by a prognostic factor $X$, with the conditional treatment
effect constant on both the OR and the RD scales, but not on the RR scale.}
\label{tab:collapsibility-nem}
\centering
\begin{tabular}{@{}crrrrr@{}}
\toprule
\textbf{$X$ (prog.)} & \hspace{1.1cm}\textbf{Control risk} & \textbf{Treated risk} & \textbf{Odds ratio} & \textbf{Risk diff.} & \textbf{Risk ratio} \\
\midrule
0 & 0.30 & 0.50 & 2.33 & $0.20$ & 1.67 \\
1 & 0.50 & 0.70 & 2.33 & $0.20$ & 1.40 \\
\addlinespace
\multicolumn{1}{@{}l}{\emph{Marginal, at $P(X{=}1)=0.5$}} &0.40 &0.60 & 2.25 & $0.20$ & 1.50 \\
\bottomrule
\end{tabular}
}

\medskip
\textbf{Collapsibility with effect modification.} Now consider two binary covariates: a prognostic factor $X$ that shifts the baseline (control) risk, and an effect
modifier $Z$ for the conditional OR. We assume that $X$ and $Z$ are independent such that $P(X, Z)=P(X)P(Z)$. In a marginally randomized trial, $P(X, Z \mid T)=P(X)P(Z)$. The two
covariates define four subgroups (Table~\ref{tab:collapsibility}). The conditional OR
depends only on $Z$ ($0.20$ at $Z=0$, $0.50$ at $Z=1$) and is identical at both levels of
the prognostic factor $X$; the control risk depends only on $X$ ($0.10$ at $X=0$, $0.60$
at $X=1$).

\medskip
{\footnotesize
\captionof{table}{Four subgroups defined by a prognostic factor $X$ and
an effect modifier $Z$. The conditional OR varies only with $Z$ and the control risk only
with $X$, but the RD and the RR vary with both.}
\label{tab:collapsibility}
\centering
\begin{tabular}{@{}ccrrrrr@{}}
\toprule
\textbf{$X$ (prog.)} & \textbf{$Z$ (eff.\ mod.)} & \textbf{Control risk} & \textbf{Treated risk} & \textbf{Odds ratio} & \textbf{Risk diff.} & \textbf{Risk ratio} \\
\midrule
0 & 0 & 0.10 & 0.022 & 0.20 & $-0.078$ & 0.22 \\
0 & 1 & 0.10 & 0.053 & 0.50 & $-0.047$ & 0.53 \\
1 & 0 & 0.60 & 0.231 & 0.20 & $-0.369$ & 0.38 \\
1 & 1 & 0.60 & 0.429 & 0.50 & $-0.171$ & 0.71 \\
\addlinespace
\multicolumn{2}{@{}l}{\emph{Marginal, at $P(X{=}1)=P(Z=1)=0.5$}}& 0.35 & 0.183 & $0.42$ & $-0.167$ & $0.52$ \\
\bottomrule
\end{tabular}
}

The contrast is now less clean. Despite the conditional OR being constant across $X$, $X$ is an effect modifier on the RD scale. Hence, the marginal RD varies with $P(X)$. Also, $X$ remains an effect modifier on the RR scale and modifies the marginal RR. 

At $P(X{=}1)=P(Z{=}1)=0.5$, $P(X=x, Z=z)=0.25$: the marginal treated risk is $0.25\times (0.0217+0.0526+0.2308+0.4286)=0.183$ and the marginal control risk is $0.25\times (0.10+0.10+0.60+0.60)=0.350$. Due to direct collapsibility, the marginal RD ($0.183-0.350=-0.167$) is equal to a simple weighted average of conditional RDs (the population-average conditional RD), where the weights are given by the covariate proportions $0.25\times(-0.0783-0.0474-0.3692-0.1714)=-0.167$. The RR is
collapsible but not directly collapsible: the marginal RR ($0.183/0.350=0.52$) can still be expressed as a baseline-risk-weighted average of conditional RRs $
(0.10\times(0.22+0.53)+0.60\times(0.38+0.71))/(1.40)=0.52$, but the weights are not simply given by the covariate proportions, which would yield $0.25\times (0.22+0.53+0.38+0.71)=0.46$. Even where there is effect modification on the RD scale, the marginal RD is equal to the population-average conditional RD. Conversely, the marginal RR is not generally equal to the population-average conditional RR when there is effect modification on the RR scale. 

Due to non-collapsibility, the marginal OR cannot be recovered by a simple or baseline-risk-weighted average of conditional ORs. There is no weighting scheme applied to conditional ORs that can recover the marginal OR, which must be computed directly from the marginal risks. 

\medskip
\textbf{Transportability.} Table~\ref{tab:collapsibility-transport} gives the marginal
OR over a grid of populations that differ in the distribution of the prognostic
factor $X$ and the effect modifier $Z$, holding the two independent. The grid is only valid for $X \indep Z$, $P(X, Z)=P(X)P(Z)$, as the marginal OR depends on the joint covariate distribution. Starting from the trial population with $P(X{=}1)=P(Z{=}1)=0.5$ (marginal OR $0.42$), moving only the prognostic
factor to $P(X{=}1)=0.8$ shifts the marginal OR to $0.37$, and moving only the effect modifier
to $P(Z{=}1)=0.8$ shifts the marginal OR to $0.52$. Even in the scenarios where a single conditional OR applies to all patients ($P(Z{=}0)=1$ or $P(Z{=}1)=1$), the marginal OR still varies with $P(X{=}1)$, with values of $0.20$ at the extremes and $0.27$ at $P(X{=}1)=0.5$.

Table~\ref{tab:collapsibility-compare} sets
each marginal measure against the simple weighted average of the four  subgroup-level conditional measures.
For the RD, the two agree exactly: at $P(X{=}1)=0.5$ and $P(Z{=}1)=0$ the only subgroups
with positive weight are the two with $Z=0$, giving $0.5(-0.078)+0.5(-0.369)=-0.224$,
which is the marginal value. The consistency between these marginal and population-average conditional measures is expected due to direct collapsibility. As discussed, the mismatch for the RR and OR is due to the lack of direct collapsibility. 

\medskip
{\footnotesize
\captionof{table}{Marginal OR over the full grid of populations defined by the
distribution of the prognostic factor $X$ (rows) and the effect modifier $Z$ (columns). We have set $X$ and $Z$ to be independent, but interior entries will generally depend on the association between $X$ and $Z$ as well as on the margins. For example, at $P(X{=}1)=P(Z{=}1)=0.5$ the marginal OR ranges from
$0.33$ to $0.51$ as the correlation between covariates varies from $-0.8$ to $+0.8$. The first and last columns depend only on $P(X=1)$ because $Z$ is constant across the population; the first and last rows depend only on $P(Z=1)$.}
\label{tab:collapsibility-transport}
\centering
\begin{tabular}{@{}lcccc@{}}
\toprule
& \multicolumn{4}{c}{\textbf{$P(Z{=}1)$ (effect modifier)}} \\
\cmidrule(l){2-5}
\textbf{$P(X{=}1)$ (prognostic)} & \textbf{0} & \textbf{0.5} & \textbf{0.8} & \textbf{1} \\
\midrule
0   & 0.20 & 0.35 & 0.44 & 0.50 \\
0.5 & 0.27 & 0.42 & 0.52 & 0.59 \\
0.8 & 0.23 & 0.37 & 0.47 & 0.55 \\
1   & 0.20 & 0.33 & 0.42 & 0.50 \\
\bottomrule
\end{tabular}
}

\medskip
{\footnotesize
\captionof{table}{Marginal effect measures compared with the simple weighted average of subgroup-level conditional effect measures, for selected populations. The RD agrees exactly, being
directly collapsible; the RR and the OR do not. In the last row, $P(X=1)=1, P(Z=1)=1$, such that the population is homogeneous, so marginal and population-average conditional effects necessarily coincide for all measures.}
\label{tab:collapsibility-compare}
\centering
\begin{tabular}{@{}lrrrrrr@{}}
\toprule
& \multicolumn{2}{c}{\textbf{Risk difference}} & \multicolumn{2}{c}{\textbf{Risk ratio}} & \multicolumn{2}{c}{\textbf{Odds ratio}} \\
\cmidrule(lr){2-3}\cmidrule(lr){4-5}\cmidrule(l){6-7}
\textbf{Population} & Marginal & Wtd.\ avg. & Marginal & Wtd.\ avg. & Marginal & Wtd.\ avg. \\
\midrule
$P(X{=}1)=0.5$, $P(Z{=}1)=0$   & $-0.224$ & $-0.224$ & 0.36 & 0.30 & 0.27 & 0.20 \\
$P(X{=}1)=0.5$, $P(Z{=}1)=0.5$ & $-0.167$ & $-0.167$ & 0.52 & 0.46 & 0.42 & 0.35 \\
$P(X{=}1)=0.8$, $P(Z{=}1)=0.8$ & $-0.180$ & $-0.180$ & 0.64 & 0.61 & 0.47 & 0.44 \\
$P(X{=}1)=1$, $P(Z{=}1)=1$     & $-0.171$ & $-0.171$ & 0.71 & 0.71 & 0.50 & 0.50 \\
\bottomrule
\end{tabular}
}

\section{Aggregation effect in a state-transition model}\label{app:cohorttransitionmodel}
In a state-transition model, non-linearity arises from the recursive application of the transition matrix. Specifically, at each cycle, individuals transition between health states according to per-cycle transition probabilities that are functions of individual covariates $\mathbf{x}$. A vector of state occupancy at cycle $t$ is obtained by repeated matrix multiplication:
\begin{equation*}
S_t(\mathbf{x}) = S_0 \cdot \mathbf{T}(\mathbf{x})^t
\end{equation*}
where $\mathbf{T}(\mathbf{x})$ is the transition matrix for an individual with covariates $\mathbf{x}$ and $S_0$ is the initial state occupancy vector. Since matrix exponentiation is a non-linear operation, applying the model to a single population-average transition matrix is not equivalent to averaging the resulting state occupancies across individuals:
\begin{equation*}
\int_{\mathfrak{X}} S_0 \cdot \mathbf{T}(\mathbf{x})^t f_{(P)}(\mathbf{x}) d\mathbf{x} \neq 
S_0 \cdot \left(\int_{\mathfrak{X}} \mathbf{T}(\mathbf{x}) f_{(P)}(\mathbf{x}) d\mathbf{x}\right)^t. 
\end{equation*}
That is, a cohort model uses a single averaged transition matrix applied repeatedly across cycles, whereas the, more appropriate, individual-level simulation approach would average the state occupancies across individuals with different covariate profiles to estimate the marginal NHB. This discrepancy compounds across cycles, growing with the length of the time horizon.

\section{Exception for the aggregation effect with partitioned survival models}\label{app:aggreffectpsm}

In oncology,  a partitioned survival cost-effectiveness model is frequently used. Such a cohort model can be constructed directly from marginal survival or state-occupancy curves. Those curves are themselves obtained by averaging individual trajectories over the target population:
\begin{equation*}
\bar{S}_t^{(k)} = \int_{\mathfrak{X}} S_t^{(k)}(\mathbf{x}) f_{(P)}(\mathbf{x}) d\mathbf{x}
\end{equation*}
so that the averaging takes place before the transformation $\varphi()$ is applied. NHB is then computed as a linear functional of the average state occupancy $\bar{S}_t^{(k)}$. In this case, the result is equivalent to averaging the state occupancies across the individuals in the target population: 
\begin{equation*}
\int_0^T \mathbf{c}_k^\top \bar{S}_t^{(k)} \, dt = 
\int_0^T \mathbf{c}_k^\top \int_{\mathfrak{X}} S_t^{(k)}(\mathbf{x}) 
f_{(P)}(\mathbf{x}) d\mathbf{x} \, dt = 
\int_{\mathfrak{X}} \int_0^T \mathbf{c}_k^\top S_t^{(k)}(\mathbf{x}) \, dt \, 
f_{(P)}(\mathbf{x}) d\mathbf{x},
\end{equation*}
where $\mathbf{c}_k$ is a vector of cycle-level weights (e.g., utilities and costs) for intervention $k$ derived from $\bm{\theta}_{k}$, introduced here to illustrate the linearity of NHB as a functional of state occupancy, and $T$ is the model time horizon. The last step follows by exchanging the order of integration, which is valid by linearity, guaranteeing that the operations commute.

The aggregation effect is therefore avoided, not because the cohort structure is absent, but because: the marginal survival curves that are input to the cohort model already encode the population-level averaging of individual trajectories; and the linearity of NHB as a functional of state occupancy ensures that this is equivalent to directly averaging individual-level NHB over the target population.

\section{Mismatch between population-average conditional and marginal measures}\label{app:cond_marg_mismatch}

The mismatch between population-average conditional and marginal measures depends on the outcome regression model and on whether we consider the baseline risk or the treatment effect. We investigate two commonly used outcome models in health economic modeling: a log-linear model with a log link, and a logistic model with a logit link, both falling within the generalized linear regression family. The treatment effect measure imposed by the log link is collapsible but not directly collapsible (log relative risk). That imposed by the logit link (log odds ratio) is non-collapsible. Our findings for the log link also apply for count outcomes and the log rate ratio, assuming person time is constant across subjects; for instance, in a study of non-fatal recurrent events where all subjects remain at risk throughout the follow-up period. In this case, (log) rate ratios are collapsible and can be interpreted as (log) risk ratios, as rates are risks per unit time. Otherwise, (log) rate ratios do not necessarily inherit collapsibility~\citep{sjolander2016note}.  

Similarly to the main text, we assume a linear predictor with correctly specified functional form that is linear in the covariates, with conditional treatment effects on the linear predictor scale varying linearly with any effect modifiers.

\subsection{Baseline risk}\label{sec:baseline}

\subsubsection{Log link}

For the log link, $g(\pi) = \log(\pi)$ and $g^{-1}(\eta) = \exp(\eta)$. The convexity of the exponential inverse link function can be investigated by examining its second derivative:
\begin{equation*}
\frac{d^2}{d\eta^2}\exp(\eta) = \exp(\eta) > 0 \quad \text{for all } \eta \in \mathbb{R}
\end{equation*}
which is strictly positive for all values of the linear predictor, confirming that $g^{-1}()$ is strictly convex (in Figure~\ref{fig:log_link}, any chord connecting two points on the curve lies strictly above the curve). By Jensen's inequality, this implies:
\begin{equation*}
g^{-1}\!\left(m_{0(P)}\right) = \exp\!\left(m_{0(P)}\right) \leq \int_{\mathfrak{X}} \exp\!\left(\mu + \mathbf{x}\beta_{1}\right) f_{(P)}\!\left(\mathbf{x}\right) d\mathbf{x} = {\overline{\pi}}_{0(P)}
\end{equation*}
so the population-average conditional baseline risk underestimates the marginal baseline risk under the log link, irrespective of the value of the linear predictor. 

\subsubsection{Logit link}

For the logit link, $g(\pi) = \log\!\left(\frac{\pi}{1-\pi}\right)$ and $g^{-1}(\eta) = \frac{1}{1+\exp(-\eta)}$. The convexity of the logistic inverse link function can be investigated by examining its second derivative:
\begin{equation*}
\frac{d^2}{d\eta^2}\frac{1}{1+\exp(-\eta)} = \frac{\exp(-\eta)\left(1-\exp(-\eta)\right)}{\left(1+\exp(-\eta)\right)^3}
\end{equation*}
which is: 
\begin{enumerate}
\item Positive when $\exp(-\eta) > 1$ -- that is $\eta < 0$, corresponding to probabilities below 0.5 -- such that $g^{-1}()$ is convex in this region;
\item Zero when $\exp(-\eta) =1$ -- that is $\eta = 0$, corresponding to a probability of 0.5 -- such that there is an inflection point at this value; 
\item Negative when $\exp(-\eta) < 1$ -- that is $\eta > 0$, corresponding to probabilities above 0.5 -- such that $g^{-1}()$ is concave in this region. 
\end{enumerate}

\begin{figure}[!htb]
\centering
\begin{subfigure}{.5\textwidth}
  \centering
\begin{tikzpicture}
\begin{axis}[
    xlabel={Linear predictor $\eta$},
    ylabel={$g^{-1}(\eta) = \exp(\eta)$},
    xmin=-3, xmax=3,
    ymin=0, ymax=8,
    grid=major,
    width=7.5cm,
    height=6cm,
    legend pos=north west
]
\addplot[blue, thick, domain=-3:3, samples=100] {exp(x)};
\addplot[red, dashed, thick] coordinates {(-2, 0.135) (2, 7.389)};
\addplot[black, dotted, thick] coordinates {(0, 1) (0, 0)};
\node at (axis cs:0.2, 0.5) [anchor=west] {$\eta = 0$};
\legend{$\exp(\eta)$}
\end{axis}
\end{tikzpicture}
  \caption{Exponential inverse link function}
  \label{fig:log_link}
\end{subfigure}%
\begin{subfigure}{.5\textwidth}
  \centering
\begin{tikzpicture}
\begin{axis}[
    xlabel={Linear predictor $\eta$},
    ylabel={$g^{-1}(\eta) = \frac{1}{1+\exp(-\eta)}$},
    xmin=-6, xmax=6,
    ymin=0, ymax=1,
    grid=major,
    width=7.5cm,
    height=6cm,
    ytick={0, 0.25, 0.5, 0.75, 1.0},
    legend pos=north west
]
\addplot[blue, thick, domain=-6:6, samples=100] {1/(1+exp(-x))};
\addplot[red, dashed, thick] coordinates {(-6, 0.002) (0, 0.5)};
\addplot[green!60!black, dashed, thick] coordinates {(0, 0.5) (6, 0.998)};
\addplot[black, dotted, thick] coordinates {(-6, 0.5) (6, 0.5)};
\node at (axis cs:-4, 0.15) [anchor=west] {\small Convex ($\eta < 0$)};
\node at (axis cs:0.5, 0.85) [anchor=west] {\small Concave ($\eta > 0$)};
\legend{$\frac{1}{1+\exp(-\eta)}$}
\end{axis}
\end{tikzpicture}
  \caption{Logistic inverse link function}
  \label{fig:logit_link}
\end{subfigure}
\caption{Demonstration of convexity (or concavity) for the exponential (left-hand side) and logistic (right-hand side) inverse link functions.}
\end{figure}
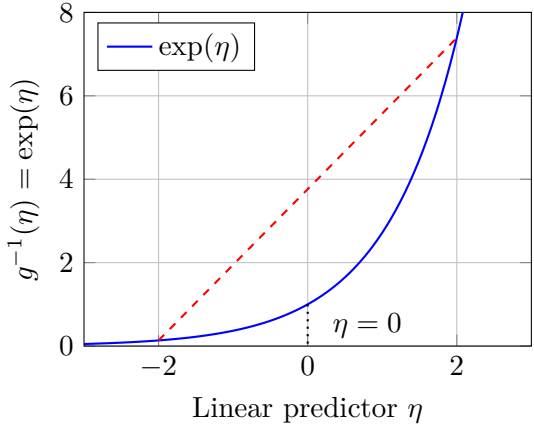
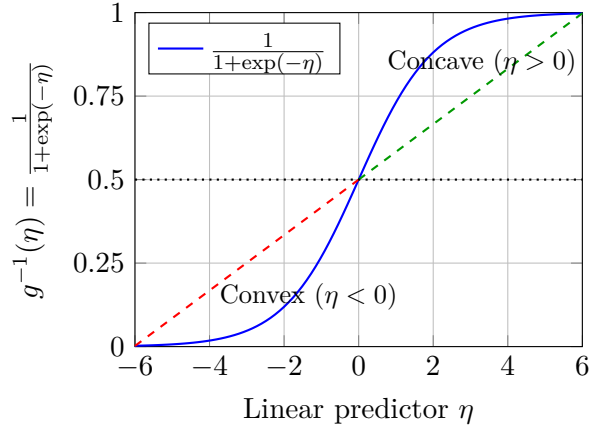

This is illustrated in Figure~\ref{fig:logit_link}. By Jensen's inequality:
\begin{itemize}
\item If $m_{0(P)} < 0$ (probability below 0.5): $g^{-1}\!\left(m_{0(P)}\right) \leq {\overline{\pi}}_{0(P)}$, so the population-average conditional baseline risk underestimates the marginal baseline risk; 
\item If $m_{0(P)} > 0$ (probability above 0.5): $g^{-1}\!\left(m_{0(P)}\right) \geq {\overline{\pi}}_{0(P)}$, so that the population-average conditional baseline risk overestimates the marginal baseline risk. 
\end{itemize}

For many events used in cost-effectiveness models (e.g.~transition probabilities between health states), the baseline probability is typically below 0.5, suggesting that the population-average conditional baseline risk will tend to underestimate the marginal baseline risk. 

\subsection{Treatment effect}\label{sec:trt_effect}

\subsubsection{Log link}\label{sec:trt_effect_log}

We work on the linear predictor scale. For the log link, the treatment effect on the linear predictor scale is the log relative risk. The population-average conditional log relative risk is
\begin{equation*}
d_{k(P)} = \int_{\mathfrak{X}} \left(\log\pi_k(\mathbf{x}) - \log\pi_0(\mathbf{x})\right) f_{(P)}\!\left(\mathbf{x}\right) d\mathbf{x}
\end{equation*}
and the marginal log relative risk is:
\begin{equation*}
\Delta_{k(P)} = \log\!\left(\int_{\mathfrak{X}} \pi_k(\mathbf{x})  f_{(P)}\!\left(\mathbf{x}\right) d\mathbf{x}\right) - \log\!\left(\int_{\mathfrak{X}} \pi_0(\mathbf{x}) f_{(P)}\!\left(\mathbf{x}\right) d\mathbf{x}\right).
\end{equation*}

The second derivative of $g() = \log()$ is: 
\begin{equation*}
\frac{d^2}{d\pi^2}\log(\pi) = -\frac{1}{\pi^2} < 0 \quad \text{for all } \pi > 0
\end{equation*}
so that $g()$ is strictly concave -- as expected, since $g^{-1}()=\exp()$ in Online Supplement~\ref{sec:baseline} is strictly convex. By Jensen's inequality, for each treatment arm $j \in \{0, k\}$:
\begin{equation*}
\log\!\left(\int_{\mathfrak{X}} \pi_j(\mathbf{x}) f_{(P)}\!\left(\mathbf{x}\right)  d\mathbf{x}\right) \geq \int_{\mathfrak{X}} \log\!\left(\pi_j(\mathbf{x})\right)  f_{(P)}\!\left(\mathbf{x}\right) d\mathbf{x}
\end{equation*}

The 
discrepancy between $d_{k(P)}$ and $\Delta_{k(P)}$ can therefore be written as:
\begin{equation*}
d_{k(P)} - \Delta_{k(P)} =\underbrace{\left(\int_{\mathfrak{X}} \log\pi_k(\mathbf{x}) f_{(P)}(\mathbf{x}) d\mathbf{x} - \log\bar{\pi}_{k(P)}\right)}_{\leq\, 0} - \underbrace{\left(\int_{\mathfrak{X}} \log\pi_0(\mathbf{x}) f_{(P)}(\mathbf{x}) d\mathbf{x} - \log\bar{\pi}_{0(P)}\right)}_{\leq\, 0}
\end{equation*}
where $\bar{\pi}_j = \int_{\mathfrak{X}} \pi_j(\mathbf{x}) f_{(P)}\!\left(\mathbf{x}\right)  d\mathbf{x}$ and both bracketed terms are non-positive by Jensen's inequality. Each term can be approximated by a second-order Taylor expansion:
\begin{equation*}
\int_{\mathfrak{X}} \log\pi_j(\mathbf{x}) f_{(P)}(\mathbf{x}) d\mathbf{x} - \log\bar{\pi}_{j(P)} \approx -\frac{\text{Var}(\pi_j(\mathbf{x}))}{2\bar{\pi}_{j(P)}^2}
\end{equation*}
so the discrepancy between $d_{k(P)}$ and $\Delta_{k(P)}$ is approximately:
\begin{equation*}
d_{k(P)} - \Delta_{k(P)} \approx \frac{1}{2}\left(\text{CV}_0^2 - \text{CV}_k^2\right)
\end{equation*}
where $\text{CV}_j = \sqrt{\text{Var}(\pi_j(\mathbf{x}))}/\bar{\pi}_{j(P)}$ is the coefficient of variation of individual-level risks in arm $j$, capturing the variability of risks relative to the mean in the corresponding arm. We note:
\begin{itemize}
\item When $\text{CV}_0 > \text{CV}_k$, $d_{k(P)} > \Delta_{k(P)}$; 
\item When $\text{CV}_0 < \text{CV}_k$, $d_{k(P)} < \Delta_{k(P)}$.
\end{itemize}
Under the specified outcome regression model, the individual-level conditional risks in each treatment arm are:
\begin{equation*}
\pi_k(\mathbf{x}) = \exp\!\left(\mu + \mathbf{x}(\beta_1 + \beta_{2,k}) + \gamma_k\right), \quad \pi_0(\mathbf{x}) = \exp\!\left(\mu + \mathbf{x}\beta_1\right).
\end{equation*} 
Using the delta method: $\text{Var}(\pi_j(\mathbf{x})) \approx \left(\frac{d\pi_j}{d\eta_j}\right)^2 \text{Var}(\eta_j(\mathbf{x})) = \bar{\pi}_{j(P)}^2 \cdot \text{Var}(\eta_j(\mathbf{x}))$, where $\eta_j(\mathbf{x})$ is the linear predictor in arm $j$, i.e., $\eta_k(\mathbf{x}) = \mu + \mathbf{x}(\beta_1 + \beta_{2,k}) + \gamma_k$ and  $\eta_0(\mathbf{x}) = \mu + \mathbf{x}\beta_1$, and where $\frac{d\pi_j}{d\eta_j} = \exp(\eta_j) = \pi_j$. Hence, the coefficient of variation in each arm is:
\begin{equation*}
\text{CV}_j = \frac{\sqrt{\text{Var}(\pi_j(\mathbf{x}))}}{\bar{\pi}_{j(P)}} \approx \frac{\bar{\pi}_{j(P)} \cdot \sqrt{\text{Var}(\eta_j(\mathbf{x}))}}{\bar{\pi}_{j(P)}} = \sqrt{\text{Var}(\eta_j(\mathbf{x}))}.
\end{equation*}

Therefore, the discrepancy between $d_{k(P)}$ and $\Delta_{k(P)}$ is approximately:
\begin{equation}
d_{k(P)} - \Delta_{k(P)} \approx \frac{1}{2}\left(\text{Var}(\eta_0(\mathbf{x})) - \text{Var}(\eta_k(\mathbf{x}))\right) = \frac{1}{2}\left(\text{Var}(\mathbf{x}\beta_1) - \text{Var}(\mathbf{x}(\beta_1 + \beta_{2,k}))\right)
\label{eqn:log-discrepancy}
\end{equation}
since constant terms do not contribute to the variance. Two cases are worth noting:

\textbf{Without effect modification} ($\beta_{2,k} = 0$): $\text{Var}(\mathbf{x}(\beta_1 + \beta_{2,k})) = \text{Var}(\mathbf{x}\beta_1)$, so the discrepancy $d_{k(P)} - \Delta_{k(P)} \approx 0$, which is consistent with the (log) relative risk being a collapsible effect measure: in the absence of effect modification, the population-average conditional and marginal (log) relative risks coincide, regardless of the distribution of prognostic factors in the target population, because the marginal measure is a weighted average of constant subgroup-level conditional measures.

\textbf{With effect modification} ($\beta_{2,k} \neq 0$): the discrepancy is determined by the difference in variances of the linear predictors across the two arms. Since $\text{Var}(\mathbf{x}(\beta_1 + \beta_{2,k})) = (\beta_1 + \beta_{2,k})^\top 
\text{Cov}(\mathbf{x})(\beta_1 + \beta_{2,k})= \beta_1^\top\text{Cov}(\mathbf{x})\beta_1 + 
2\beta_{2,k}^\top\text{Cov}(\mathbf{x})\beta_1 + 
\beta_{2,k}^\top\text{Cov}(\mathbf{x})\beta_{2,k}$ and $\text{Var}(\mathbf{x}\beta_1) = \beta_1^\top\text{Cov}(\mathbf{x})\beta_1$: 
\begin{equation}
\text{Var}(\mathbf{x}(\beta_1 + \beta_{2,k})) - \text{Var}(\mathbf{x}\beta_1) = \beta_{2,k}^\top \text{Cov}(\mathbf{x})\beta_{2,k} + 2\beta_{2,k}^\top \text{Cov}(\mathbf{x})\beta_1
\label{eqn:log-discrepancy-reverse}
\end{equation}
where the first term $\beta_{2,k}^\top \text{Cov}(\mathbf{x})\beta_{2,k} \geq 0$. The second term $2\beta_{2,k}^\top \text{Cov}(\mathbf{x})\beta_1$ depends on effect modification $\beta_{2,k}$ and its relation with prognostic effects $\beta_1$ through the covariance matrix $\text{Cov}(\mathbf{x})$, and can negative. For instance, for an effective intervention on a harmful outcome, if individuals with higher baseline (larger $\mathbf{x}\beta_1$) benefit more from treatment and have more negative $\mathbf{x}\beta_{2,k}$ (larger log relative risk reduction); or for an effective intervention on a beneficial outcome, if individuals with lower baseline benefit more from treatment and have more positive $\mathbf{x}\beta_{2,k}$. If the negative cross-term is of large enough magnitude to outweigh the positive quadratic term, then the difference in eq.~\ref{eqn:log-discrepancy-reverse} is negative, meaning that the difference in eq.~\ref{eqn:log-discrepancy} is positive (for an effective intervention, the population-average conditional log relative risk underestimates benefit). Otherwise, the difference in eq.~\ref{eqn:log-discrepancy} is negative meaning that, for an effective intervention, the population-average conditional log relative risk overestimates benefit.  

It is widely understood that, when effect modification is present, treatment rankings may conflict between population-average conditional and marginal estimates for non-collapsible measures~\citep{phillippo2025effect}. The potentially changing signs of the expressions in eq.~\ref{eqn:log-discrepancy} and eq.~\ref{eqn:log-discrepancy-reverse} suggest that effect modification can also result in conflicting treatment recommendations for collapsible (but not directly collapsible) treatment effect measures, such as the (log) risk ratio scale with the log link.

\subsubsection{Logit link}

We work on the linear predictor scale. For the logit link, the treatment effect on the linear predictor scale is the log odds ratio. The population-average conditional log odds ratio is
\begin{equation*}
d_{k(P)} = \int_{\mathfrak{X}} \left(g(\pi_k(\mathbf{x})) - g(\pi_0(\mathbf{x}))\right) f_{(P)}\!\left(\mathbf{x}\right) d\mathbf{x}
\end{equation*}
and the marginal log odds ratio is:
\begin{equation*}
\Delta_{k(P)} = g\!\left(\int_{\mathfrak{X}} \pi_k(\mathbf{x}) f_{(P)}\!\left(\mathbf{x}\right) d\mathbf{x}\right) - g\!\left(\int_{\mathfrak{X}} \pi_0(\mathbf{x}) f_{(P)}\!\left(\mathbf{x}\right) d\mathbf{x}\right).
\end{equation*}
The second derivative of the logit function $g() = \log\!\left(\frac{\pi}{1-\pi}\right)$ is:
\begin{equation*}
\frac{d^2}{d\pi^2}\log\!\left(\frac{\pi}{1-\pi}\right) = 
\frac{2\pi - 1}{\pi^2(1-\pi)^2}
\end{equation*}
which is negative for $\pi < 0.5$ (concave), zero at $\pi = 0.5$ (inflection point), and positive for $\pi > 0.5$ (convex), as expected from Online Supplement~\ref{sec:baseline}. As for the log link, the discrepancy between $d_{k(P)}$ and $\Delta_{k(P)}$ can be written as:
\begin{equation*}
d_{k(P)} - \Delta_{k(P)} =\left(\int_{\mathfrak{X}} g(\pi_k(\mathbf{x})) f_{(P)}(\mathbf{x}) d\mathbf{x} - g(\bar{\pi}_{k(P)})\right) - \left(\int_{\mathfrak{X}} g(\pi_0(\mathbf{x})) f_{(P)}(\mathbf{x}) d\mathbf{x} - g(\bar{\pi}_{0(P)})\right)
\end{equation*}
where $\bar{\pi}_j = \int_{\mathfrak{X}} \pi_j(\mathbf{x}) f_{(P)}\!\left(\mathbf{x}\right)  d\mathbf{x}$. Each term can be approximated by a second-order Taylor expansion. For each treatment arm $j \in \{0, k\}$:
\begin{equation*}
\int_{\mathfrak{X}} g(\pi_j(\mathbf{x})) f_{(P)}(\mathbf{x}) 
d\mathbf{x} - g(\bar{\pi}_{j(P)})  \approx \frac{1}{2} \cdot \frac{2\bar{\pi}_{j(P)}-1}
{\bar{\pi}_{j(P)}^2(1-\bar{\pi}_{j(P)})^2} \cdot \text{Var}(\pi_j(\mathbf{x})).
\end{equation*}
so the discrepancy between $d_{k(P)}$ and $\Delta_{k(P)}$ is approximately:
\begin{equation}
d_{k(P)} - \Delta_{k(P)} \approx \frac{1}{2}\left(  \frac{2\bar{\pi}_{k(P)}-1}{\bar{\pi}_{k(P)}^2(1-\bar{\pi}_{k(P)})^2} \cdot \text{Var}(\pi_k(\mathbf{x})) - 
\frac{2\bar{\pi}_{0(P)}-1}{\bar{\pi}_{0(P)}^2(1-\bar{\pi}_{0(P)})^2} \cdot \text{Var}(\pi_0(\mathbf{x}))
\right).
\label{eqn:discrepancy-logodds-general}
\end{equation}
Two cases are worth noting:
\paragraph{Without effect modification ($\beta_{2,k} = 0$):} We have
\begin{equation}
d_{k(P)} - \Delta_{k(P)} \approx \frac{1}{2}\left(
\frac{2\bar{\pi}_{k(P)}-1}{\bar{\pi}_{k(P)}^2(1-\bar{\pi}_{k(P)})^2}
-
\frac{2\bar{\pi}_{0(P)}-1}{\bar{\pi}_{0(P)}^2(1-\bar{\pi}_{0(P)})^2} 
\right) 
\cdot \text{Var}(\pi_0(\mathbf{x}))
\label{eqn:discrepancy-logodds}
\end{equation}
because $\text{Var}(\pi_k(\mathbf{x})) = \text{Var}(\pi_0(\mathbf{x}))$ when $\beta_{2,k} = 0$. The above expression is non-zero whenever $\bar{\pi}_{k(P)} \neq \bar{\pi}_{0(P)}$, confirming the non-collapsibility of the log odds ratio: $d_{k(P)} \neq \Delta_{k(P)}$ even in the absence of effect modification. The direction of the discrepancy in eq.~\ref{eqn:discrepancy-logodds} depends on the sign of:
\begin{equation}
\frac{2\bar{\pi}_{k(P)}-1}{\bar{\pi}_{k(P)}^2(1-\bar{\pi}_{k(P)})^2}
-
\frac{2\bar{\pi}_{0(P)}-1}{\bar{\pi}_{0(P)}^2(1-\bar{\pi}_{0(P)})^2}
\label{eqn:sign}
\end{equation}
which is negative when the intervention arm has a lower risk than the reference comparator arm, giving $\Delta_{k(P)} > d_{k(P)}$. Conversely, the expression in eq.~\ref{eqn:sign} is positive when the reference comparator arm has a lower risk than the intervention arm, giving $\Delta_{k(P)} < d_{k(P)}$. In both situations, the population-average conditional log odds ratio overestimates benefit for an effective intervention: for a harmful outcome, $d_{k(P)}$ is more negative than $\Delta_{k(P)}$; and for a beneficial outcome, $d_{k(P)}$ is more positive than $\Delta_{k(P)}$. Of note, there is no discrepancy in the null scenario where the intervention has no effect, so that all conditional and marginal log odds ratios are zero ($\bar{\pi}_{k(P)}=\bar{\pi}_{0(P)}$). 

\paragraph{With effect modification ($\beta_{2,k} \neq 0$):} Using the delta method analogously to the log link case: $\text{Var}(\pi_j(\mathbf{x})) \approx \left(\frac{d\pi_j}{d\eta_j}\right)^2 \text{Var}(\eta_j(\mathbf{x})) = \bar{\pi}_{j(P)}^2(1-\bar{\pi}_{j(P)})^2 \cdot \text{Var}(\eta_j(\mathbf{x}))$, where $\eta_j(\mathbf{x})$ is the linear predictor in arm $j$, i.e., $\eta_k(\mathbf{x}) = \mu + \mathbf{x}(\beta_1 + \beta_{2,k}) + \gamma_k$ and  $\eta_0(\mathbf{x}) = \mu + \mathbf{x}\beta_1$, and where $\frac{d\pi_j}{d\eta_j} = \pi_j(1-\pi_j)$. Hence, for each arm: 
\begin{equation*}
\frac{2\bar{\pi}_{j(P)}-1}{\bar{\pi}_{j(P)}^2(1-\bar{\pi}_{j(P)})^2} \cdot \text{Var}(\pi_j(\mathbf{x}))\approx (2\bar{\pi}_{j(P)}-1) \cdot \text{Var}(\eta_j(\mathbf{x})).
\end{equation*}
Substituting this expression at $\bar{\pi}_{0(P)}$ and $\bar{\pi}_{k(P)}$ into the discrepancy approximation in eq.~\ref{eqn:discrepancy-logodds-general}:
\begin{equation*}
d_{k(P)} - \Delta_{k(P)} \approx \frac{1}{2}\left(
(2\bar{\pi}_{k(P)}-1) \cdot \text{Var}(\eta_k(\mathbf{x})) -
(2\bar{\pi}_{0(P)}-1) \cdot \text{Var}(\eta_0(\mathbf{x}))  
\right).
\end{equation*}
As for the log link, the discrepancy is related to the difference in variances of the linear predictors across the two arms. However, for the logit link, the weights $(2\bar{\pi}_{j(P)}-1)$ introduce an additional 
dependence on the marginal risks in each arm. As constant terms do not contribute to the variance: 
\begin{equation*}
d_{k(P)} - \Delta_{k(P)} \approx \frac{1}{2}\left(
(2\bar{\pi}_{k(P)}-1) \cdot \text{Var}(\mathbf{x}(\beta_1 + \beta_{2,k}))
-
(2\bar{\pi}_{0(P)}-1) \cdot \text{Var}(\mathbf{x}\beta_1) 
\right).
\end{equation*}
Similarly to the log link, eq.~\ref{eqn:log-discrepancy-reverse} holds so the sign of the discrepancy depends on effect modification $\beta_{2,k}$ and its relation with prognostic effects $\beta_1$ through the covariance matrix $\text{Cov}(\mathbf{x})$. Here the discrepancy is more specifically determined by the differential magnitude of weighted variance terms, with $(2\bar{\pi}_{j(P)}-1)$ introducing an additional dependence on the marginal risks in each arm. 

\subsubsection{A note on proportional hazards models and the (log) hazard ratio}

Proportional hazards models -- and the corresponding (log) hazard ratio scale -- are widely used for time-to-event data in health economic modeling. For the generalized linear models with a log or logit link previously discussed, the discrepancy between population-average conditional and marginal treatment effects arises from the non-linearity of link function $g()$. While proportional hazards models use non-linear links to map hazards to the linear predictor, there is an additional mechanism resulting in the discrepancy between population-average conditional and marginal measures (and non-collapsibility) for the (log) hazard ratio. 

This is conditioning on past survival, leading to the selective depletion of the at-risk population over time. As time progresses, individuals with higher baseline risk are depleted from the at-risk population more rapidly, progressively changing the covariate composition of the at-risk population. This causes the marginal (log) hazard ratio to be time-varying, even when the corresponding conditional (log) hazard ratio is constant, and introduces a systematic discrepancy between the population-average conditional and marginal (log) hazard ratios. The discrepancy has been previously investigated by Phillippo et al~\citep{phillippo2025effect}. While population-average conditional (log) hazard ratios overestimate the benefit of interventions relative to marginal (log) hazard ratios in the absence of effect modification, the mismatch can go in either direction when there is effect modification. Proportional hazards models and the (log) hazard ratio scale feature in the illustrative example in Section~\ref{sec:example} and population-average conditional versus marginal hazard ratios are discussed in the context of the motivating example in Online Supplement~\ref{app:tvhr}.

\section{Additional output for the illustrative example}\label{illustrative-example-output}

\subsection{Health-state occupancy over time}\label{hsoccupancy}

\begin{figure}[!htb]
\centering
\includegraphics[width=\linewidth]{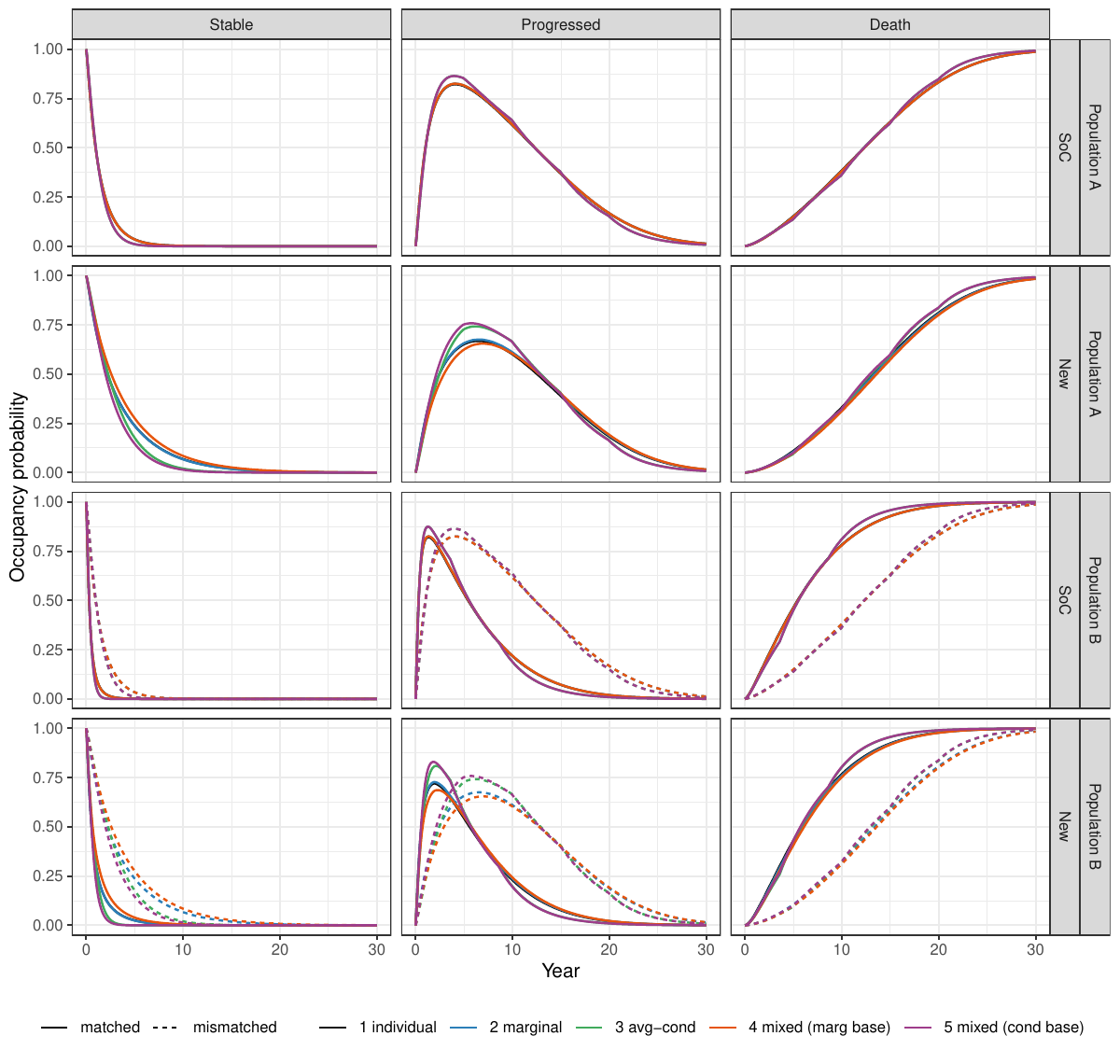}
\caption{Health-state occupancy (Stable, Progressed, Death) for standard of care (SoC)
and the new treatment, in both target populations (A and B). With the individual-level modeling approach, we obtain appropriate occupancy estimates for both target populations (solid black curves). With the cohort approaches, we can use correct-population inputs (matched) for both target populations, but still misestimate occupancy (solid colored curves). Using incorrect-population inputs with the cohort approaches (i.e., population A estimates used for a population B target population) results in much larger discrepancies (dashed colored curves).}
\label{fig:example-stateocc}
\end{figure}

Figure~\ref{fig:example-stateocc} shows the health-state occupancy (the share of the cohort in each state) over time for every modeling scenario in both target populations. It is the state-occupancy counterpart of the survival curves in the main text (Figure~\ref{fig:example-survival}), and carries the same message: the individual-level approach recovers the correct occupancy for both target populations, whereas the cohort approaches produce important discrepancies even with correct-population (matched) inputs. These discrepancies are exacerbated with incorrect-population (mismatched) inputs.

\FloatBarrier

\subsection{The time-varying marginal hazard ratio}\label{app:tvhr}

\begin{figure}[htbp]
\centering
\includegraphics[width=0.8\linewidth]{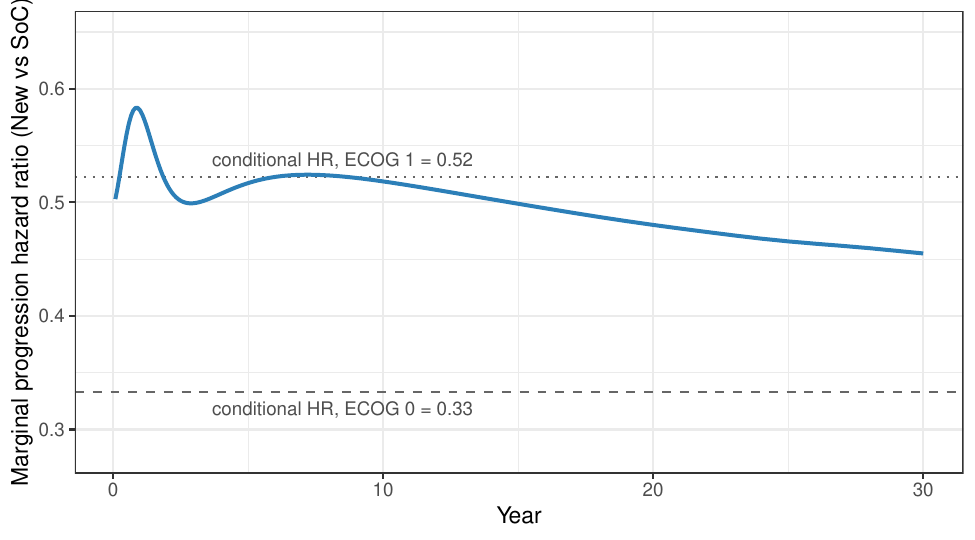}
\caption{The marginal HR (new treatment ``New'' vs.\ SoC) for progression for Population~B
varies over time. It stays close to the weaker ECOG~1 conditional hazard ratio (0.52,
dotted), well above the more favorable ECOG~0 value (0.33, dashed).}
\label{fig:example-mhr}
\end{figure}

Figure~\ref{fig:example-mhr} shows the marginal hazard ratio (new treatment vs.\ SoC) for progression for Population~B over time. Each patient's conditional hazard ratio is constant ($0.33$ for ECOG~0, $0.52$ for ECOG~1), yet the marginal hazard ratio still changes over time. The marginal hazard in each arm is the average of the patient-level hazards, weighted by the patients in that arm still at risk (still in the Stable state). Higher-hazard patients (ECOG~1, and older ages) progress sooner and leave the at-risk set first, so that set is progressively enriched with lower-hazard ECOG~0 patients, and this depletion runs faster under standard of care than under the more effective new treatment. Two effects follow. The shift toward ECOG~0 patients, whose conditional hazard ratio is the more favorable $0.33$, pulls the marginal hazard ratio down; by year~5, almost all patients still in the Stable state are ECOG~0. The slower depletion under the new treatment pulls it up, because that arm keeps more of its higher-risk patients than the SoC arm. The second effect largely offsets the first. The marginal hazard ratio starts at $0.503$, rises to $0.583$ at about 10 months, above both conditional values, stays between $0.499$ and $0.524$ from year~2 to year~10, and falls only to $0.455$ by year~30. This time dependence, a form of frailty selection (depletion of susceptibles), is intrinsic to the non-collapsible hazard ratio, and is why no single value can stand in for the marginal effect measure.

\FloatBarrier

\clearpage
\section{Implemention of illustrative example with R}
\noindent The following sections reproduce the
oncology illustrative example of Section~\ref{sec:example} using R. All three alternative implementations use the same model structure, data, and target population; they differ only in the modeling engine.

\subsection{Hand-coded discrete-time state-transition models}
\label{app:handcoded}
This section is the ``hand-coded'' R implementation of the oncology
illustrative example.

\subsubsection{Setup}

\begin{lstlisting}[language=R]
library("data.table")
library("ggplot2")
theme_set(theme_bw())
\end{lstlisting}

\subsubsection{Model parameters}

All model inputs: conditional Weibull PH progression coefficients, the
mortality-while-stable treatment hazard ratio, the age-banded background
mortality table, utilities, costs, and model settings.

\begin{lstlisting}[language=R]
# Weibull PH coefficients for progression (Stable -> Progressed)
prog_coef <- c(
  lngamma = 0.15,          # shape: ln(gamma), gamma ~ 1.16 (mildly increasing hazard)
  cons = -5.5,             # intercept (baseline log-scale, slower progression)
  age = 0.08,              # older patients progress faster (prognostic, HR=1.08/yr)
  ecog1 = 1.10,            # ECOG 1 patients progress faster (prognostic, HR = 3.00)
  trt = -1.10,             # treatment effect at ECOG 0 (conditional HR = 0.33)
  trt_ecog1 = 0.45         # interaction: treatment less effective in ECOG 1
  # ECOG 0: HR = exp(-1.10) = 0.33
  # ECOG 1: HR = exp(-1.10 + 0.45) = 0.52
)

# Treatment effect on mortality while stable (HR for death from Stable)
hr_death_trt <- 0.45       # Conditional HR for death while stable

# Background mortality (annual rates by age, converted to monthly below)
mort_annual <- data.table(
  age_lower = c(50, 55, 60, 65, 70, 75, 80),
  age_upper = c(55, 60, 65, 70, 75, 80, Inf),
  rate = c(0.004, 0.007, 0.012, 0.020, 0.035, 0.060, 0.100)
)

# Post-progression mortality multiplier
mr_prog_multiplier <- 3.0  # 3x background mortality after progression

# Utility values
u_stable <- 0.75
u_progressed <- 0.45

# Monthly costs
c_soc_drug <- 1000         # SoC drug cost per month (while Stable)
c_new_drug <- 3000          # New treatment cost per month (while Stable)
c_prog_care <- 1500         # Post-progression care per month
c_death <- 0

# Model settings
n_cycles <- 360             # 30 years in monthly cycles
cycle_length <- 1/12        # 1 month in years
dr_qalys <- 0.035           # Annual discount rate for QALYs
dr_costs <- 0.035           # Annual discount rate for costs
\end{lstlisting}

\subsubsection{Two populations}

Population A (trial: younger, 30\% ECOG 1) and Population B (target:
older, 70\% ECOG 1), each an age (Beta-shaped) by ECOG distribution.

\begin{lstlisting}[language=R]
# Fine age grid
ages_A <- 50:70
ages_B <- 50:80

# Population A: Trial population (younger, less ECOG 1)
age_dens_A <- dbeta((ages_A - 50) / 20, 3, 3)
prop_ecog1_A <- 0.30
pop_A <- data.table(
  age = rep(ages_A, 2),
  ecog1 = rep(c(0, 1), each = length(ages_A)),
  patient_wt = c((1 - prop_ecog1_A) * age_dens_A / sum(age_dens_A),
                 prop_ecog1_A * age_dens_A / sum(age_dens_A))
)

# Population B: Target population (older, more ECOG 1)
age_dens_B <- dbeta((ages_B - 50) / 30, 5, 2)
prop_ecog1_B <- 0.70
pop_B <- data.table(
  age = rep(ages_B, 2),
  ecog1 = rep(c(0, 1), each = length(ages_B)),
  patient_wt = c((1 - prop_ecog1_B) * age_dens_B / sum(age_dens_B),
                 prop_ecog1_B * age_dens_B / sum(age_dens_B))
)
\end{lstlisting}

\includegraphics{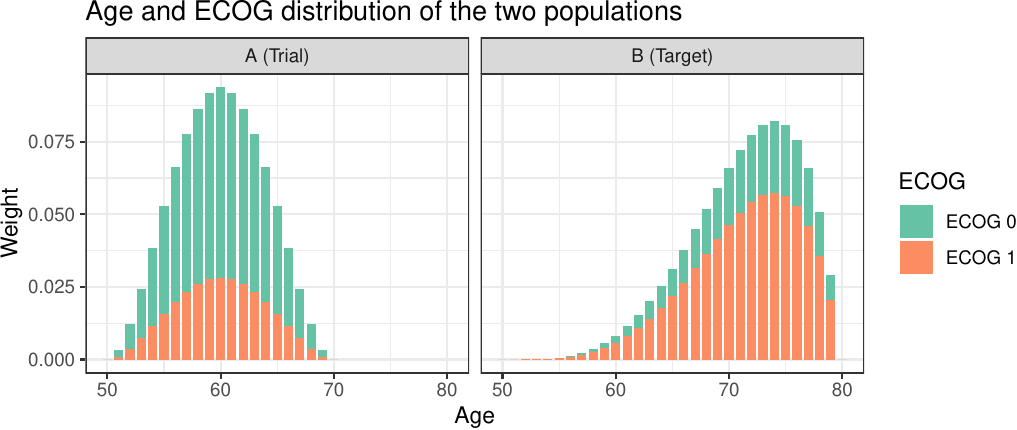}

\subsubsection{Helper functions}

Monthly transition probabilities from the conditional model, the
three-state cohort trace, discounted outcomes, and the
population-marginal inputs used by the cohort scenarios.

\begin{lstlisting}[language=R]
# Compute monthly progression probability for given covariates at each cycle
compute_prog <- function(age, ecog1, trt, times = 1:n_cycles) {
  shape <- exp(prog_coef["lngamma"])
  scale <- exp(prog_coef["cons"] + prog_coef["age"] * age +
               prog_coef["ecog1"] * ecog1 + prog_coef["trt"] * trt +
               prog_coef["trt_ecog1"] * trt * ecog1)
  # Convert from annual Weibull PH to monthly transition probability
  # Time in years: t_years = times * cycle_length
  t_yr <- times * cycle_length
  t_yr_prev <- (times - 1) * cycle_length
  # P(event in cycle) = 1 - S(t)/S(t-1) = 1 - exp(scale * (t_prev^shape - t^shape))
  tp <- 1 - exp(scale * (t_yr_prev^shape - t_yr^shape))
  pmin(pmax(tp, 0), 0.999)
}

# Compute monthly background mortality for given age at each cycle
# trt: 0 = SoC, 1 = new treatment (applies hr_death_trt to Stable mortality)
compute_mr_stable <- function(age, trt = 0, times = 1:n_cycles) {
  mr <- numeric(length(times))
  for (i in seq_along(times)) {
    age_now <- age + times[i] * cycle_length
    row <- mort_annual[age_now >= age_lower & age_now < age_upper]
    if (nrow(row) == 0) row <- mort_annual[.N]
    # Convert annual rate to monthly probability, apply treatment HR
    rate <- row$rate * ifelse(trt == 1, hr_death_trt, 1)
    mr[i] <- 1 - exp(-rate / 12)
  }
  mr
}

# Monthly mortality after progression (elevated, no treatment effect)
compute_mr_prog <- function(age, times = 1:n_cycles) {
  mr_base <- compute_mr_stable(age, trt = 0, times)  # no trt effect post-progression
  1 - (1 - mr_base)^mr_prog_multiplier
}

# Run cohort trace: 3 states (Stable, Progressed, Death)
run_cohort_trace <- function(prog, mr_s, mr_p) {
  n_states <- 3
  trace <- matrix(0, nrow = n_cycles + 1, ncol = n_states)
  trace[1, 1] <- 1  # Start in Stable

  for (t in 1:n_cycles) {
    tp <- matrix(0, n_states, n_states)
    tp[1, 1] <- max(0, 1 - prog[t] - mr_s[t])  # Stay stable
    tp[1, 2] <- prog[t]                          # Stable -> Progressed
    tp[1, 3] <- mr_s[t]                          # Stable -> Death
    tp[2, 2] <- max(0, 1 - mr_p[t])              # Stay progressed
    tp[2, 3] <- mr_p[t]                          # Progressed -> Death
    tp[3, 3] <- 1                                 # Absorbing
    # Ensure valid probabilities
    tp[tp < 0] <- 0
    tp <- tp / rowSums(tp)
    trace[t + 1, ] <- trace[t, ] %*% tp
  }
  trace
}

# Compute discounted QALYs and costs from trace
compute_outcomes <- function(trace, trt) {
  # Monthly discount factors
  disc_q <- (1 + dr_qalys)^(-(1:n_cycles) * cycle_length)
  disc_c <- (1 + dr_costs)^(-(1:n_cycles) * cycle_length)

  # QALYs (monthly state probs * utility * cycle length)
  qalys <- sum((trace[2:(n_cycles + 1), 1] * u_stable +
                trace[2:(n_cycles + 1), 2] * u_progressed) *
               cycle_length * disc_q)

  # Costs
  drug_cost <- if (trt == 1) c_new_drug else c_soc_drug
  costs <- sum((trace[2:(n_cycles + 1), 1] * drug_cost +
                trace[2:(n_cycles + 1), 2] * c_prog_care) *
               disc_c)

  list(qalys = qalys, costs = costs)
}

# Population-marginal inputs: eq. 4 applied to the cumulative risk (see below)
get_marginal_inputs <- function(pop, strategy) {
  w <- pop$patient_wt
  # Average the SURVIVAL function over the covariate distribution and difference
  # it (eq. 4 on the cumulative risk). Averaging each cycle's transition
  # probability with fixed weights instead keeps the baseline covariate mix for
  # the whole horizon; the two agree only in the first cycle.
  dif <- function(M) {
    S  <- t(apply(1 - M, 1, cumprod))
    Sb <- colSums(w * S)
    1 - Sb / c(1, Sb[-length(Sb)])
  }
  list(
    prog = dif(t(sapply(1:nrow(pop), function(i)
             compute_prog(pop$age[i], pop$ecog1[i], strategy)))),
    mr_s = dif(t(sapply(1:nrow(pop), function(i)
             compute_mr_stable(pop$age[i], trt = strategy)))),
    mr_p = dif(t(sapply(1:nrow(pop), function(i)
             compute_mr_prog(pop$age[i]))))
  )
}
\end{lstlisting}

\subsubsection{Approach 1: individual-level simulation (eq.~1)}

Conditional inputs per individual, individual traces, outcomes averaged
over the target population (marginalize late).

\begin{lstlisting}[language=R]
run_individual_sim <- function(pop) {
  results <- list()
  for (strategy in 0:1) {
    qalys_all <- costs_all <- numeric(nrow(pop))
    for (i in 1:nrow(pop)) {
      prog_i <- compute_prog(pop$age[i], pop$ecog1[i], strategy)
      mr_s_i <- compute_mr_stable(pop$age[i], trt = strategy)
      mr_p_i <- compute_mr_prog(pop$age[i])
      trace_i <- run_cohort_trace(prog_i, mr_s_i, mr_p_i)
      out_i <- compute_outcomes(trace_i, strategy)
      qalys_all[i] <- out_i$qalys
      costs_all[i] <- out_i$costs
    }
    results[[strategy + 1]] <- data.table(
      strategy = ifelse(strategy == 0, "SoC", "New"),
      qalys = weighted.mean(qalys_all, pop$patient_wt),
      costs = weighted.mean(costs_all, pop$patient_wt))
  }
  rbindlist(results)
}

res1_A <- run_individual_sim(pop_A)
res1_B <- run_individual_sim(pop_B)
\end{lstlisting}

\subsubsection{Approach 2: cohort-model with marginal inputs (eq.~3)}

Conditional survival averaged over the target covariate distribution and
differenced into per-cycle inputs (eq.~4 on the cumulative risk scale), then
a single cohort trace.

\begin{lstlisting}[language=R]
run_cohort_marginal <- function(pop) {
  results <- list()
  for (strategy in 0:1) {
    inp <- get_marginal_inputs(pop, strategy)
    trace <- run_cohort_trace(inp$prog, inp$mr_s, inp$mr_p)
    out <- compute_outcomes(trace, strategy)
    results[[strategy + 1]] <- data.table(
      strategy = ifelse(strategy == 0, "SoC", "New"),
      qalys = out$qalys, costs = out$costs)
  }
  rbindlist(results)
}

res2_A <- run_cohort_marginal(pop_A)
res2_B <- run_cohort_marginal(pop_B)
\end{lstlisting}

\subsubsection{Approach 3: cohort with population-average conditional inputs (eq.~6)}

The conditional model evaluated at the population's mean covariate
values, equivalent to the population-average conditional hazard ratio because the linear predictor is assumed to vary linearly with the only effect modifier. 

\begin{lstlisting}[language=R]
run_cohort_avg_conditional <- function(pop) {
  mean_age <- weighted.mean(pop$age, pop$patient_wt)
  prop_ecog1 <- sum(pop$patient_wt[pop$ecog1 == 1])
  results <- list()
  for (strategy in 0:1) {
    prog_avg <- compute_prog(mean_age, prop_ecog1, strategy)
    mr_s_avg <- compute_mr_stable(mean_age, trt = strategy)
    mr_p_avg <- compute_mr_prog(mean_age)
    trace <- run_cohort_trace(prog_avg, mr_s_avg, mr_p_avg)
    out <- compute_outcomes(trace, strategy)
    results[[strategy + 1]] <- data.table(
      strategy = ifelse(strategy == 0, "SoC", "New"),
      qalys = out$qalys, costs = out$costs)
  }
  rbindlist(results)
}

res3_A <- run_cohort_avg_conditional(pop_A)
res3_B <- run_cohort_avg_conditional(pop_B)
\end{lstlisting}

\subsubsection{Approach 4: cohort with mixed inputs (marginal baseline, conditional effect) (eq.~8)}

The population-average conditional hazard ratio \(\exp(d_k)\) (derived from the
conditional model evaluated at the population's mean covariates) applied to the marginal SoC baseline.

\begin{lstlisting}[language=R]
run_cohort_mixed <- function(pop) {
  # Step 1: Marginal baseline inputs from target population (SoC)
  inp_soc <- get_marginal_inputs(pop, strategy = 0)

  # Step 2: Apply the population-average conditional (at-mean) HR to the marginal baseline
  # Progression: conditional HR evaluated at the population's mean covariate (d_k, eq. 5)
  pe <- sum(pop$patient_wt[pop$ecog1 == 1])
  hr_prog_conditional <- exp(prog_coef["trt"] + prog_coef["trt_ecog1"] * pe)
  h_prog_soc <- -log(1 - pmin(inp_soc$prog, 0.999))
  prog_mixed_new <- 1 - exp(-h_prog_soc * hr_prog_conditional)

  # Mortality: conditional HR
  h_mr_soc <- -log(1 - pmin(inp_soc$mr_s, 0.999))
  mr_s_mixed_new <- 1 - exp(-h_mr_soc * hr_death_trt)

  results <- list()
  trace_soc <- run_cohort_trace(inp_soc$prog, inp_soc$mr_s, inp_soc$mr_p)
  results[[1]] <- data.table(strategy = "SoC",
    qalys = compute_outcomes(trace_soc, 0)$qalys,
    costs = compute_outcomes(trace_soc, 0)$costs)
  trace_new <- run_cohort_trace(prog_mixed_new, mr_s_mixed_new, inp_soc$mr_p)
  results[[2]] <- data.table(strategy = "New",
    qalys = compute_outcomes(trace_new, 1)$qalys,
    costs = compute_outcomes(trace_new, 1)$costs)
  rbindlist(results)
}

res4_A <- run_cohort_mixed(pop_A)
res4_B <- run_cohort_mixed(pop_B)
\end{lstlisting}

\subsubsection{Approach 5: cohort with reverse mixed inputs (conditional baseline, marginal effect) (eq.~9)}

The marginal hazard ratio applied to the population-average
conditional SoC baseline (the mirror image of Approach 4).

\begin{lstlisting}[language=R]
run_cohort_mixed_rev <- function(pop) {
  # Baseline: population-average conditional inputs (model at mean covariates), SoC
  mean_age <- weighted.mean(pop$age, pop$patient_wt)
  prop_ecog1 <- sum(pop$patient_wt[pop$ecog1 == 1])
  prog_soc <- compute_prog(mean_age, prop_ecog1, 0)
  mr_s_soc <- compute_mr_stable(mean_age, trt = 0)
  mr_p_soc <- compute_mr_prog(mean_age)

  # Treatment effect: per-cycle marginal hazard ratio (from marginal SoC/New inputs)
  m0 <- get_marginal_inputs(pop, strategy = 0)
  m1 <- get_marginal_inputs(pop, strategy = 1)
  hr_prog <- -log(1 - pmin(m1$prog, 0.999)) / -log(1 - pmin(m0$prog, 0.999))
  hr_mr_s <- -log(1 - pmin(m1$mr_s, 0.999)) / -log(1 - pmin(m0$mr_s, 0.999))

  # New arm: apply the marginal HR to the conditional baseline hazards
  prog_new <- 1 - exp(-(-log(1 - pmin(prog_soc, 0.999))) * hr_prog)
  mr_s_new <- 1 - exp(-(-log(1 - pmin(mr_s_soc, 0.999))) * hr_mr_s)

  trace_soc <- run_cohort_trace(prog_soc, mr_s_soc, mr_p_soc)
  trace_new <- run_cohort_trace(prog_new, mr_s_new, mr_p_soc)
  rbindlist(list(
    data.table(strategy = "SoC", qalys = compute_outcomes(trace_soc, 0)$qalys,
               costs = compute_outcomes(trace_soc, 0)$costs),
    data.table(strategy = "New", qalys = compute_outcomes(trace_new, 1)$qalys,
               costs = compute_outcomes(trace_new, 1)$costs)))
}

res5_A <- run_cohort_mixed_rev(pop_A)
res5_B <- run_cohort_mixed_rev(pop_B)
\end{lstlisting}

\subsubsection{Cost-effectiveness results}

Incremental QALYs, incremental costs, and the ICER (new treatment vs.\ SoC) for each scenario and
target population.

\begin{lstlisting}[language=R]
incr <- function(res) data.table(
  inc_qalys = res$qalys[2] - res$qalys[1],
  inc_costs = res$costs[2] - res$costs[1],
  icer      = (res$costs[2] - res$costs[1]) / (res$qalys[2] - res$qalys[1]))

ce_summary <- rbindlist(list(
  cbind(scenario = "1 individual",   population = "A", incr(res1_A)),
  cbind(scenario = "1 individual",   population = "B", incr(res1_B)),
  cbind(scenario = "2 marginal",     population = "A", incr(res2_A)),
  cbind(scenario = "2 marginal",     population = "B", incr(res2_B)),
  cbind(scenario = "3 avg-cond",     population = "A", incr(res3_A)),
  cbind(scenario = "3 avg-cond",     population = "B", incr(res3_B)),
  cbind(scenario = "4 mixed (marg)", population = "A", incr(res4_A)),
  cbind(scenario = "4 mixed (marg)", population = "B", incr(res4_B)),
  cbind(scenario = "5 mixed (cond)", population = "A", incr(res5_A)),
  cbind(scenario = "5 mixed (cond)", population = "B", incr(res5_B))))
ce_summary[, `:=`(inc_qalys = round(inc_qalys, 3),
                  inc_costs = round(inc_costs), icer = round(icer))]
ce_summary[]
\end{lstlisting}

\begin{lstlisting}
##           scenario population inc_qalys inc_costs   icer
##             <char>     <char>     <num>     <num>  <num>
##  1:   1 individual          A     0.781     73538  94164
##  2:   1 individual          B     0.275     24167  87777
##  3:     2 marginal          A     0.804     74277  92426
##  4:     2 marginal          B     0.297     24903  83991
##  5:     3 avg-cond          A     0.663     64026  96639
##  6:     3 avg-cond          B     0.208     19107  91880
##  7: 4 mixed (marg)          A     0.963     82744  85944
##  8: 4 mixed (marg)          B     0.414     30902  74632
##  9: 5 mixed (cond)          A     0.554     58196 104975
## 10: 5 mixed (cond)          B     0.155     16339 105580
\end{lstlisting}

\clearpage
\subsection{Discrete-time state-transition models in \texttt{hesim}}
\label{app:hesim}
This code implements the oncology example in the
\passthrough{\lstinline!hesim!} package. Note here, the \passthrough{\lstinline!hesim!} \passthrough{\lstinline!CohortDtstm!} functions (purposefully designed for cohort discrete-time state-transition models) are also used for the individual-level simulation model to capture the discrete-time transitions of the individual model in the example. In the next section, the hesim fucntions specifically designed for individual-level simulations with continuous time are used.

\subsubsection{Setup}

\begin{lstlisting}[language=R]
library("hesim")
library("data.table")
library("ggplot2")
library("kableExtra")
theme_set(theme_bw())
\end{lstlisting}

\subsubsection{Model parameters}

Conditional Weibull PH progression coefficients, background mortality,
utilities, costs, and model settings.

\begin{lstlisting}[language=R]
# Weibull PH coefficients for progression (Stable -> Progressed)
prog_coef <- c(
  lngamma   = 0.15,   # ln(shape); shape ~ 1.16
  cons      = -5.5,   # intercept (log scale)
  age       = 0.08,   # prognostic: HR = 1.08 / year
  ecog1     = 1.10,   # prognostic: HR = 3.00 for ECOG 1
  trt       = -1.10,  # treatment effect at ECOG 0 (conditional HR = 0.33)
  trt_ecog1 = 0.45    # interaction: weaker effect in ECOG 1 (HR = 0.52)
)
hr_death_trt <- 0.45  # conditional treatment HR for mortality while Stable

# Background mortality (annual rate by age band)
mort_annual <- data.table(
  age_lower = c(50, 55, 60, 65, 70, 75, 80),
  age_upper = c(55, 60, 65, 70, 75, 80, Inf),
  rate      = c(0.004, 0.007, 0.012, 0.020, 0.035, 0.060, 0.100)
)
mr_prog_multiplier <- 3.0     # post-progression mortality = 3x background

# Utilities and costs
u_stable <- 0.75; u_progressed <- 0.45
c_soc_drug <- 1000; c_new_drug <- 3000; c_prog_care <- 1500

# Model settings
n_cycles <- 360                # 30 years, monthly cycles
cycle_length <- 1/12           # cycle length in years
dr_annual <- 0.035             # annual discount rate (QALYs and costs)
dr_m <- (1 + dr_annual)^(1/12) - 1   # equivalent monthly discount rate
\end{lstlisting}

\subsubsubsection{A note on time units}

hesim's cDTSTM measures time in model cycles. Because we use
\textbf{monthly} cycles, we (i) discount with the equivalent
\textbf{monthly} rate \passthrough{\lstinline!dr\_m!}, and (ii) express
utilities per cycle by scaling annual utilities by
\passthrough{\lstinline!cycle\_length!} so that QALYs accrue in
life-years. Monthly costs need no scaling. These conventions make the
hesim output directly comparable to the hand-coded supplement.

\subsubsection{Two populations}

Population A (trial: younger, 30\% ECOG 1) and Population B (target:
older, 70\% ECOG 1) as hesim patient tables.

\begin{lstlisting}[language=R]
# Population A: trial population (younger, less ECOG 1)
ages_A <- 50:70
age_dens_A <- dbeta((ages_A - 50) / 20, 3, 3)
prop_ecog1_A <- 0.30
pop_A <- data.table(
  age = rep(ages_A, 2),
  ecog1 = rep(c(0, 1), each = length(ages_A)),
  patient_wt = c((1 - prop_ecog1_A) * age_dens_A / sum(age_dens_A),
                 prop_ecog1_A * age_dens_A / sum(age_dens_A))
)

# Population B: target population (older, more ECOG 1)
ages_B <- 50:80
age_dens_B <- dbeta((ages_B - 50) / 30, 5, 2)
prop_ecog1_B <- 0.70
pop_B <- data.table(
  age = rep(ages_B, 2),
  ecog1 = rep(c(0, 1), each = length(ages_B)),
  patient_wt = c((1 - prop_ecog1_B) * age_dens_B / sum(age_dens_B),
                 prop_ecog1_B * age_dens_B / sum(age_dens_B))
)
\end{lstlisting}

\subsubsection{Conditional statistical model (input
builders)}

Functions that turn the conditional coefficients into monthly transition
probabilities.

\begin{lstlisting}[language=R]
# Monthly progression probability from the Weibull PH model
prog_prob <- function(age, ecog1, trt, t) {
  shape <- exp(prog_coef["lngamma"])
  scale <- exp(prog_coef["cons"] + prog_coef["age"] * age +
               prog_coef["ecog1"] * ecog1 + prog_coef["trt"] * trt +
               prog_coef["trt_ecog1"] * trt * ecog1)
  tp <- 1 - exp(scale * (((t - 1) / 12)^shape - (t / 12)^shape))
  pmin(pmax(tp, 0), 0.999)
}

# Age-band background annual mortality rate
mr_rate_age <- function(age_now) {
  r <- rep(mort_annual$rate[nrow(mort_annual)], length(age_now))
  for (i in seq_len(nrow(mort_annual))) {
    r[age_now >= mort_annual$age_lower[i] & age_now < mort_annual$age_upper[i]] <-
      mort_annual$rate[i]
  }
  r
}

# Monthly mortality while Stable (treatment applies hr_death_trt)
mr_stable_prob <- function(age, trt, t) {
  rate <- mr_rate_age(age + t / 12) * ifelse(trt == 1, hr_death_trt, 1)
  1 - exp(-rate / 12)
}

# Monthly mortality after progression (elevated, no treatment effect)
mr_prog_prob <- function(age, t) {
  base <- 1 - exp(-mr_rate_age(age + t / 12) / 12)   # trt = 0
  1 - (1 - base)^mr_prog_multiplier
}
\end{lstlisting}

\subsubsection{hesim scaffolding}

Helpers that build a \passthrough{\lstinline!CohortDtstm!} from a
patient table + transition-probability vectors and summarize it into
per-strategy QALYs and costs.

\begin{lstlisting}[language=R]
strategies <- data.table(strategy_id = 1:2, strategy_name = c("SoC", "New"))
states <- data.table(state_id = 1:2, state_name = c("Stable", "Progressed"))

# Build a CohortDtstm from patient table + aligned transition-prob vectors
build_econ <- function(patients_dt, prog_v, mr_s_v, mr_p_v, tpdata) {
  tpmat <- tpmatrix(
    C, prog_v, mr_s_v,
    0, C,      mr_p_v,
    0, 0,      1)
  transmod <- CohortDtstmTrans$new(
    params = tparams_transprobs(tpmat, tpmatrix_id(tpdata, 1)))

  hd <- hesim_data(strategies = strategies, patients = patients_dt, states = states)
  # Utilities scaled by cycle_length so QALYs accrue in life-years
  u_tbl <- stateval_tbl(
    data.table(state_id = 1:2, est = c(u_stable, u_progressed) * cycle_length),
    dist = "fixed")
  # Drug cost accrues while Stable and differs by strategy
  cdrug_tbl <- stateval_tbl(
    data.table(strategy_id = rep(1:2, each = 2), state_id = rep(1:2, 2),
               est = c(c_soc_drug, 0, c_new_drug, 0)), dist = "fixed")
  # Post-progression care accrues while Progressed
  ccare_tbl <- stateval_tbl(
    data.table(state_id = 1:2, est = c(0, c_prog_care)), dist = "fixed")

  CohortDtstm$new(
    trans_model   = transmod,
    utility_model = create_StateVals(u_tbl, hesim_data = hd, n = 1),
    cost_models   = list(drug = create_StateVals(cdrug_tbl, hesim_data = hd, n = 1),
                         care = create_StateVals(ccare_tbl, hesim_data = hd, n = 1)))
}

# Simulate and summarize a model into per-strategy QALYs and costs
run_ce <- function(econ, by_grp = FALSE) {
  econ$sim_stateprobs(n_cycles = n_cycles)
  econ$sim_qalys(dr = dr_m, integrate_method = "riemann_right")
  econ$sim_costs(dr = dr_m, integrate_method = "riemann_right")
  ce <- econ$summarize(by_grp = by_grp)
  q   <- ce$qalys[, .(qalys = sum(qalys)), by = strategy_id]
  cst <- ce$costs[category == "total", .(costs = sum(costs)), by = strategy_id]
  merge(q, cst, by = "strategy_id")[order(strategy_id)]
}

incr <- function(res) data.table(
  inc_qalys = res$qalys[2] - res$qalys[1],
  inc_costs = res$costs[2] - res$costs[1],
  icer = (res$costs[2] - res$costs[1]) / (res$qalys[2] - res$qalys[1]))

# Expanded (strategy x patient x time) grid; time = cycle index used by the builders
single_tpdata <- function(patients_dt) {
  hd <- hesim_data(strategies = strategies, patients = patients_dt, states = states)
  tpd <- expand(hd, by = c("strategies", "patients"), times = 1:n_cycles)
  tpd[, time := time_start + 1][]
}

# Single representative cohort (weight 1) and per-cycle time index for the cohort approaches
pat_single <- data.table(patient_id = 1, grp_id = 1, patient_wt = 1)
tpd_s <- single_tpdata(pat_single)   # SoC rows (time 1..n+1), then New rows
tt <- 1:(n_cycles + 1)               # hesim uses n_cycles+1 intervals per strategy

# Population-marginal per-cycle inputs: eq. 4 applied to the cumulative risk
marg_inputs <- function(pop, trt) {
  w <- pop$patient_wt
  # Average the SURVIVAL function over the covariate distribution and difference
  # it (eq. 4 on the cumulative risk). Averaging each cycle's transition
  # probability with fixed weights instead keeps the baseline covariate mix for
  # the whole horizon; the two agree only in the first cycle.
  dif <- function(M) {
    S  <- t(apply(1 - M, 1, cumprod))
    Sb <- colSums(w * S)
    1 - Sb / c(1, Sb[-length(Sb)])
  }
  list(
    prog = dif(t(sapply(seq_len(nrow(pop)), function(i)
             prog_prob(pop$age[i], pop$ecog1[i], trt, tt)))),
    mrs  = dif(t(sapply(seq_len(nrow(pop)), function(i)
             mr_stable_prob(pop$age[i], trt, tt)))),
    mrp  = dif(t(sapply(seq_len(nrow(pop)), function(i)
             mr_prog_prob(pop$age[i], tt))))
  )
}
\end{lstlisting}

\subsubsection{Approach 1: individual-level simulation (eq.~1)}

One patient profile per covariate combination; hesim averages outcomes
over the population last (marginalize late).

\begin{lstlisting}[language=R]
run_individual <- function(pop) {
  pat <- copy(pop)[, `:=`(patient_id = .I, grp_id = .I)]
  tpd <- single_tpdata(pat)
  trt <- ifelse(tpd$strategy_name == "New", 1, 0)
  econ <- build_econ(
    pat,
    prog_prob(tpd$age, tpd$ecog1, trt, tpd$time),
    mr_stable_prob(tpd$age, trt, tpd$time),
    mr_prog_prob(tpd$age, tpd$time),
    tpd)
  run_ce(econ, by_grp = FALSE)
}
res1_A <- run_individual(pop_A)
res1_B <- run_individual(pop_B)
\end{lstlisting}

\subsubsection{Approach 2: cohort-model with marginal inputs (eq.~3)}

Conditional survival averaged over the target covariate distribution and
differenced into per-cycle inputs (eq.~4 on the cumulative risk scale), then
a single cohort trace.

\begin{lstlisting}[language=R]
run_marginal <- function(pop) {
  m_soc <- marg_inputs(pop, 0)
  m_new <- marg_inputs(pop, 1)
  econ <- build_econ(pat_single,
    c(m_soc$prog, m_new$prog), c(m_soc$mrs, m_new$mrs), c(m_soc$mrp, m_new$mrp), tpd_s)
  run_ce(econ)
}
res2_A <- run_marginal(pop_A)
res2_B <- run_marginal(pop_B)
\end{lstlisting}

\subsubsection{Approach 3: cohort with population-average conditional inputs (eq.~6)}

\begin{lstlisting}[language=R]
run_avg_conditional <- function(pop) {
  mean_age <- weighted.mean(pop$age, pop$patient_wt)
  prop_e1  <- sum(pop$patient_wt[pop$ecog1 == 1])
  econ <- build_econ(pat_single,
    c(prog_prob(mean_age, prop_e1, 0, tt), prog_prob(mean_age, prop_e1, 1, tt)),
    c(mr_stable_prob(mean_age, 0, tt),     mr_stable_prob(mean_age, 1, tt)),
    c(mr_prog_prob(mean_age, tt),          mr_prog_prob(mean_age, tt)),
    tpd_s)
  run_ce(econ)
}
res3_A <- run_avg_conditional(pop_A)
res3_B <- run_avg_conditional(pop_B)
\end{lstlisting}

\subsubsection{Approach 4: cohort with mixed inputs (marginal baseline, conditional effect) (eq.~8)}

The population-average conditional hazard ratio applied to the marginal SoC baseline. 

\begin{lstlisting}[language=R]
run_mixed <- function(pop) {
  m_soc <- marg_inputs(pop, 0)
  prop_e1 <- sum(pop$patient_wt[pop$ecog1 == 1])
  hr_prog_cond <- exp(prog_coef["trt"] + prog_coef["trt_ecog1"] * prop_e1)  # at-mean conditional HR (d_k)
  h_prog_soc <- -log(1 - pmin(m_soc$prog, 0.999))       # marginal SoC hazards
  h_mr_soc   <- -log(1 - pmin(m_soc$mrs,  0.999))
  prog_new <- 1 - exp(-h_prog_soc * hr_prog_cond)
  mrs_new  <- 1 - exp(-h_mr_soc * hr_death_trt)
  econ <- build_econ(pat_single,
    c(m_soc$prog, prog_new), c(m_soc$mrs, mrs_new), c(m_soc$mrp, m_soc$mrp), tpd_s)
  run_ce(econ)
}
res4_A <- run_mixed(pop_A)
res4_B <- run_mixed(pop_B)
\end{lstlisting}

\subsubsection{Approach 5: cohort with reverse mixed inputs (conditional baseline, marginal effect) (eq.~9)}

The marginal hazard ratio applied to the population-average
conditional SoC baseline (the mirror image of Approach 4).

\begin{lstlisting}[language=R]
run_mixed_rev <- function(pop) {
  # Baseline: population-average conditional (at-mean) inputs, SoC arm
  mean_age <- weighted.mean(pop$age, pop$patient_wt)
  prop_e1  <- sum(pop$patient_wt[pop$ecog1 == 1])
  prog_soc <- prog_prob(mean_age, prop_e1, 0, tt)
  mrs_soc  <- mr_stable_prob(mean_age, 0, tt)
  mrp_soc  <- mr_prog_prob(mean_age, tt)
  # Treatment effect: per-cycle population-marginal hazard ratio (from marginal inputs)
  m_soc <- marg_inputs(pop, 0)
  m_new <- marg_inputs(pop, 1)
  hr_prog <- -log(1 - pmin(m_new$prog, 0.999)) / -log(1 - pmin(m_soc$prog, 0.999))
  hr_mrs  <- -log(1 - pmin(m_new$mrs,  0.999)) / -log(1 - pmin(m_soc$mrs,  0.999))
  # New arm: apply the marginal HR to the conditional baseline hazards
  prog_new <- 1 - exp(-(-log(1 - pmin(prog_soc, 0.999))) * hr_prog)
  mrs_new  <- 1 - exp(-(-log(1 - pmin(mrs_soc,  0.999))) * hr_mrs)
  econ <- build_econ(pat_single,
    c(prog_soc, prog_new), c(mrs_soc, mrs_new), c(mrp_soc, mrp_soc), tpd_s)
  run_ce(econ)
}
res5_A <- run_mixed_rev(pop_A)
res5_B <- run_mixed_rev(pop_B)
\end{lstlisting}

\subsubsection{Cost-effectiveness results}

Incremental QALYs, incremental costs, and the ICER (new treatment vs.\ SoC) for each scenario and
target population, reusing the \texttt{incr()} helper defined in the scaffolding.

\begin{lstlisting}[language=R]
ce_summary <- rbindlist(list(
  cbind(scenario = "1 individual",   population = "A", incr(res1_A)),
  cbind(scenario = "1 individual",   population = "B", incr(res1_B)),
  cbind(scenario = "2 marginal",     population = "A", incr(res2_A)),
  cbind(scenario = "2 marginal",     population = "B", incr(res2_B)),
  cbind(scenario = "3 avg-cond",     population = "A", incr(res3_A)),
  cbind(scenario = "3 avg-cond",     population = "B", incr(res3_B)),
  cbind(scenario = "4 mixed (marg)", population = "A", incr(res4_A)),
  cbind(scenario = "4 mixed (marg)", population = "B", incr(res4_B)),
  cbind(scenario = "5 mixed (cond)", population = "A", incr(res5_A)),
  cbind(scenario = "5 mixed (cond)", population = "B", incr(res5_B))))
ce_summary[, `:=`(inc_qalys = round(inc_qalys, 3),
                  inc_costs = round(inc_costs), icer = round(icer))]
ce_summary[]
\end{lstlisting}

\begin{lstlisting}
##           scenario population inc_qalys inc_costs   icer
##             <char>     <char>     <num>     <num>  <num>
##  1:   1 individual          A     0.781     73522  94176
##  2:   1 individual          B     0.275     24164  87784
##  3:     2 marginal          A     0.803     74261  92438
##  4:     2 marginal          B     0.296     24900  83997
##  5:     3 avg-cond          A     0.662     64016  96650
##  6:     3 avg-cond          B     0.208     19105  91886
##  7: 4 mixed (marg)          A     0.962     82725  85955
##  8: 4 mixed (marg)          B     0.414     30898  74638
##  9: 5 mixed (cond)          A     0.554     58188 104987
## 10: 5 mixed (cond)          B     0.155     16338 105587
\end{lstlisting}

\clearpage
\subsection{Individual-level simulation in \texttt{hesim} (IndivCtstm)}
\label{app:hesim-indiv}
The code in this section implements the oncology example as an individual-level,
continuous-time model in the \passthrough{\lstinline!hesim!} package
(eq.~1, the reference approach).

\subsubsection{Setup}

\begin{lstlisting}[language=R]
library("hesim")
library("data.table")
library("flexsurv")
library("ggplot2")
library("kableExtra")
theme_set(theme_bw())
set.seed(123)
\end{lstlisting}

\subsubsection{The clinical model}

The conditional model is the same as in the other supplements. Two
things are adapted for continuous time:

\begin{itemize}
\tightlist
\item
  \textbf{Progression (Stable \(\rightarrow\) Progressed)} uses the
  identical Weibull proportional-hazards model with covariates for age,
  ECOG, treatment, and the treatment-by-ECOG interaction. Age and ECOG
  are prognostic; ECOG is also an effect modifier (conditional
  progression HR = 0.33 at ECOG 0 vs 0.52 at ECOG 1).
\item
  \textbf{Background mortality} is specified in continuous time. The
  age-banded life-table rates used in the discrete supplements are
  almost perfectly log-linear in age, i.e., a \textbf{Gompertz} hazard
  \(h(\text{age}) = e^{a + b\cdot \text{age}}\) (fit below,
  \(R^2 > 0.999\)). This lets a single continuous hazard capture
  mortality that rises as patients age. Treatment reduces mortality
  while Stable (conditional HR = 0.45); post-progression mortality is 3x
  background.
\end{itemize}

\begin{lstlisting}[language=R]
# Weibull PH progression coefficients (identical to other supplements)
prog_coef <- c(lngamma = 0.15, cons = -5.5, age = 0.08,
               ecog1 = 1.10, trt = -1.10, trt_ecog1 = 0.45)
hr_death_trt <- 0.45          # treatment HR for mortality while Stable
mr_prog_multiplier <- 3.0     # post-progression mortality multiplier

# Gompertz fit to the age-banded background mortality rates
mort_bands <- data.table(age = c(50,55,60,65,70,75,80),
                         rate = c(.004,.007,.012,.020,.035,.060,.100))
gomp_fit <- lm(log(rate) ~ age, data = mort_bands)
mort_a <- unname(coef(gomp_fit)[1]); mort_b <- unname(coef(gomp_fit)[2])
c(intercept = mort_a, slope = mort_b, R2 = summary(gomp_fit)$r.squared)
\end{lstlisting}

\begin{lstlisting}
##   intercept       slope          R2 
## -10.8734826   0.1073138   0.9998727
\end{lstlisting}

\begin{lstlisting}[language=R]
# Utilities and costs (costs are monthly in the other supplements -> annualize)
u_stable <- 0.75; u_progressed <- 0.45
c_soc_drug_yr  <- 1000 * 12
c_new_drug_yr  <- 3000 * 12
c_prog_care_yr <- 1500 * 12
dr <- 0.035                   # annual discount rate (QALYs and costs)
\end{lstlisting}

\subsubsection{Target population as
individuals}

The defining feature of an individual-level simulation is that the
target population is an explicit set of \textbf{individual patients}. We
sample both target populations (A and B) from their covariate
distributions; adapting the analysis to a different population requires
only re-sampling this table and re-running the model.

\begin{lstlisting}[language=R]
n_patients <- 20000
sample_pop <- function(n, ages, age_prob, p_ecog1) {
  data.table(patient_id = 1:n,
             age   = sample(ages, n, replace = TRUE, prob = age_prob),
             ecog1 = rbinom(n, 1, p_ecog1))
}

# Population A (trial: younger, 30% ECOG 1)
ages_A <- 50:70
age_dens_A <- dbeta((ages_A - 50) / 20, 3, 3); age_dens_A <- age_dens_A / sum(age_dens_A)
# Population B (target: older, 70% ECOG 1)
ages_B <- 50:80
age_dens_B <- dbeta((ages_B - 50) / 30, 5, 2); age_dens_B <- age_dens_B / sum(age_dens_B)

set.seed(123)
patients_A <- sample_pop(n_patients, ages_A, age_dens_A, 0.30)
patients_B <- sample_pop(n_patients, ages_B, age_dens_B, 0.70)
\end{lstlisting}

\subsubsection{Building the individual-level
model}

\subsubsubsection{Scaffolding and transition
structure}

\begin{lstlisting}[language=R]
strategies <- data.table(strategy_id = 1:2, strategy_name = c("SoC", "New"))
states <- data.table(state_id = 1:2, state_name = c("Stable", "Progressed"))
n_samples <- 1   # deterministic run (no PSA); see note at the end

# Transition matrix: 1 = Stable, 2 = Progressed, 3 = Death (absorbing)
tmat <- rbind(c(NA, 1, 2),
              c(NA, NA, 3),
              c(NA, NA, NA))
colnames(tmat) <- rownames(tmat) <- c("Stable", "Progressed", "Death")
\end{lstlisting}

\subsubsubsection{Multi-state parametric
model}

Each transition is a survival model. Progression is Weibull PH; the two
mortality transitions are Gompertz, with the log-rate linear in age
(slope = the Gompertz shape) so that a patient starting at a given age
experiences an age-increasing hazard over follow-up. The treatment HR
enters transition 2 (Stable death) only.

\begin{lstlisting}[language=R]
# helper: repeat a one-row coefficient table for each PSA sample
cf <- function(...) { d <- data.table(...); d[rep(1, n_samples)] }

transmod_params <- params_surv_list(
  # 1. Stable -> Progressed : Weibull PH
  params_surv(
    coefs = list(
      shape = cf(cons = prog_coef["lngamma"]),
      scale = cf(cons = prog_coef["cons"], age = prog_coef["age"],
                 ecog1 = prog_coef["ecog1"], trt = prog_coef["trt"],
                 trt_ecog1 = prog_coef["trt_ecog1"])),
    dist = "weibullPH"),

  # 2. Stable -> Death : Gompertz background mortality (+ treatment HR)
  params_surv(
    coefs = list(
      shape = cf(cons = mort_b),
      rate  = cf(cons = mort_a, age = mort_b, trt = log(hr_death_trt))),
    dist = "gompertz"),

  # 3. Progressed -> Death : Gompertz x 3 (no treatment effect)
  params_surv(
    coefs = list(
      shape = cf(cons = mort_b),
      rate  = cf(cons = mort_a + log(mr_prog_multiplier), age = mort_b)),
    dist = "gompertz")
)
\end{lstlisting}

\subsubsubsection{Utility and cost models}

\begin{lstlisting}[language=R]
utility_tbl <- stateval_tbl(
  data.table(state_id = 1:2, est = c(u_stable, u_progressed)), dist = "fixed")
drugcost_tbl <- stateval_tbl(
  data.table(strategy_id = rep(1:2, each = 2), state_id = rep(1:2, 2),
             est = c(c_soc_drug_yr, 0, c_new_drug_yr, 0)), dist = "fixed")
carecost_tbl <- stateval_tbl(
  data.table(state_id = 1:2, est = c(0, c_prog_care_yr)), dist = "fixed")
\end{lstlisting}

\subsubsection{Assembling and simulating for both
populations}

For a given patient table, \passthrough{\lstinline!run\_indiv()!}
expands the covariate data, assembles the transition, utility, and cost
models, and simulates. \passthrough{\lstinline!sim\_disease()!}
simulates a unique trajectory for each patient x strategy;
\passthrough{\lstinline!sim\_qalys()!}/\passthrough{\lstinline!sim\_costs()!}
accrue discounted outcomes; \passthrough{\lstinline!summarize()!}
averages over the population (the marginalization step). We run it for
both target populations.

\begin{lstlisting}[language=R]
run_indiv <- function(patients) {
  hesim_dat <- hesim_data(strategies = strategies, patients = patients, states = states)

  # Input data with covariates (one row per strategy x patient)
  transmod_data <- expand(hesim_dat, by = c("strategies", "patients"))
  transmod_data[, `:=`(cons = 1, trt = ifelse(strategy_name == "New", 1, 0))]
  transmod_data[, trt_ecog1 := trt * ecog1]

  # Transition, utility, and cost models
  transmod <- create_IndivCtstmTrans(
    transmod_params, input_data = transmod_data, trans_mat = tmat,
    clock = "forward", start_age = patients$age)
  utilmod <- create_StateVals(utility_tbl, n = n_samples, hesim_data = hesim_dat)
  costmods <- list(
    drug = create_StateVals(drugcost_tbl, n = n_samples, method = "wlos", hesim_data = hesim_dat),
    care = create_StateVals(carecost_tbl, n = n_samples, method = "wlos", hesim_data = hesim_dat))
  econmod <- IndivCtstm$new(trans_model = transmod,
                            utility_model = utilmod, cost_models = costmods)

  # Simulate and marginalize over the population
  econmod$sim_disease(max_t = 30, max_age = 200)   # 30-year horizon
  econmod$sim_qalys(dr = dr)
  econmod$sim_costs(dr = dr)
  ce <- econmod$summarize()
  res <- merge(ce$qalys[, .(qalys = mean(qalys)), by = strategy_id],
               ce$costs[category == "total", .(costs = mean(costs)), by = strategy_id],
               by = "strategy_id")[order(strategy_id)]
  res[, strategy := c("SoC", "New")][]
}

res_A <- run_indiv(patients_A)
res_B <- run_indiv(patients_B)
\end{lstlisting}

\subsubsection{Cost-effectiveness results}

Incremental QALYs, incremental costs, and the ICER (new treatment vs.\ SoC) for each target
population.

\begin{lstlisting}[language=R]
incr <- function(res) data.table(
  inc_qalys = res$qalys[2] - res$qalys[1],
  inc_costs = res$costs[2] - res$costs[1],
  icer      = (res$costs[2] - res$costs[1]) / (res$qalys[2] - res$qalys[1]))

ce_summary <- rbindlist(list(
  cbind(population = "A", incr(res_A)),
  cbind(population = "B", incr(res_B))))
ce_summary[, `:=`(inc_qalys = round(inc_qalys, 3),
                  inc_costs = round(inc_costs), icer = round(icer))]
ce_summary[]
\end{lstlisting}

\begin{lstlisting}
##    population inc_qalys inc_costs  icer
##        <char>     <num>     <num> <num>
## 1:          A     0.783     74783 95496
## 2:          B     0.278     25318 90941
\end{lstlisting}

\end{document}